\documentclass[aps,prl,superscriptaddress]{revtex4-1}
\usepackage[latin9]{inputenc}
\usepackage{graphicx}
\usepackage{textcomp}
\usepackage{amsmath}
\usepackage{esint}
\usepackage[unicode=true,
 bookmarks=false,
 breaklinks=false,pdfborder={0 0 1},backref=section,colorlinks=false]
 {hyperref}
\usepackage{breakurl}

\usepackage{dcolumn}% Align table columns on decimal point
\usepackage{physics}
\usepackage{color}

\makeatletter

\DeclareFontEncoding{LGR}{}{}
\DeclareRobustCommand{\greektext}{%
  \fontencoding{LGR}\selectfont\def\encodingdefault{LGR}}
\DeclareRobustCommand{\textgreek}[1]{\leavevmode{\greektext #1}}
\ProvideTextCommand{\~}{LGR}[1]{\char126#1}

\usepackage{fixmath}
\usepackage{accents}

\allowdisplaybreaks

\makeatother

\begin{document}

\title{ Optical properties and electromagnetic modes of Weyl semimetals}

\author{ Qianfan Chen}

\affiliation{ Department of Physics and Astronomy, Texas A\&M University,
College Station, TX, 77843 USA}

\author{ A. Ryan Kutayiah}

\affiliation{ Department of Physics and Astronomy, Texas A\&M University,
College Station, TX, 77843 USA}

\author{ Ivan Oladyshkin}

\affiliation{ Institute of Applied Physics, Russian Academy of Sciences, Nizhny Novgorod, 603950, Russia }

\author{ Mikhail Tokman}

\affiliation{ Institute of Applied Physics, Russian Academy of Sciences, Nizhny Novgorod, 603950, Russia }

\author{ Alexey Belyanin}

\affiliation{ Department of Physics and Astronomy, Texas A\&M University,
College Station, TX, 77843 USA}

\begin{abstract}
We present systematic theoretical studies of both bulk and surface electromagnetic eigenmodes, or polaritons,  in  Weyl semimetals. We derive the tensors of bulk and surface conductivity taking into account all possible combinations of the optical transitions involving bulk and surface electron states.  We show how information about electronic structure of Weyl semimetals, such as position and separation of Weyl nodes, Fermi energy, and Fermi arc surface states, can be unambiguously extracted from measurements of the dispersion, transmission, reflection, and polarization of electromagnetic waves. 
\end{abstract}

\date{ \today}

\maketitle

\section{Introduction}

Weyl semimetals (WSMs) have attracted a lot of interest as a new class of gapless three-dimensional topological materials. Their Brillouin zone contains an even number of band-touching points, or Weyl nodes, that can be described by topological invariants defined as integrals over the two-dimensional Fermi surface. For each pair of Weyl nodes, these invariants  can be viewed as topological chiral charges of opposite sign of chirality \cite{volovik2007}. The electron dispersion near each Weyl node corresponds to three-dimensional massless Weyl fermions. For crystals with broken time-reversal or inversion symmetry (or both), the Weyl nodes of opposite chirality are separated in momentum space. The separation makes them stable against small perturbations and also gives rise to surface states with Fermi arcs. For reviews of WSMs discovered so far and their properties, see \cite{hosur2013, yan2017, hasan2017, armitage2018, burkov2018, kotov2018}. 

So far, the bulk of the research has been focused on measuring and modeling the electronic structure of WSMs and topological signatures in electron transport. However, it is becoming increasingly clear that optical studies in the terahertz to mid-infrared spectral regions (e.g. \cite{shushkov2015}) can provide a sensitive and sometimes more selective probe into the unique properties of these materials as compared to other methods. For a WSM in a magnetic field several proposals explored the signatures of the chiral anomaly in the interband optical absorption and plasmon mode properties; see e.g.~the calculations of the magnetooptical conductivity in the quasiclassical limit \cite{spivak2016, zhou2015,pell2015,tabert2016,ashby2014,ma2015,gorbar2017} and the quantum-mechanical theory in a strong magnetic field  \cite{ashby2013, long2018}. Note that these studies did not include finite separation of Weyl nodes in a microscopic Hamiltonian. 

Here we study electromagnetic eigenmodes of WSMs in the presence of finite separation between Weyl nodes in momentum space and without an external magnetic field.  To calculate the optical response, one needs to determine a realistic low-energy Hamiltonian that captures the essential topological
structure of WSMs. While many WSMs discovered in experiment  have a complicated
arrangement of several pairs of Weyl nodes, essential physics and
electronic properties of WSMs are already revealed in a model containing 
only two Weyl nodes separated in momentum space. Such models serve as a usual starting point for theoretical studies of transport and optical phenomena. Probably the simplest approach is to add a Zeeman-like constant shift term to the Hamiltonian for  a Dirac semimetal, which preserves the linear form of the Hamiltonian with respect to momentum operators \cite{chen2015}. The bulk optical conductivity for this model was calculated in \cite{tabert2016-2}.  In another approach, developed in \cite{zyuzin2012} and used in many optical response studies to date, a phenomenological axion $\theta$-term is introduced in the action for the electromagnetic field. This gives rise to the gyrotropic terms in the dielectric permittivity tensor and associated effects of Faraday and Kerr rotation, linear dichroism, modification of surface plasmon dispersion etc.; see e.g. \cite{kargarian2015, hofmann2016, ukhtary2017, kotov2018}. 

In yet another approach, Okugawa and Murakami \cite{okugawa2014} derived
a minimal 2x2 Hamiltonian (one conduction and one valence band)
containing one parameter which describes the transition from the normal
insulator to the WSM with two Weyl nodes separated in momentum space and eventually to a topological insulator
in the bulk. In the WSM phase, this Hamiltonian allows for surface state solutions
with Fermi arcs. Therefore, a single microscopic Hamiltonian
can be used to describe optical transitions between the bulk states,
surface states, and surface-to-bulk states. As a result, both bulk
and surface tensors of the optical conductivity can be derived. The Hamiltonian of \cite{okugawa2014} has been recently used to develop a quantum-mechanical theory of surface plasmons (Fermi arc plasmons) and their dissipation \cite{andolina2018}.  

Here we use a slightly more general Hamiltonian, which is free of certain surface state pathologies, to perform quantum-mechanical derivation of the tensors of both bulk and surface conductivity. We  take into account all possible combinations of  transitions between bulk and surface electron states. 
We then proceed to determine the properties of bulk and surface electromagnetic eigenmodes, or polaritons. 
We show how information about the electronic structure of WSMs, such as position and separation of Weyl nodes, Fermi energy, surface states, Fermi arcs,  etc. can be extracted from the transmission, dispersion, reflection, and polarization of electromagnetic modes. We identify the most sensitive optical signatures of the electronic properties of WSMs and discuss the potential use of WSM thin films for optoelectronic applications. 

Section II describes the effective Hamiltonian, or rather a family of Hamiltonians used in this study and derives the properties of corresponding bulk and surface electron states. Section III gives the classification of possible optical transitions and outlines all steps in the derivation of tensors of bulk and surface optical conductivity. The explicit expressions for the tensor elements are given in the Appendix. Section IV provides a detailed description of the electromagnetic normal modes (polaritons)  in bulk WSMs.  Section V provides boundary conditions which are then used in Sec. VI to calculate the reflection of incident radiation from the surface of a WSM. Section VII describes surface electromagnetic eigenmodes, i.e.~surface plasmon-polaritons. Conclusions are in Sec. VIII. The Appendix contains matrix elements of the current density operator, general expressions for elements of the bulk and surface conductivity tensor,  their low-frequency limit and the limit of small Weyl node separation.

\section{Effective Hamiltonian}

In this section we describe the family of Hamiltonians that serve as a microscopic basis in this study. We derive the properties of bulk and surface electron states and use them to calculate the optical conductivity. 
Consider a family of Hamiltonians of the type
\begin{equation}
\hat{H}=v_{F}\left(\frac{\hat{Q}^{2}-\hbar^{2}m(z)}{2\hbar b}\hat{\sigma}_{x}+\hat{p}_{y}\hat{\sigma}_{y}+\hat{p}_{z}\hat{\sigma}_{z}\right),\label{Eq:Hamiltonian}
\end{equation}
 where the function $m(z)$ takes
into account that the system may be nonuniform along $z$ and, in
particular, has boundaries. Here $\hat{\sigma}_{x,y,z}$ are Pauli matrices and the operator $\ensuremath{\hat{Q}^{2}}\textbf{ }$
is defined by one of the following three expressions:

\begin{eqnarray}
 (1) \;  \hat{Q}^{2} &=& \hat{p}_{x}^{2} 
\nonumber \\
(2) \; \hat{Q}^{2} &=& \hat{p}_{x}^{2}+\hat{p}_{y}^{2} \nonumber \\
(3) \;  \hat{Q}^{2} &=& \hat{p}_{x}^{2}+\hat{p}_{y}^{2}+\hat{p}_{z}^{2}  \nonumber
\end{eqnarray}
The first option is the original Hamiltonian in \cite{okugawa2014}.

 To make the derivation of surface states more convenient
\cite{okugawa2014}, we apply the unitary transformation $\hat{H}\Longrightarrow\hat{S}^{-1}\hat{H}\hat{S}$
to Eq.~(\ref{Eq:Hamiltonian}), where $\hat{S}=\frac{1}{\sqrt{2}}\left(1-i\hat{\sigma}_{x}\right).$
This gives
\begin{equation}
\hat{H}=v_{F}\left(\frac{\hat{Q}^{2}-\hbar^{2}m(z)}{2\hbar b}\hat{\sigma}_{x}+\hat{p}_{z}\hat{\sigma}_{y}-\hat{p}_{y}\hat{\sigma}_{z}\right),\label{Eq:Transformed Hamiltonian}
\end{equation}
One can check that this Hamiltonian violates time-reversal symmetry
due to the term proportional to $\hat{\sigma}_x$. The gyrotropy
axis is the $x$-axis. In $\boldsymbol{k}$-representation the Hamiltonian
of Eq.~(\ref{Eq:Transformed Hamiltonian}) becomes
\begin{equation}
\hat{H_{\boldsymbol{k}}}=\hbar v_{F}\left(K_{x}\left(\boldsymbol{k}\right)\hat{\sigma}_{x}+k_{z}\hat{\sigma}_{y}-k_{y}\hat{\sigma}_{z}\right),\label{Eq:Transformed Hamiltonian in k representation}
\end{equation}
where $K_{x}\left(\boldsymbol{k}\right)$ for the same three Hamiltonians
is given by

\begin{eqnarray}
 (1) \; K_{x} & =& \frac{k_{x}^{2}-m}{2b}   \nonumber \\
(2) \; K_{x} &=& \frac{k_{x}^{2}+k_{y}^{2}-m}{2b} \nonumber \\
(3) \; K_{x} &=& \frac{k_{x}^{2}+k_{y}^{2}+k_{z}^{2}-m}{2b}  \nonumber
\end{eqnarray}

In all three cases the Weyl nodes are located at $k_{x}=\pm\sqrt{m}$
assuming that $m>0$. We have found bulk and surface eigenstates for
all three Hamiltonians. Below is a summary of main results related
to electron states.
%%%%%%%%%%%%%%%%%%%%%%%%%%%%%%%

\subsection{Hamiltonians 1 and 2}

\subsubsection{Bulk states}

 The stationary spinor eigenstate of the Hamiltonian in Eq.~(\ref{Eq:Transformed Hamiltonian in k representation})
is
\begin{equation}
\left|\Psi_{\boldsymbol{k}}\right\rangle =\left(\begin{array}{c}
\Psi_{1}\\
\Psi_{2}
\end{array}\right)e^{i\boldsymbol{kr}-i\frac{E}{\hbar}t},\label{Eq:spinor eigenstate}
\end{equation}
where the components are determined from
\begin{equation}
\begin{pmatrix}-k_{y}-\frac{E}{\hbar v_{F}} & K_{x}\left(\boldsymbol{k}\right)-ik_{z}\\
K_{x}\left(\boldsymbol{k}\right)+ik_{z} & k_{y}-\frac{E}{\hbar v_{F}}
\end{pmatrix}\ensuremath{\left(\begin{array}{c}
\Psi_{1}\\
\Psi_{2}
\end{array}\right)=0.},\label{Eq:eigen-equation of bulk state in Hamiltonians 2 and 3}
\end{equation}
From Eq.~(\ref{Eq:eigen-equation of bulk state in Hamiltonians 2 and 3}) one can get the eigenenergy of the bulk states $E\left(\boldsymbol{k}\right)$
\begin{equation}
E=s\hbar v_{F}\sqrt{K_{x}^{2}+k_{y}^{2}+k_{z}^{2}},\label{Eq:eigenvalue}
\end{equation}
and corresponding components of the spinor eigenstate in Eq.~(\ref{Eq:spinor eigenstate}):
\begin{equation}
\left(\begin{array}{c}
\Psi_{1}\\
\Psi_{2}
\end{array}\right)=\frac{1}{\sqrt{2V}}\left(\begin{array}{c}
\sqrt{1-s\cos\theta_{\boldsymbol{k}}}e^{-i\phi_{\boldsymbol{k}}}\\
s\sqrt{1+s\cos\theta_{\boldsymbol{k}}}
\end{array}\right),\label{Eq:components of spinor eigenstate}
\end{equation}
where $\cos\theta_{\boldsymbol{k}}=\frac{k_{y}}{\sqrt{K_{x}^{2}+k_{y}^{2}+k_{z}^{2}}}$,
$e^{i\phi_{\boldsymbol{k}}}=\frac{K_{x}+ik_{z}}{\sqrt{K_{x}^{2}+k_{z}^{2}}}$
; $s=\pm1$ denotes the conduction and valence bands, and $V$ is
the quantization volume.

 To visualize the dispersion of electron states, we take
for simplicity $m=b^{2}$. The 3D plot for one projection of 3D dispersion
of the Hamiltonian 2 is shown in Fig. 1. For small energies $\vert\frac{E}{\hbar v_{F}}\vert\ll b$
the constant energy surface consists of two disconnected spheres,
each of them enclosing a corresponding Weyl point; see Fig.~2. At $\vert\frac{E}{\hbar v_{F}}\vert=\frac{b}{2}$
a separatrix isoenergy surface is a 3D ``figure
of eight''. For $\vert\frac{E}{\hbar v_{F}}\vert>\frac{b}{2}$
the constant energy surface is simply connected and encloses both
Weyl points. Figures 2a  and 2b shows contours of constant energy surfaces 
 on the plane $k_{z}=0$  for the Hamiltonians 2 and 1, respectively. The electron dispersion is strongly anisotropic. This  will result in different values for the diagonal elements of the bulk dielectric permittivity tensor, as in two-axial crystals. 
The dotted circle in Fig.~2a is the boundary of a region
that contains surface states. For Hamiltonian 1 in Fig.~2b the surface states exist between the dotted lines. 

%%%%%%%%%%%%%%%%%%%%%%%%%%%%%%%%%

\begin{figure}[htb]
\begin{center}
\includegraphics[scale=0.3]{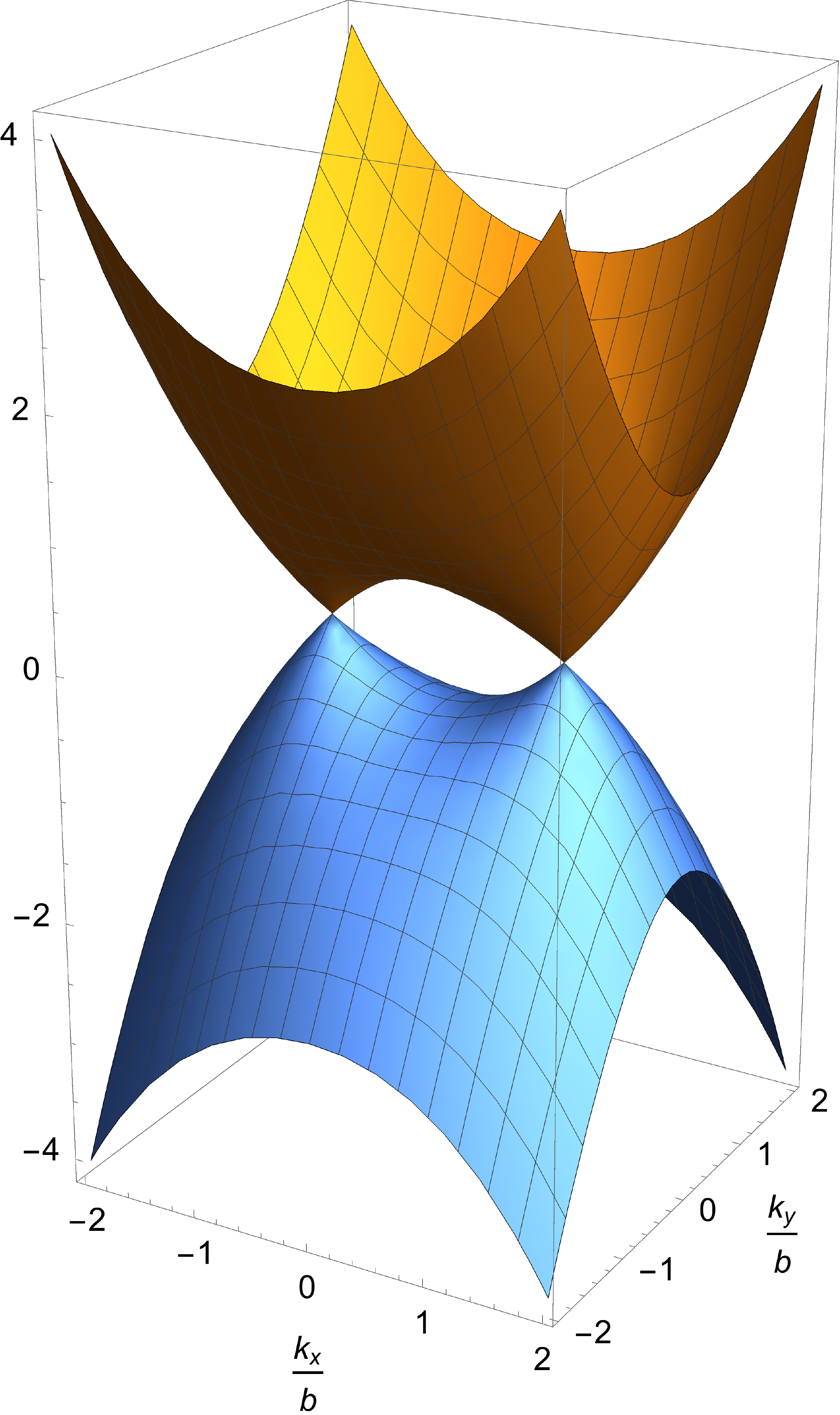}
\caption{Bulk energy dispersion for Hamiltonian 2 on the surface $k_z = 0$. Here the energy is normalized by $\hbar v_F b$ and $k_{x,y}$ are normalized by $b$.   }
\label{Fig:1a-bulkE}
\end{center}
\end{figure}

\begin{figure}[htb]
\begin{center}
\includegraphics[scale=0.5]{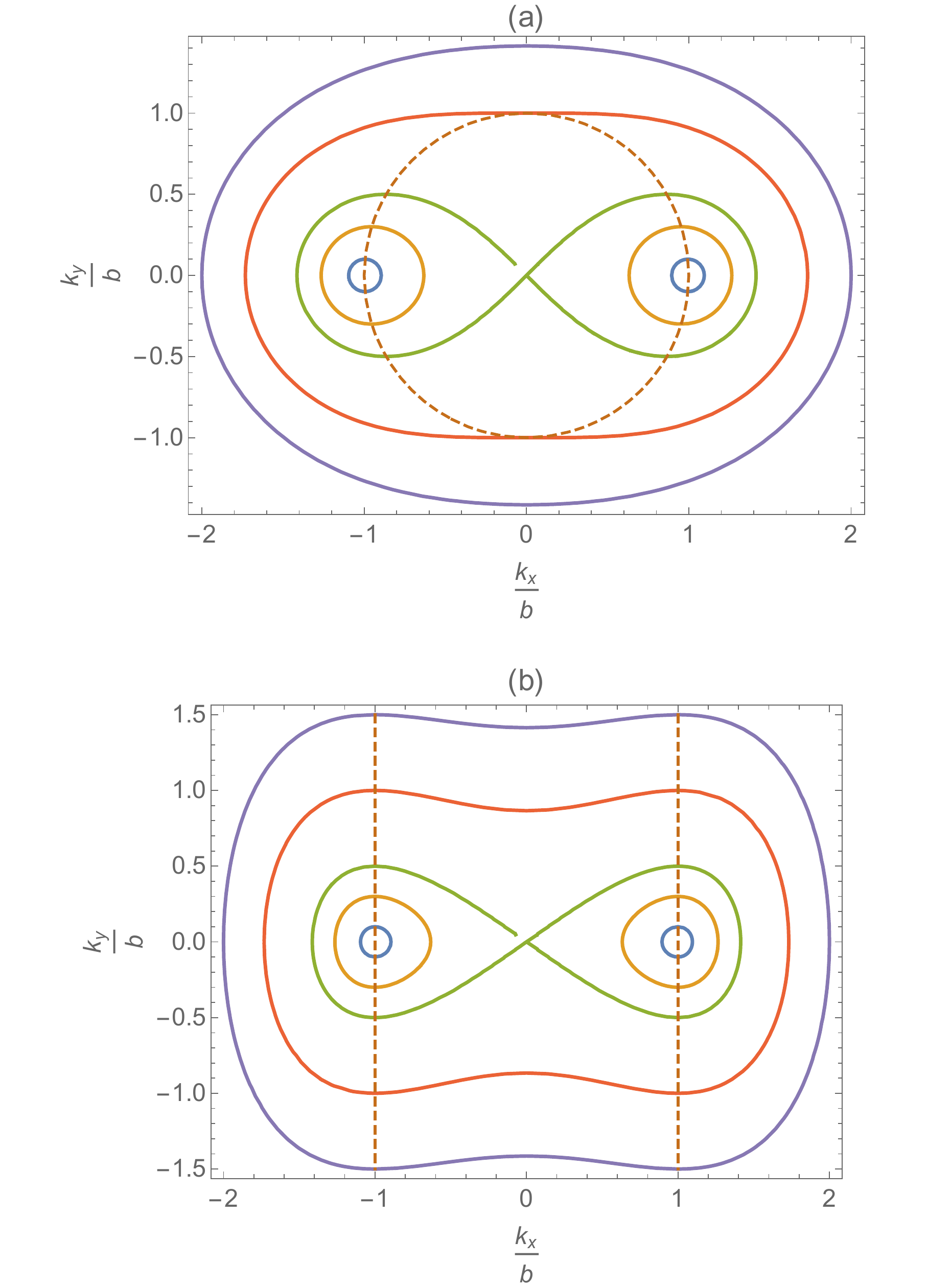}
\caption{(a) Contours of constant energy surfaces for Hamiltonian 2 on the surface $k_z = 0$.   The dotted circle is the boundary of a region $k_x^2 + k_y^2 \leq b^2$ 
where surface states exist.  (b) Contours of constant energy surfaces for Hamiltonian 1 on the surface $k_z = 0$. Here $x,y = k_{x,y}/b$.  The dotted lines indicate the boundary of a region $k_x^2  \leq b^2$ 
where surface states exist. }
\label{Fig:2}
\end{center}
\end{figure}

%%%%%%%%%%%%%%%%%%%%%%%%%%%%%

\subsubsection{Reflection from the boundary. Surface states and Fermi arcs}

 Following \cite{okugawa2014}, we define the boundary as a jump
in the parameter $m$, so that $m=b^{2}$ inside the WSM and $m=-m_{\infty}$
outside. Then Eqs.~(\ref{Eq:Transformed Hamiltonian in k representation}) and (\ref{Eq:eigen-equation of bulk state in Hamiltonians 2 and 3}) will contain the parameter $\textit{m}$
as a function of the coordinate $r_{j}$ orthogonal to the boundary,
and the corresponding component of the quasimomentum $k_{j}$ is replaced
by $k_{j}\Longrightarrow-i\frac{\partial}{\partial r_{j}}$.

 For the boundary parallel to the gyrotropy axis $\textit{x}$,
we assume that it coincides with the surface $\textit{z}=0$ and the
WSM fills the halfspace $z<0$. In this case $m=b^{2}$ for
$\ensuremath{z<0}$ and $m=-m_{\infty}$, $m_{\infty}\rightarrow\infty$
for $\ensuremath{z>0}$.

 For Hamiltonian 3, the Schr\"{o}dinger equation given by Eq.~(\ref{Eq:eigen-equation of bulk state in Hamiltonians 2 and 3}) is
a 4th order differential equation, since its matrix elements contain
$\frac{\partial^{2}}{\partial z^{2}}$ . For Hamiltonians 1 and 2
we get a 2nd order set of equations. The velocity operator $\hat{v}_{z}=\frac{i}{\hbar}\left[H,z\right]$
for Hamiltonian 3 is $\hat{v}_{z}=-i\frac{v_{F}}{b}\hat{\sigma}_{x}\frac{\partial}{\partial z}+v_{F}\hat{\sigma}_{y}$,
i.e. it depends on the coordinate derivative. In contrast, the velocity operator $\hat{v}_{z}=v_{F}\hat{\sigma}_{y}$ for Hamiltonians 1 and 2 
does not depend on the coordinate derivative.
Therefore, for Hamiltonian 3 at  $z=0$, the continuity of both the eigenstate and its derivative is required, whereas  one only needs 
the continuity of the eigenstates for Hamiltonians 1 and 2. 

 Using Eq.~(\ref{Eq:eigen-equation of bulk state in Hamiltonians 2 and 3}) one can find that the eigenstate of Hamiltonians 1 and 2 in the region $\textit{z}>0$ at
$m_{\infty}\rightarrow\infty$ is $\left|\Psi_{\infty}\right\rangle \propto\left(\begin{array}{c}
1\\
0
\end{array}\right)e^{ik_{x}x+ik_{y}y-\frac{m_{\infty}}{2b}z}$. In the region $\textit{z}<0$ we take the eigenstate $\left|\Psi_{B}\right\rangle $ which is 
given by Eq.~(\ref{Eq:components of spinor eigenstate}). Stitching together these two eigenstates $\left|\Psi_{\infty}\right\rangle $
and $\ensuremath{\left|\Psi_{B}\right\rangle }$ at the boundary yields
the following expression for the bulk state:

\begin{equation}
\ensuremath{\left|\Psi_{B}\right\rangle =\frac{e^{ik_{x}x+ik_{y}y}}{2\sqrt{V}}\left[\left(\begin{array}{c}
\sqrt{1-s\cos\theta_{\boldsymbol{k}}}e^{-i\phi_{\boldsymbol{k}}}\\
s\sqrt{1+s\cos\theta_{\boldsymbol{k}}}
\end{array}\right)e^{ik_{z}z}-\left(\begin{array}{c}
\sqrt{1-s\cos\theta_{\boldsymbol{k}}}e^{i\phi_{\boldsymbol{k}}}\\
s\sqrt{1+s\cos\theta_{\boldsymbol{k}}}
\end{array}\right)e^{-ik_{z}z}\right]},\label{Eq:bulk state in Hamiltonians 2 and 3}
\end{equation}
where the quantization volume is limited from one side by the $\textit{z}=0$
plane. The eigenenergy is still given by Eq.~(\ref{Eq:eigenvalue}), and the angles $\ensuremath{\theta_{\boldsymbol{k}}}$
and $\phi_{\boldsymbol{k}}$ are defined below Eq.~(\ref{Eq:components of spinor eigenstate}).

 If $\ensuremath{\frac{E^{2}}{\hbar^{2}v_{F}^{2}}<k_{y}^{2}+K_{x}^{2}}$
the value of $\ensuremath{k_{z}}$ in Eq.~(\ref{Eq:eigenvalue}) is imaginary: $k_{z}=\pm i\kappa$.
In order to connect the eigenstate $\left|\Psi_{\infty}\right\rangle \propto\left(\begin{array}{c}
1\\
0
\end{array}\right)$ in $\textit{z}>0$ with the eigenstate localized at $\textit{z}<0$
which is $\ensuremath{e^{\kappa z}}$ , the localized eigenstate should
be also a spinor $\left(\begin{array}{c}
1\\
0
\end{array}\right)$. After replacing $k_{z}\Rightarrow-i\kappa$ in Eq.~(\ref{Eq:eigen-equation of bulk state in Hamiltonians 2 and 3}), we obtain
the following eigenenergies and eigenvectors for surface states in
the limit $m_{\infty}\rightarrow\infty$:
\begin{equation}
\ensuremath{\frac{E}{\hbar v_{F}}=-k_{y}},\ \ensuremath{\left|\Psi_{S}\right\rangle =\sqrt{\frac{2\kappa}{S}}\left(\begin{array}{c}
1\\
0
\end{array}\right)\Theta\left(-z\right)e^{\kappa z+ik_{x}x+ik_{y}y}},\label{Eq:surface state at z<0 in Hamiltonians 2 and 3}
\end{equation}
where $\Theta$ is a step function, $\textit{S}$ is the quantization
area, $\kappa=-K_{x}>0$. For Hamiltonian 2 the surface states
exist inside a dashed circle $b^{2}>k_{x}^{2}+k_{y}^{2}$ in Fig.~2a. For Hamiltonian 1 the surface states exist in the region $b^{2}>k_{x}^{2}$ in Fig.~2b.

 If a WSM occupies the region $\textit{z}>0$, instead
of Eqs.~(\ref{Eq:surface state at z<0 in Hamiltonians 2 and 3}) we obtain
\begin{equation}
\ensuremath{\frac{E}{\hbar v_{F}}=+k_{y}},\ \ensuremath{\left|\Psi_{S}\right\rangle =\sqrt{\frac{2\kappa}{S}}\left(\begin{array}{c}
0\\
1
\end{array}\right)\Theta\left(z\right)e^{-\vert\kappa\vert z+ik_{x}x+ik_{y}y}},\label{Eq:surface state at z>0 in Hamiltonians 2 and 3}
\end{equation}
where $\kappa=+K_{x}<0$. Equations (\ref{Eq:surface state at z<0 in Hamiltonians 2 and 3}),(\ref{Eq:surface state at z>0 in Hamiltonians 2 and 3}) can be easily generalized
to the case of a parameter $m(z)$ which varies continuously
between the values $b^{2}$ and $\ensuremath{-m_{\infty}}$ \cite{okugawa2014}.
For example, instead of Eqs.~(\ref{Eq:surface state at z<0 in Hamiltonians 2 and 3}) we get
\begin{equation}
\frac{E}{\hbar v_{F}}=-k_{y},\ \left|\Psi_{S}\right\rangle =N\left(\begin{array}{c}
1\\
0
\end{array}\right)e^{ik_{x}x+ik_{y}y}  \left\{   \begin{array}{ll}
e^{\int_{0}^{z}\frac{m\left(z\right)-k_x^2}{2b}dz} & {\rm for\;Hamiltonian\;1}\\
e^{\int_{0}^{z}~\frac{m\left(z\right)-k_x^2 -k_y^2}{2b}~dz} & {\rm for\;Hamiltonian\;2,}
\end{array} \right.
\end{equation}
 where $N$ is a normalization factor.

 Note that the constant surface energy lines $k_{y}=$ const
are tangent to the points where the bulk-state
constant energy surface intersects the boundary of the surface states,
shown as dotted lines in Fig.~2a and 2b. The union of these $k_{y}=$ const
lines and the bulk-state constant energy surface is a set of bulk
and surface energy states with the same energy. In particular, at
the energy equal to the Fermi energy $E_F$ the $k_{y}= E_F/(\hbar v_F)$ line forms
a Fermi arc.

%%%%%%%%%%%%%%%%%%%%%%%

\subsection{Hamiltonian 3}

 For a 4th order set of differential equations the construction
of electron states including their interaction with a boundary is
more complicated.

 First, we use Eq.~(\ref{Eq:eigenvalue}) to find the value of $\ensuremath{k_{z}}$
for given $k_{x,y}$ and $E$. Consider the parameter range $m\leq b^{2}$,
including both positive and negative values of $m$. If $\frac{E^{2}}{\hbar^{2}v_{F}^{2}}>k_{y}^{2}+\frac{\left(k_{x}^{2}+k_{y}^{2}-m\right)^{2}}{4b^{2}}$,
one always has two real solutions $k_{z1}=-k_{z2}>0$ together with
two imaginary solutions corresponding to evanescent states: $k_{z3,4}=i\kappa_{3,4}$,
where $0<\kappa_{3}=-\kappa_{4}$ . If $\frac{E^{2}}{\hbar^{2}v_{F}^{2}}<k_{y}^{2}+\frac{\left(k_{x}^{2}+k_{y}^{2}-m\right)^{2}}{4b^{2}}$
, all four solutions are imaginary and correspond to evanescent states:
$k_{z1,2,3,4}=i\kappa_{1,2,3,4}$, where $ 0 < \kappa_{1}=-\kappa_{3}$, $0<\kappa_{2}=-\kappa_{4}$.
In the region $z>0$ (i.e. outside the sample, where $m=-m_{\infty}$
) it is reasonable to take the solution as a superposition of two localized
modes $e^{-\vert\kappa_{3,4}\vert z}$. In this case for $z<0$, i.e.
inside the sample where $m=b^{2}$, there could be two options:

(i) A superposition of two counterpropagating waves 
with quasimomenta $\ensuremath{k_{z1}=-k_{z2}}$ together with a
localized wave $e^{\kappa_{3}z}$. The localized solution cannot be
discarded, since without it the number of constants becomes smaller
than the number of the boundary conditions.

 (ii) A superposition of two localized waves 
i.e.~the surface state. In this option the number of constants is
always smaller than the number of the boundary conditions, so such
a state can exist only at certain values of energy.

 The procedure of stitching the spinor components and their
derivatives is simplified if $\ensuremath{m_{\infty}\rightarrow\infty}$
since in this limit the continuity of the derivative is equivalent
to setting both components of a spinor $\Psi_{1,2}$ equal
to zero in the cross section $z=0$.

\subsubsection{Bulk states near the boundary}

 In case (i) we obtain
\begin{align}
 & \left|\Psi_{B}\right\rangle \approx\frac{e^{ik_{x}x+ik_{y}y}}{2\sqrt{V}}\nonumber \\
\times & \ensuremath{\left[\left(\begin{array}{c}
\sqrt{1-s\cos\theta_{\boldsymbol{k}}}e^{-i\phi_{\boldsymbol{k}}}\\
s\sqrt{1+s\cos\theta_{\boldsymbol{k}}}
\end{array}\right)e^{ik_{z}z}+r\left(\begin{array}{c}
\sqrt{1-s\cos\theta_{\boldsymbol{k}}}e^{i\phi_{\boldsymbol{k}}}\\
s\sqrt{1+s\cos\theta_{\boldsymbol{k}}}
\end{array}\right)e^{-ik_{z}z}+l\left(\begin{array}{c}
\sqrt{1-s\cos\theta_{\boldsymbol{k}}}e^{\alpha_{\kappa}}\\
-s\sqrt{1+s\cos\theta_{\boldsymbol{k}}}
\end{array}\right)e^{\kappa z}\right]}\label{Eq:bulk state near the boundary in Hamiltonian 1}
\end{align}
where
\[
\ k_{z}=\sqrt{2b\sqrt{\frac{E^{2}}{\hbar^{2}v_{F}^{2}}+k_{x}^{2}}-\left(k_{x}^{2}+k_{y}^{2}+b^{2}\right)},\ \kappa=\sqrt{2b\sqrt{\frac{E^{2}}{\hbar^{2}v_{F}^{2}}+k_{x}^{2}}+\left(k_{x}^{2}+k_{y}^{2}+b^{2}\right)},
\]
\[
\ensuremath{r=-\frac{e^{\alpha_{\kappa}}+e^{-i\phi_{\boldsymbol{k}}}}{e^{\alpha_{\kappa}}+e^{i\phi_{\boldsymbol{k}}}}}, \;
\ensuremath{\sinh\alpha_{\kappa}=\frac{\kappa}{\sqrt{\frac{E^{2}}{\hbar^{2}v_{F}^{2}}-k_{y}^{2}}}},\;  \ensuremath{l=2i\frac{\sin\phi_{\boldsymbol{k}}}{e^{\alpha_{\kappa}}+e^{i\phi_{\boldsymbol{k}}}}}.
\]
Clearly, $\vert r\vert^{2}=1$, which corresponds, as expected, to
the total reflection from the boundary. The quantization volume in
Eq.~(\ref{Eq:bulk state near the boundary in Hamiltonian 1}) is chosen in such a way that its length along the $z$ axis
is much larger than $k_{z}^{-1}>\kappa^{-1}$. Therefore, the last
term in the brackets in Eq.~(\ref{Eq:bulk state near the boundary in Hamiltonian 1}) is unimportant in a sense that it
does not affect the eigenstate normalization or the matrix elements.

%%%%%%%%%%%%%%%%%%%%%%%

\subsubsection{Surface states}

 To construct the surface states (option (ii))
it is convenient to to go back to Eq.~(\ref{Eq:eigen-equation of bulk state in Hamiltonians 2 and 3}), use  
$m=b^{2}$, and make the substitution $k_{z}=-i\kappa$:
\begin{equation}
\begin{pmatrix}-k_{y}-\frac{E}{\hbar v_{F}} & \frac{k_{x}^{2}+k_{y}^{2}-\kappa^{2}-b^{2}}{2b}-\kappa\\
\frac{k_{x}^{2}+k_{y}^{2}-\kappa^{2}-b^{2}}{2b}+\kappa & k_{y}-\frac{E}{\hbar v_{F}}
\end{pmatrix}\ensuremath{\left(\begin{array}{c}
\Psi_{1}\\
\Psi_{2}
\end{array}\right)=0}\label{Eq:eigen-equation of surface state in Hamiltonian 1}
\end{equation}

 Consider the solution of Eq.~(\ref{Eq:eigen-equation of surface state in Hamiltonian 1}), corresponding to different
positive values of $\kappa_{1,2}$ but the same spinor constant $\left(\begin{array}{c}
a\\
b
\end{array}\right)$. One can build a nontrivial localized solution $\left|\Psi_{S}\right\rangle \propto\left(\begin{array}{c}
a\\
b
\end{array}\right)\Theta\left(-z\right)\left(e^{\kappa_{1}z}-e^{\kappa_{2}z}\right)$, which corresponds to the null boundary conditions at the surface
$\textit{z}=0$. Such a solution of Eq.~(\ref{Eq:eigen-equation of surface state in Hamiltonian 1}) is possible under the
following conditions:

 $\ensuremath{-k_{y}-\frac{E}{\hbar v_{F}}=\frac{k_{x}^{2}+k_{y}^{2}-\kappa^{2}-b^{2}}{2b}+\kappa=0}$,
or $k_{y}-\frac{E}{\hbar v_{F}}=\frac{k_{x}^{2}+k_{y}^{2}-\kappa^{2}-b^{2}}{2b}-\kappa=0$,
or $k_{y}-\frac{E}{\hbar v_{F}}=\frac{k_{x}^{2}+k_{y}^{2}-\kappa^{2}-b^{2}}{2b}-\kappa=0$,
where$\left(\begin{array}{c}
a\\
b
\end{array}\right)=\left(\begin{array}{c}
1\\
0
\end{array}\right)$ or $\ensuremath{\left(\begin{array}{c}
a\\
b
\end{array}\right)=\left(\begin{array}{c}
0\\
1
\end{array}\right)}$respectively. It is easy to see that the first option corresponds
to the surface state when the WSM occupies the halfspace
$z<0$, whereas the second option corresponds to the WSM
in the region $z>0$, since in this case the values of $\kappa_{1,2}$
are negative. The resulting states are as follows.

(i) WSM in $z<0$:

\begin{equation}
\ensuremath{\frac{E}{\hbar v_{F}}=-k_{y}},\ \ensuremath{\left|\Psi_{S}\right\rangle =\sqrt{\frac{2}{S\left(\frac{1}{\kappa_{1}}+\frac{1}{\kappa_{2}}-\frac{4}{\kappa_{1}+\kappa_{2}}\right)}}\left(\begin{array}{c}
1\\
0
\end{array}\right)\Theta\left(-z\right)\left(e^{\kappa_{1}z}-e^{\kappa_{2}z}\right)\cdot e^{ik_{x}x+ik_{y}y}};\label{Eq:surface state at z<0 in Hamiltonian 1}
\end{equation}

(ii) WSM in $z>0$:

\begin{equation}
\ensuremath{\frac{E}{\hbar v_{F}}=k_{y}},\ \left|\Psi_{S}\right\rangle =\sqrt{\frac{2}{S\left(\frac{1}{\kappa_{1}}+\frac{1}{\kappa_{2}}-\frac{4}{\kappa_{1}+\kappa_{2}}\right)}}\left(\begin{array}{c}
1\\
0
\end{array}\right)\Theta\left(z\right)\left(e^{-\kappa_{1}z}-e^{-\kappa_{2}z}\right)\cdot e^{ik_{x}x+ik_{y}y}.\label{Eq:surface state at z>0 in Hamiltonians 1}
\end{equation}
Here $\kappa_{1,2}=b\mp\sqrt{k_{x}^{2}+k_{y}^{2}}$ . 

In the region
$b^{2}<k_{x}^{2}+k_{y}^{2}$ there is only one localized evanescent
solution for any fixed value of energy, which is not enough to satisfy
the boundary conditions. Therefore, the region $\ensuremath{b^{2}>k_{x}^{2}+k_{y}^{2}}$,
where the surface states exist, is the same in the models described
by the Hamiltonian 2 and Hamiltonian 3 (see the dotted circle in Fig.~2a).

 Taking into account a finite value of $m_{\infty}$ modifies
the above expression, but their general structure remains the same.
For example, when a WSM fills the halfspace $z<0$, then the
eigenstate in Eq.~(\ref{Eq:surface state at z<0 in Hamiltonian 1}) is replaced by
\[
\ensuremath{\left|\Psi_{S;z<0}\right\rangle \propto\left(\begin{array}{c}
1\\
0
\end{array}\right)\left(e^{\kappa_{1}z}-\zeta e^{\kappa_{2}z}\right)e^{ik_{x}x+ik_{y}y}},
\]
\begin{equation}
\left|\Psi_{S;z>0}\right\rangle \propto\left(\begin{array}{c}
1\\
0
\end{array}\right)\frac{\kappa_{2}-\kappa_{1}}{\kappa_{2}+\sqrt{m_{\infty}}}e^{-\sqrt{m_{\infty}}z}e^{ik_{x}x+ik_{y}y},\label{Eq:surface state at z<0 in Hamiltonian 1 for finite m}
\end{equation}
where $\displaystyle \zeta=\frac{\kappa_{1}+\sqrt{m_{\infty}}}{\kappa_{2}+\sqrt{m_{\infty}}}$.

%%%%%%%%%%%%%%%%%%%%%%%%%

\subsection{The boundary orthogonal to the gyrotropy axis}

 Any Hamiltonian, 1, 2, or 3,
contains the second derivative $\frac{\partial^{2}}{\partial x^{2}}$.
Therefore, the analysis of the bulk and surface states near the boundary
orthogonal to the gyrotropy axis is similar to the one for the boundary
parallel to the gyrotropy axis when the Hamiltonian contains the second
derivative $\frac{\partial^{2}}{\partial z^{2}}$. Repeating the same
arguments as in the previous section, we obtain that the orthogonal
boundary is trivial and does not contain surface states.
%%%%%%%%%%%%%%%%%%%%%%%

\subsection{Comparison of Hamiltonians 1, 2, and 3}

 The only principal difference between the eigenstates of the 
effective Hamiltonians considered above is the region where the surface
states exist. Such a region is determined by the inequality $b>\sqrt{k_{x}^{2}+k_{y}^{2}}$
for Hamiltonians 2 and 3, and the inequality $b>\vert k_{x}\vert$
for Hamiltonian 1. The latter condition leads to an infinite density
of surface states, which is unphysical and would have to be restricted
artificially. Therefore, it is better to work with Hamiltonian
2 or 3. Hamiltonian 2 leads to a simpler $z$-component of the
velocity operator: $\hat{v}_{z}=v_{F}\hat{\sigma}_{y}$ instead of $\hat{v}_{z}=-i\frac{v_{F}}{b}\hat{\sigma}_{x}\frac{\partial}{\partial z}+v_{F}\hat{\sigma}_{y}$, which corresponds to Hamiltonian 3.
The velocity operator of Hamiltonian 2 makes calculations of the surface current easier without losing
any essential physics. Therefore, we will use Hamiltonian 2 for
subsequent calculations of the optical properties.

%%%%%%%%%%%%%%%%%%%%%
%%%%%%%%%%%%%%%%%%%%

\section{Optical transitions and the tensors of bulk and surface conductivity}

 In the presence of external fields one should replace $\boldsymbol{\hat{\boldsymbol{p}}}\Longrightarrow\boldsymbol{\hat{p}}-\frac{e}{c}\boldsymbol{A}$,
and also add the electrostatic potential $\hat{H}\Longrightarrow\hat{H} + e\varphi\hat{1}$ in
Eq.~(\ref{Eq:Transformed Hamiltonian}). Particles are assumed to have charge $e$ where $-e$ is the magnitude of the electron charge. If the potential has a coordinate dependence $\boldsymbol{A}(\boldsymbol{r})$
we assume symmetrized operators
\[
\ensuremath{\left(\hat{p}_{x,y,z} - \frac{e}{c}A_{x,y,z}\right)^{2}\Longrightarrow\hat{p}_{x,y,z}^{2}+\frac{e^{2}}{c^{2}}A_{x,y,z}^{2}- \frac{e}{c}\left(\hat{p}_{x,y,z}A_{x,y,z}+A_{x,y,z}\hat{p}_{x,y,z}\right)},
\]
and in the expressions for the velocity operator we need to replace
\[
\ensuremath{-i\frac{\partial}{\partial x,\partial y,\partial z}\Longrightarrow-i\frac{\partial}{\partial x,\partial y,\partial z} -\frac{e}{c\hbar}A_{x,y,z}}.
\]

Throughout the paper, we will consider the potentials corresponding to a monochromatic electromagnetic field
propagating in the arbitrary direction $\boldsymbol{r}$ with angular
frequency $\omega$ and wavevector $\boldsymbol{q}$, i.e.
\begin{align}
\text{} & \phi=\frac{1}{2}\phi(\omega)e^{-i\omega t+i\boldsymbol{q}\cdot\boldsymbol{r}}+c.c.,\\
 & \boldsymbol{A}=\frac{1}{2}[\boldsymbol{x_{0}}A_{x}(\omega)+\boldsymbol{y_{0}}A_{y}(\omega)+\boldsymbol{z_{0}}A_{z}(\omega)]e^{-i\omega t+i\boldsymbol{q}\cdot\boldsymbol{r}}+c.c.
\end{align}

Bulk-to-bulk and surface-to-surface transitions contribute to the bulk and surface conductivity tensors, respectively. The contributions are detailed in the Appendix. Surface-to-bulk transitions contribute to the surface conductivity tensor only. They have to be handled with more care, as  
we briefly describe below.

 Generally, the electron and current densities expressed in terms of the density matrix are
given by
\begin{equation}
\ensuremath{n\left(\boldsymbol{r}\right)=\sum_{\alpha\beta}^ {}n_{\beta\alpha}\left(\boldsymbol{r}\right)\rho_{\alpha\beta}},\ \ensuremath{\boldsymbol{j}\left(\boldsymbol{r}\right)=\sum_{\alpha\beta}\boldsymbol{j}_{\beta\alpha}\left(\boldsymbol{r}\right)\rho_{\alpha\beta}},\label{Eq:electron density and current density}
\end{equation}
\begin{equation}
\ensuremath{n_{\beta\alpha}=\Psi_{\beta}^{\ast}\Psi_{\alpha}},\ \ensuremath{\boldsymbol{j}_{\beta\alpha}=\frac{1}{2}\left[\Psi_{\beta}^{\ast}\left(\boldsymbol{\hat{j}}\Psi_{\alpha}\right)+\left(\boldsymbol{\hat{j}}^{\ast}\Psi_{\beta}^{\ast}\right)\Psi_{\alpha}\right]},\label{Eq:components of electron density and current density}
\end{equation}
where $\ensuremath{\boldsymbol{\hat{j}}=e\boldsymbol{\hat{v}}}\textit{.}$

The Fourier harmonics of the the electron and current densities are
\[
\ensuremath{\boldsymbol{j\left(r\right)}=\frac{1}{2}\sum_{\boldsymbol{q}}\boldsymbol{j^{\left(q\right)}}e^{i\boldsymbol{qr}}+c.c.},\ \ \ensuremath{n\left(\boldsymbol{r}\right)=\frac{1}{2}\sum_{\boldsymbol{q}}n^{\left(\boldsymbol{q}\right)}e^{i\boldsymbol{qr}}+c.c.},
\]
where
\[
\ensuremath{\frac{1}{2}\boldsymbol{j^{\left(q\right)}}=\frac{1}{V}\int_{V}\boldsymbol{j\left(r\right)}e^{-i\boldsymbol{qr}}d^{3}r},\ \ \ensuremath{\frac{1}{2}n^{\left(\boldsymbol{q}\right)}=\frac{1}{V}\int_{V}^ {}n^{\left(\boldsymbol{q}\right)}e^{-i\boldsymbol{qr}}d^{3}r}.
\]
For their matrix elements we have
\begin{equation}
\ensuremath{\boldsymbol{j^{\left(q\right)}}=\sum_{\alpha\beta}\boldsymbol{j}_{\beta\alpha}^{\left(\boldsymbol{q}\right)}\rho_{\alpha\beta}},  \quad  \ensuremath{n^{\left(\boldsymbol{q}\right)}=\sum_{\alpha\beta}^ {}n_{\beta\alpha}^{\left(\boldsymbol{q}\right)}\rho_{\alpha\beta}},\label{Eq: matrix elements of Fourier harmonics}
\end{equation}
where
\begin{equation}
\ensuremath{\boldsymbol{j}_{\beta\alpha}^{\left(\boldsymbol{q}\right)}=2\left\langle \beta\right|e^{-i\boldsymbol{qr}}\boldsymbol{\hat{j}}\left|\alpha\right\rangle },  \quad  \ensuremath{n_{\beta\alpha}^{\left(\boldsymbol{q}\right)}=2\left\langle \beta\right|e^{-i\boldsymbol{qr}}\left|\alpha\right\rangle }\label{Eq: definition of matrix elements of Fourier harmonics}
\end{equation}
To find the current without the effect of a boundary we can use the
states given by Eq.~(\ref{Eq:components of spinor eigenstate}).

 Now consider the states near the surface. We will denote
the bulk states with latin indices and surface states with greek ones.
For convenience we rewrite them, having in mind an upper boundary
$z=0$ with the WSM located at $z<0:$
\begin{equation}
\ensuremath{\left|\Psi_{m}\right\rangle =\frac{e^{ik_{x}x+ik_{y}y}}{2\sqrt{V}}\left[\left(\begin{array}{c}
\sqrt{1+s\cos\theta_{\boldsymbol{\boldsymbol{k}_{\parallel}}}}e^{-i\theta_{\boldsymbol{\boldsymbol{k}_{\perp}}}}\\
s\sqrt{1-s\cos\theta_{\boldsymbol{\boldsymbol{k}_{\parallel}}}}
\end{array}\right)e^{ik_{z}z}-\left(\begin{array}{c}
\sqrt{1-s\cos\theta_{\boldsymbol{\boldsymbol{k}_{\parallel}}}}e^{i\theta_{\boldsymbol{\boldsymbol{k}_{\perp}}}}\\
s\sqrt{1+s\cos\theta_{\boldsymbol{\boldsymbol{k}_{\parallel}}}}
\end{array}\right)e^{-ik_{z}z}\right],}\label{Eq: Psi m}
\end{equation}
 where $\ensuremath{E_{m}=s\hbar v_{F}\sqrt{\left(\frac{k_{x}^{2}+k_{y}^{2}-b^{2}}{2b}\right)^{2}+k_{y}^{2}+k_{z}^{2}}}$
is the eigenenergy, $s=\pm1$ is the band index, the values $k_{x,y}$
can be of either sign whereas $k_{z}>0$; $\ensuremath{\cos\theta_{\boldsymbol{\boldsymbol{k}_{\parallel}}}=\frac{k_{z}}{\frac{\vert E\vert}{\hbar v_{F}}}}$.
\begin{equation}
\left|\Psi_{\alpha}\right\rangle =\sqrt{\frac{2\kappa}{S}}\left(\begin{array}{c}
1\\
0
\end{array}\right)\Theta\left(-z\right)e^{ik_{x}x+ik_{y}y+\kappa z},\label{Eq: Psi alpha}
\end{equation}
where $S$ is the area; the energy of the state
is $\ensuremath{E_{\alpha}=-\hbar v_{F}k_{y}}$, $\kappa=\frac{b^{2}-k_{x}^{2}-k_{y}^{2}}{2b}$, $\sqrt{k_{x}^{2}+k_{y}^{2}}<b$.

 Let us limit the surface states by the condition $\kappa>\kappa_{min}$,
where the latter could be a typical scattering length $\ensuremath{\sim\kappa_{min}^{-1}}$.
We will assume that $\kappa_{min}^{-1}$ is much smaller than $L$,
which enters the quantization volume $V=SL$ in Eq.~(\ref{Eq: Psi m}). When we calculate
the matrix elements of the interaction Hamiltonian in the von Neumann
equation, the matrix elements $V_{mn}^{\left(int\right)}$ ,$\ensuremath{V_{\alpha\beta}^{\left(int\right)}}$ and 
$V_{m\alpha}^{\left(int\right)}$ have no peculiarities: the integration
is carried out over the whole volume. However when we calculate the
matrix elements of the density and current, and if at least one of
the indices belongs to the surface state, we will perform the integration
over $dz$:
\begin{equation}
\ensuremath{n_{\beta\alpha}=\int_{-\infty}^{0}\Psi_{\beta}^{\ast}\Psi_{\alpha}dz},  \quad \ensuremath{n_{m\alpha}=\int_{-\infty}^{0}\Psi_{m}^{\ast}\Psi_{\alpha}dz},\label{Eq: matrix elements of density}
\end{equation}
\begin{equation}
\ensuremath{\boldsymbol{j}_{\beta\alpha}=\frac{1}{2}\int_{-\infty}^{0}\left[\Psi_{\beta}^{\ast}\left(\hat{\boldsymbol{j}}\Psi_{\alpha}\right)+\left(\hat{\boldsymbol{j}}^{\ast}\Psi_{\beta}^{\ast}\right)\Psi_{\alpha}\right]dz},\ \ensuremath{\boldsymbol{j}_{m\alpha}=\frac{1}{2}\int_{-\infty}^{0}\left[\Psi_{m}^{\ast}\left(\hat{\boldsymbol{j}}\Psi_{\alpha}\right)+\left(\hat{\boldsymbol{j}}^{\ast}\Psi_{m}^{\ast}\right)\Psi_{\alpha}\right]dz}. 
\label{Eq: matrix elements of current}
\end{equation}
These quantities will depend only on $\textit{x}$ and $\textit{y}$,
and therefore determine the surface current and density.

 The following current component is nontrivial: $\sum_{\alpha\beta}^ {}\left(j_{z}\right)_{\beta\alpha}\rho_{\alpha\beta}+\sum_{m\alpha}^ {}\left(j_{z}\right)_{m\alpha}\rho_{\alpha m}$.
It determines the polarization of a thin double layer:
\begin{equation}
\ensuremath{\frac{\partial}{\partial t}p_{z}\left(x,y\right)=\sum_{\alpha\beta}^ {}\left(j_{z}\right)_{\beta\alpha}\rho_{\alpha\beta}+\sum_{m\alpha}^ {}\left(j_{z}\right)_{m\alpha}\rho_{\alpha m}},\label{Eq: Polarization}
\end{equation}
This layer radiates, but not normally to the layer, and it cannot
affect the surface density of carriers.

 With properly defined matrix elements of the current and
density the continuity equation is satisfied automatically, so we
can consider the volume current flowing toward the boundary $\left(\sum_{mn}^ {}\left(j_{z}\right)_{nm}\rho_{mn}\right)_{z=0}$
as a source in the surface continuity equation.

%%%%%%%%%%%%%%%%%%%%%%%%%%%%%%%%%%

 \subsection{Tensors of bulk and surface conductivity}

The matrix elements of the Fourier components of the current density operator are evaluated in Appendix A.  After evaluating them,
in Appendix B and C we used the Kubo-Greenwood formula to calculate the bulk and surface conductivity tensors, respectively; e.g.  
\begin{equation}
\label{kubo}
\sigma_{\alpha\beta}(\omega)=g\frac{i\hbar}{V}\sum_{mn}\left(\frac{f_{n}-f_{m}}{E_{m}-E_{n}}\right)\frac{\left\langle n\right|\hat{j}_{\alpha}\left|m\right\rangle \left\langle m\right|\hat{j}_{\beta}\left|n\right\rangle }{\hbar(\omega+i\gamma)+(E_{n}-E_{m})},
\end{equation}
 for the bulk conductivity, where $g = 2$ is the spin degeneracy factor and $\alpha,\beta$ denote Cartesian coordinate components. The surface conductivity tensor has a similar structure, but the contribution is summed over surface-to-surface and surface-to-bulk transitions, and the normalization is over the surface area $S$ instead of a volume $V$. Both interband and intraband transitions are included. Three-dimensional integrals over electron momenta in Appendix B and C  cannot be evaluated analytically, except limiting cases of small frequencies or small $b$ (see Appendix D and E). Therefore, integrals were calculated numerically at zero temperature for all plots below.

 The  bulk (3D) conductivity tensor due to low-energy electrons near Weyl points is
\begin{equation}\label{BulkConductivity}
\sigma _{ij}^{B}(\omega) = \begin{pmatrix}
\sigma _{xx}^{B}  &  0  &  0\\
0  &   \sigma _{yy}^{B}  &   \sigma _{yz}^{B}\\
0  &   \sigma _{zy}^{B}  &   \sigma _{zz}^{B}\\
\end{pmatrix}
\end{equation}
where $
\sigma _{zy}^{B}=- \sigma _{yz}^{B}$. 
The surface conductivity tensor at  $z=0$ has a similar structure, with superscript $B$ replaced by $S$ and 
 $\sigma _{zy}^{S}=- \sigma _{yz}^{S}$. 

The background bulk dielectric tensor in the most general form which corresponds to the one for a two-axial nongyrotropic crystal is 
\begin{equation}\label{BulkBackgroundDielectric}
\varepsilon _{ij}^{(0)}(\omega) = \begin{pmatrix}
\varepsilon _{xx}^{(0)}  &  0  &  0\\
0  &   \varepsilon _{yy}^{(0)}  &  0\\
0  &  0  &   \varepsilon _{zz}^{(0)}\\
\end{pmatrix}
 \end{equation}
so that the total dielectric permittivity tensor is
\begin{equation}\label{DielectricTensor}
\varepsilon _{ij}(\omega) = \varepsilon _{ij}^{(0) }(\omega) +i\frac{4 \pi  \sigma _{ij}^{B}(\omega) }{ \omega }= \begin{pmatrix}
 \varepsilon _{xx}  &  0  &  0\\
0  &   \varepsilon _{yy}  &  ig\\
0  &  -ig  &   \varepsilon _{zz}\\
\end{pmatrix}
\end{equation}
where
\begin{equation}
g=\frac{4 \pi  \sigma _{yz}^{B}}{ \omega }. \label{gFactor}
\end{equation}
 Note that for Hamiltonian 3 we would have $\sigma _{yy}^{B}= \sigma _{zz}^{B}$, whereas for Hamiltonian 2 (used in all calculations of the conductivity tensors in this paper) 
we have  $\sigma _{yy}^{B} \neq  \sigma _{zz}^{B}$. Therefore, even if the background dielectric tensor is isotropic, the contribution of massless Weyl electrons will create a two-axial anisotropy.  In the numerical plots below we will take an isotropic background dielectric tensor and neglect its frequency dependence at low frequencies, $\varepsilon _{xx}^{(0)} = \varepsilon _{yy}^{(0)} = \varepsilon _{zz}^{(0)} = 10$, so that all nontrivial effects of anisotropy and gyrotropy are due to Weyl fermions. 

The salient feature of both bulk and surface conductivity tensor is the presence of nonzero off-diagonal (gyrotropic) components due to time-reversal symmetry breaking in the Hamiltonian. These terms originate from the finite separation of the Weyl nodes in momentum space and the existence of surface states (Fermi arcs). The gyrotropic effects in the propagation, reflection, and transmission  of bulk and surface modes can serve as a definitive diagnostic of Weyl nodes,  surface states, and Fermi surface. They could also find applications in optoelectronic devices such as Faraday isolators, modulators etc. 

Figures 3-6 show spectra of $\varepsilon _{xx}(\omega)$, $\varepsilon _{yy}(\omega)$,  $\varepsilon _{zz}(\omega)$, and $g(\omega)$ for several values of the Fermi momentum $k_F$ (at zero temperature), when the Weyl node separation $2\hbar v_F b = 200$ meV. The characteristic feature in all plots is strong absorption and dispersion at the onset of interband transitions, when $\omega = 2 v_F k_F$. Another common feature is a Drude-like increase in the absolute value of all tensor components at low frequencies. Indeed, as shown in Appendix D, in the limit $\omega \ll v_F k_F \ll v_F b$ when only the intraband transitions in the vicinity of each Weyl point are important, the off-diagonal components are equal to zero and the diagonal conductivity components are reduced to the same Drude form: 
\begin{equation}
\sigma_{xx}^{intra}\left(\omega\right)=\sigma_{yy}^{intra}\left(\omega\right)=\sigma_{zz}^{intra}\left(\omega\right)=\frac{ge^{2}v_{F}k_{F}^{2}}{3\pi^{2}\hbar(-i\omega+\gamma)}.
\end{equation}

%%%%%%%%%%%%%%%%%%%%%%
\begin{figure}[htb]
\begin{center}
\includegraphics[scale=0.4]{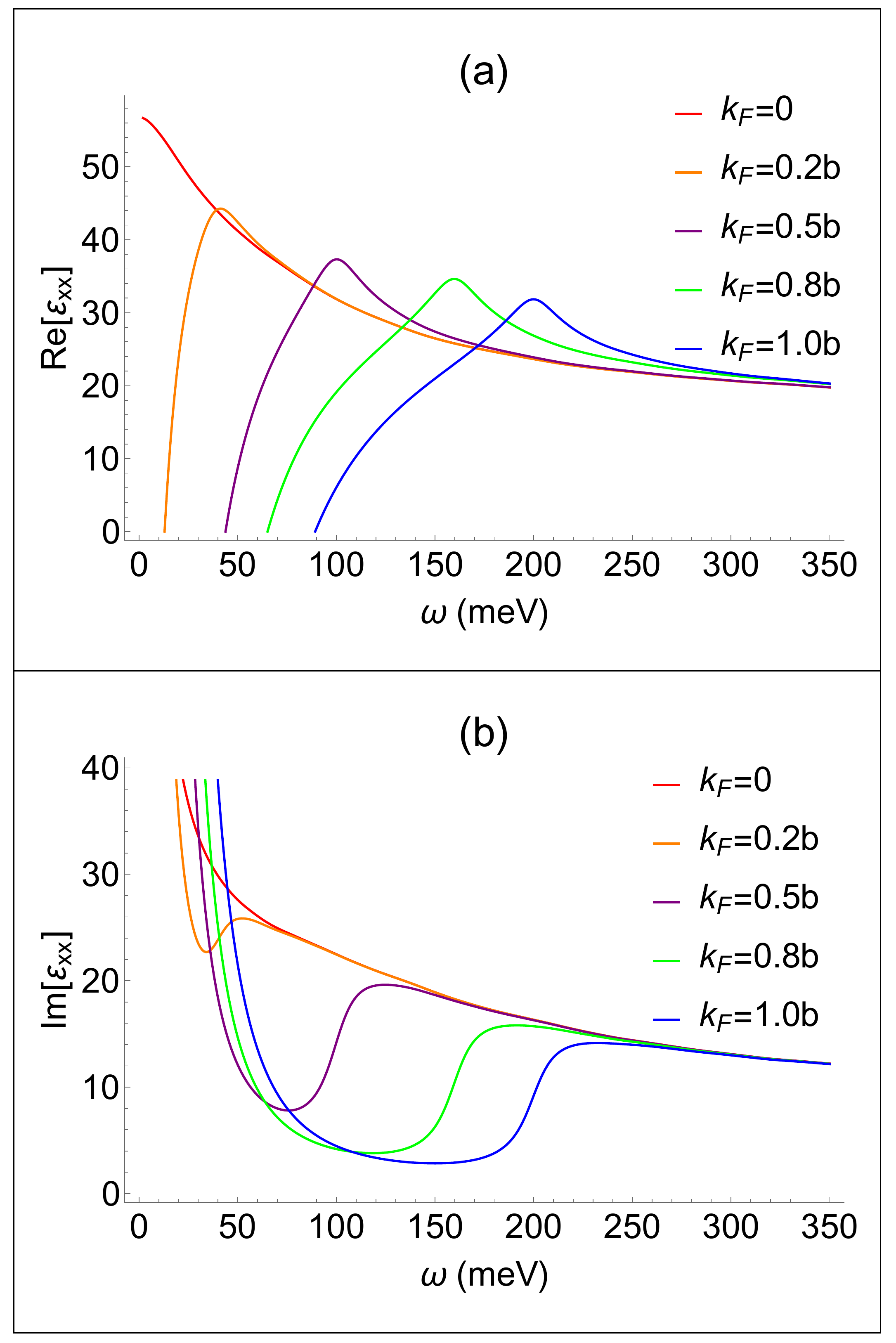}
\caption{Real and imaginary parts of the $\varepsilon _{xx}$ component of the dielectric tensor as a function of frequency for $\hbar v_F b = 100$ meV, dephasing rate $\gamma = 10$ meV,  and $\varepsilon _{xx}^{(0)} = 10$.   }
\label{Fig:3}
\end{center}
\end{figure}

\begin{figure}[htb]
\begin{center}
\includegraphics[scale=0.4]{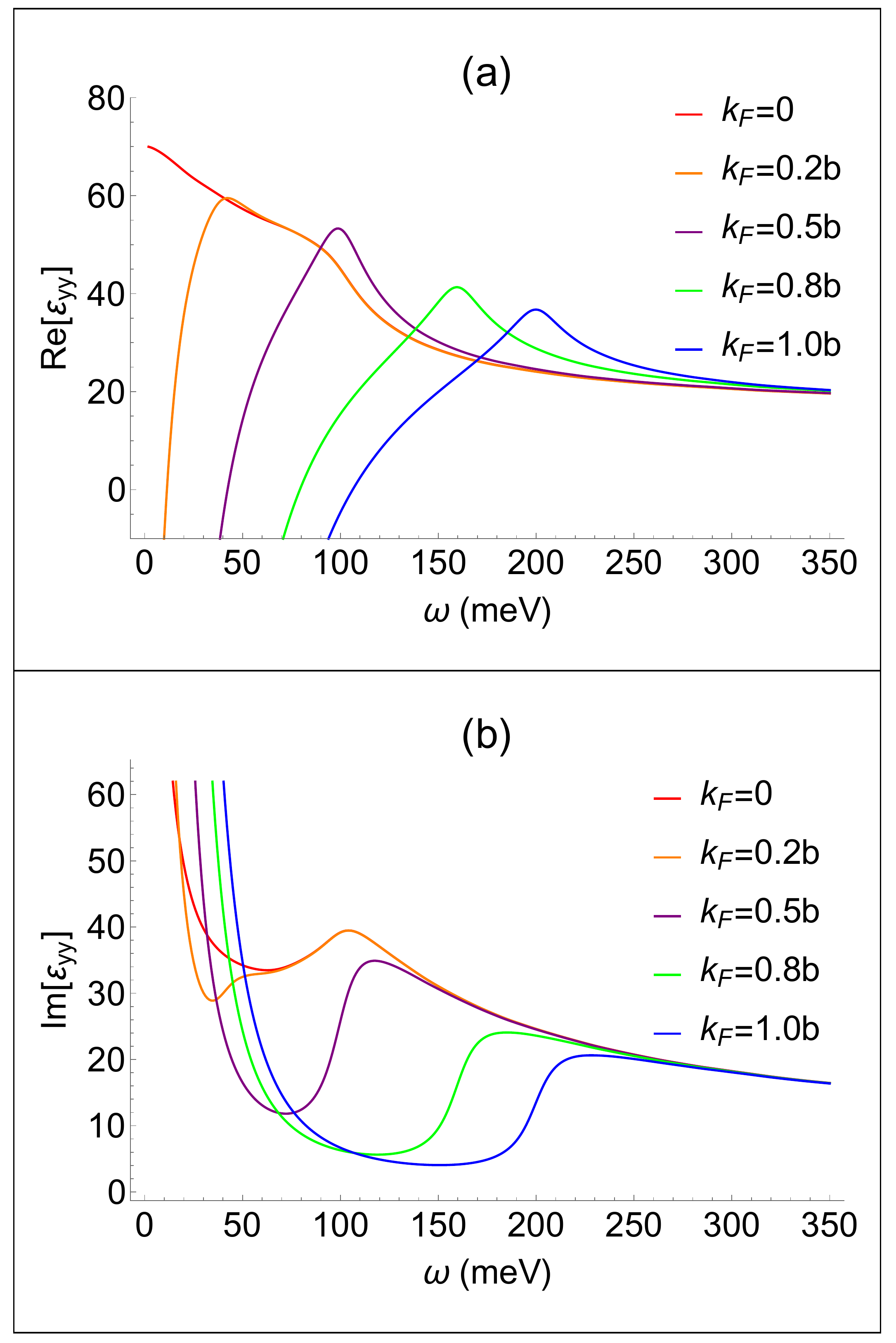}
\caption{Real and imaginary parts of the $\varepsilon _{yy}$ component of the dielectric tensor as a function of frequency for $\hbar v_F b = 100$ meV, dephasing rate $\gamma = 10$ meV,  and $\varepsilon _{yy}^{(0)} = 10$.   }
\label{Fig:4}
\end{center}
\end{figure}

\begin{figure}[htb]
\begin{center}
\includegraphics[scale=0.4]{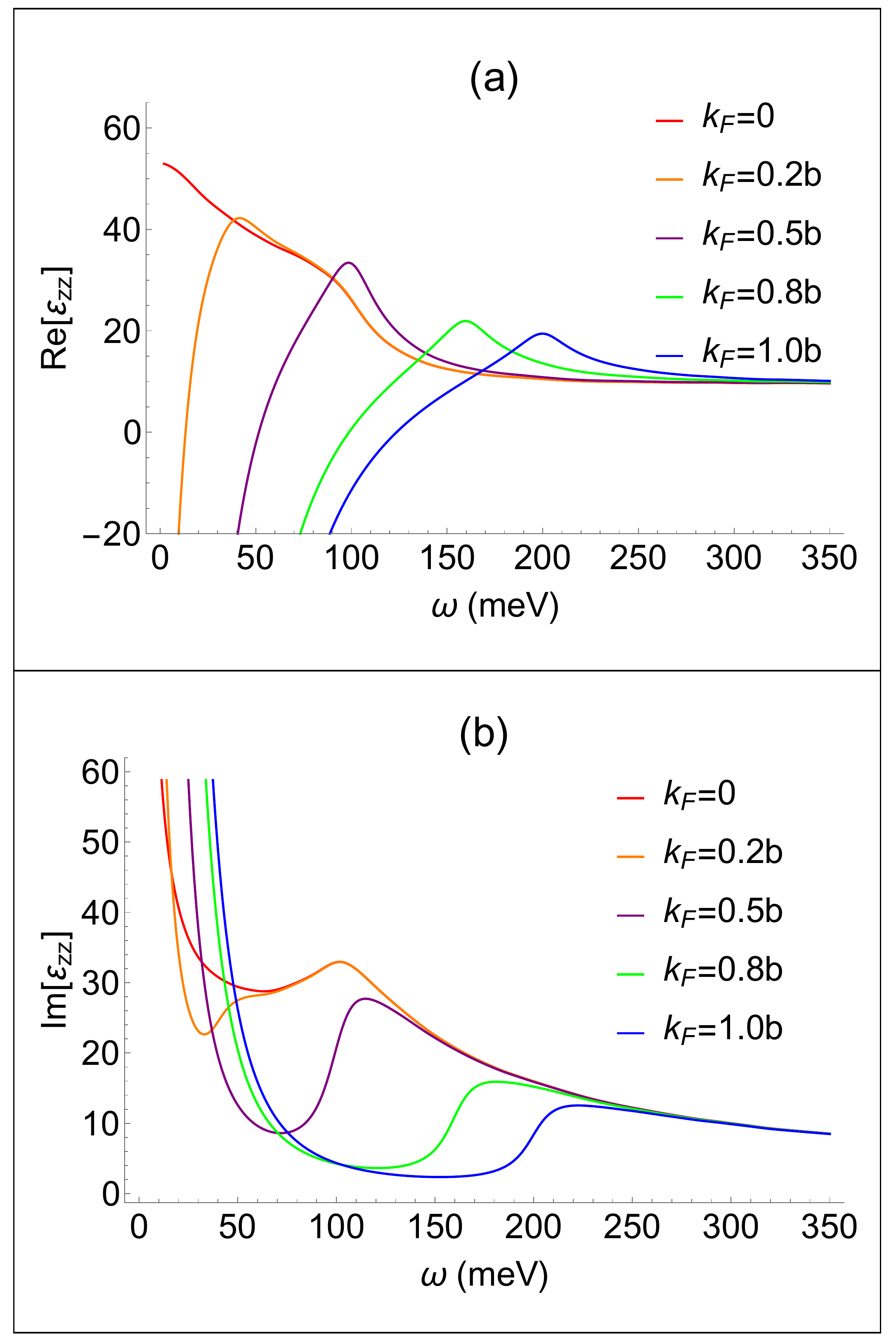}
\caption{Real and imaginary parts of the $\varepsilon _{zz}$ component of the dielectric tensor as a function of frequency for $\hbar v_F b = 100$ meV, dephasing rate $\gamma = 10$ meV,  and $\varepsilon _{zz}^{(0)} = 10$.   }
\label{Fig:3}
\end{center}
\end{figure}

\begin{figure}[htb]
\begin{center}
\includegraphics[scale=0.4]{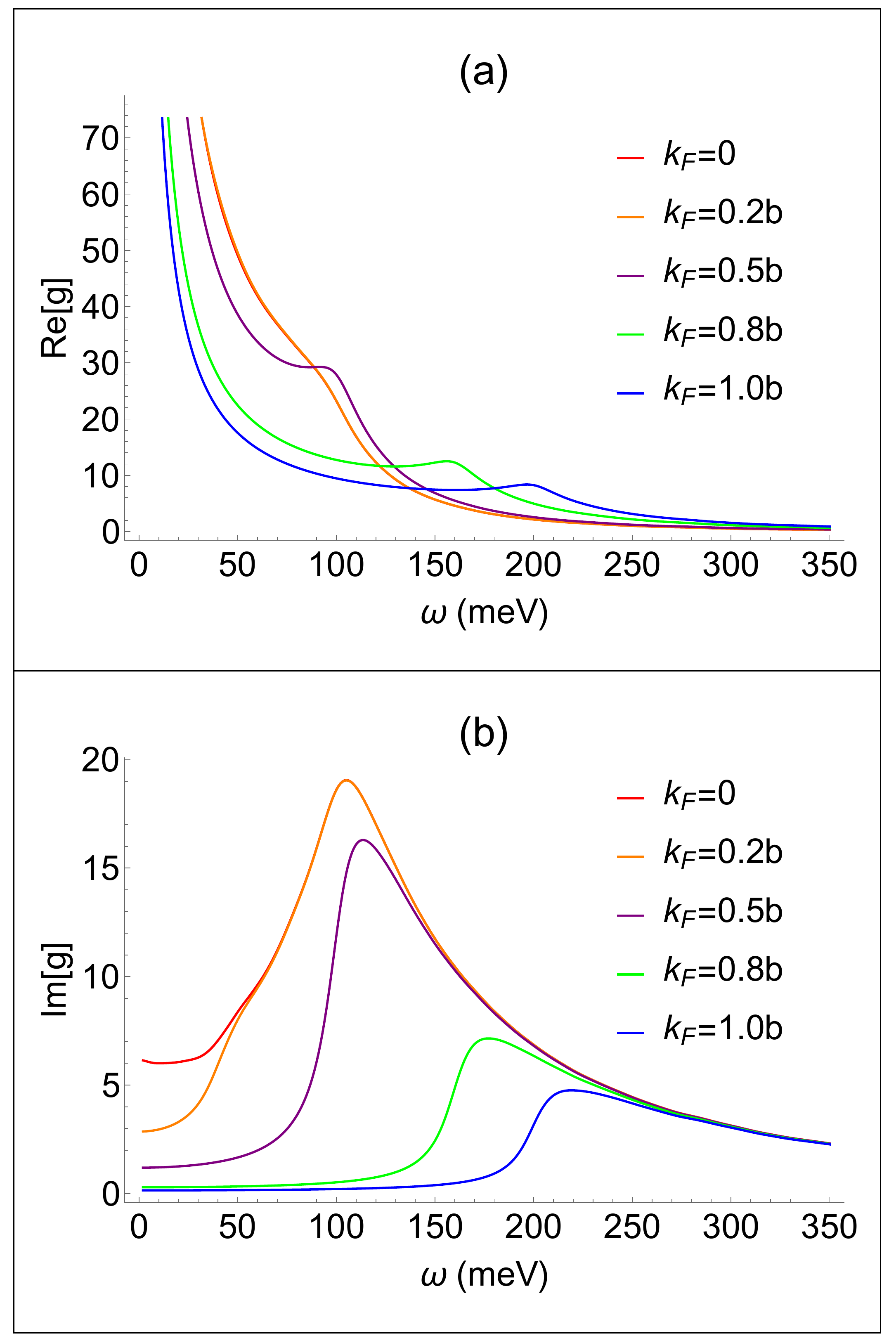}
\caption{Real and imaginary parts  of $g=\frac{4 \pi  \sigma _{yz}^{B}}{ \omega }$ as a function of frequency for $\hbar v_F b = 100$ meV and dephasing rate $\gamma = 10$ meV.   }
\label{Fig:5}
\end{center}
\end{figure}

Note an absorption peak at $\omega = 100$ meV at low Fermi momenta, which corresponds to a Van Hove singularity at the interband transitions between saddle points of conduction and valence bands at $k = 0$, i.e. in the middle between the Weyl points. 

Note also that diagonal and off-diagonal parts of the conductivity tensor are of the same order at low frequencies comparable to the Weyl node separation, which indicates that gyrotropic effects should be quite prominent. 

All figures in this paper are plotted for a relatively high dephasing rate $\gamma = 10$ meV, which smoothes out all spectral features and introduces strong losses for electromagnetic eigenmodes even below the interband transition edge.   The dephasing rate originates from electron scattering and obviously depends on the temperature and material quality in realistic materials. Its derivation is beyond the scope of the present paper.

%%%
%%%%%%%%%%%%%%%%%%%%%%%%%%%

\section{Bulk polaritons in Weyl semimetals}

Consider first the propagation of plane monochromatic waves in a bulk Weyl semimetal. For complex amplitudes of the electric field and induction, $(\boldsymbol{D,E})  e^{i\boldsymbol{kr}-i \omega t}$, where  $\boldsymbol{D} = \hat{\varepsilon} \boldsymbol{E}$ and 
  $\hat{\varepsilon}$ is a bulk dielectric tensor, Maxwell's equations  give $\boldsymbol{n\cdot D}=0$, where  $\boldsymbol{n}=\frac{c\boldsymbol{k}}{ \omega }$. The resulting dispersion equations are 
  \begin{equation}
  \label{WaveEq}
  \boldsymbol{n} \left( \boldsymbol{n\cdot E} \right) -n^{2}\boldsymbol{E}+ \hat{\varepsilon} \boldsymbol{E}=0,
  \end{equation}
  or
\begin{align}
&\begin{pmatrix}
 \varepsilon _{xx}-n^{2}+n_{x}^{2}  &  n_{x}n_{y}  &  n_{x}n_{z}\\
n_{y}n_{x}  &   \varepsilon _{yy}-n^{2}+n_{y}^{2}  &  ig+n_{y}n_{z}\\
n_{z}n_{x}  &  -ig+n_{z}n_{y}  &   \varepsilon _{zz}-n^{2}+n_{z}^{2}\\
\end{pmatrix}
 \begin{pmatrix}
E_{x}\\
E_{y}\\
E_{z}\\
\end{pmatrix}\label{WaveEqMatrixForm}
=0. 
\end{align}
%%%
%%%

The structure of these equations indicate strongly anisotropic and gyrotropic properties of bulk polaritons. The dispersion is drastically different for normal modes propagating perpendicular to the $x$-axis and to the $y$-axis. For each direction, there are furthermore two normal modes with different refractive indices. We will consider each case separately. 

\subsection{Propagation perpendicular to the anisotropy x-axis}

In this case we have  $n_{x}=0$,  $n^{2}=n_{y}^{2}+n_{z}^{2}$,  $n_{z}=n\cos \theta$,  $n_{y}=n\sin \theta$, where  $\theta$  is the angle between the wave vector and $z$-axis. From Eqs.~(\ref{WaveEqMatrixForm}) we obtain 
two normal modes that can be called an ordinary (O) and extraordinary (X) wave. An O-wave has an electric field along $x$ and the refractive index 
\begin{equation}
\label{O-modeRefraction}
n_{O}^{2}= \varepsilon _{xx}.
\end{equation} 
Therefore, its dispersion and absorption are completely described by the spectrum of $\varepsilon _{xx}(\omega)$. As  shown in Fig.~7, at low frequencies the O-mode experiences strong metallic absorption and at $\omega = 2 E_F = 160$ meV there is an onset of interband transitions. 

%%%%%%%%%%%%%%

\begin{figure}[htb]
\begin{center}
\includegraphics[scale=0.4]{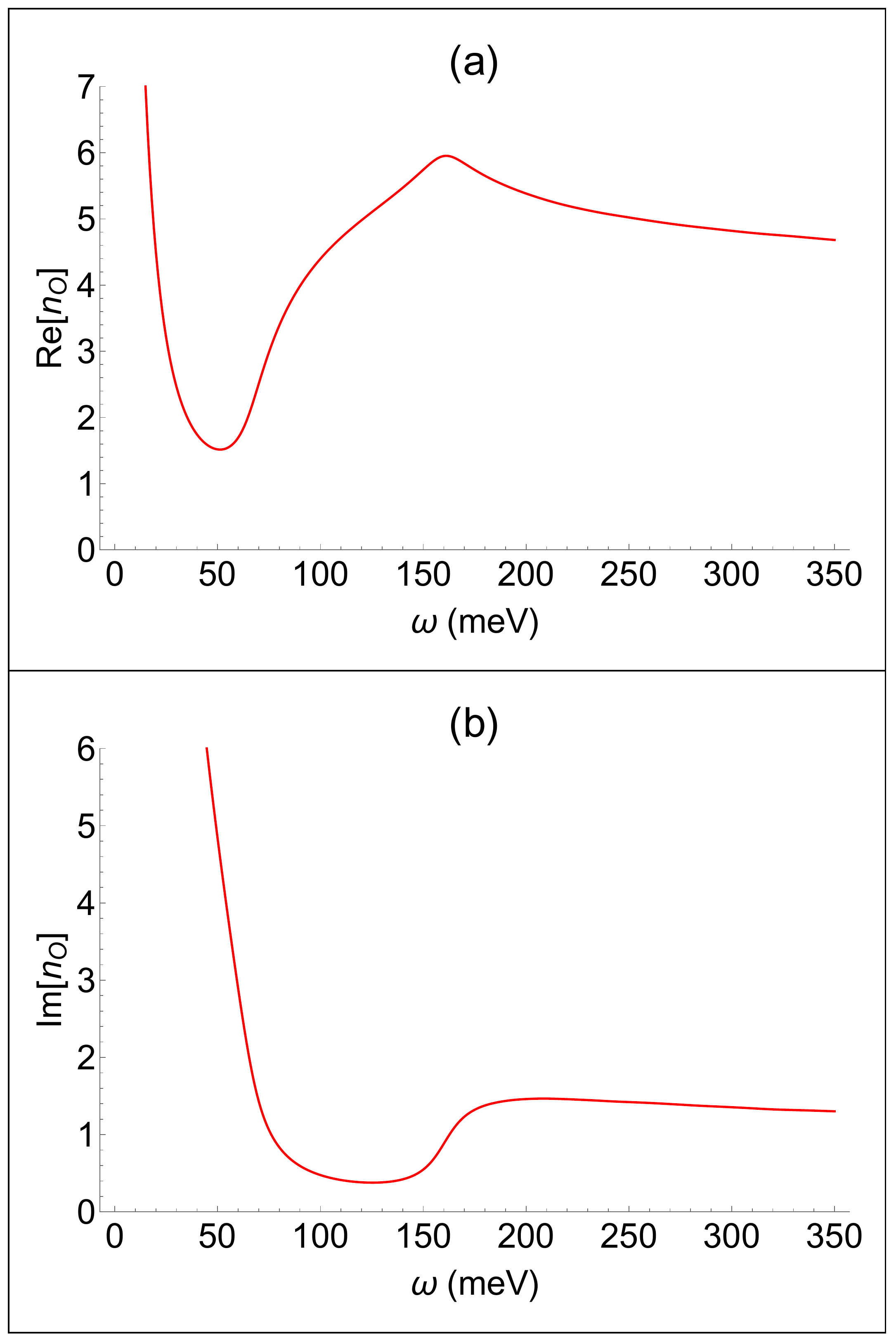}
\caption{Real and imaginary parts of the refractive index $n_O$ of an O-wave as a function of frequency for $E_F = 80$ meV, $\hbar v_F b = 100$ meV, and dephasing rate $\gamma = 10$ meV.   }
\label{Fig:nO}
\end{center}
\end{figure}

%%%%%%%%%%%%%%%

An X-wave have an electric field in the $(y,z)$ plane and the refractive index showing strong $\theta$-dependence and resonances: 
\begin{equation}\label{X-modeRefraction}
n_{X}^{2}=\frac{ \varepsilon _{yy} \varepsilon _{zz}-g^{2}}{\cos^{2} \theta  \varepsilon _{zz}+\sin^{2} \theta  \varepsilon _{yy}}.
\end{equation}
For normal incidence  $\theta =0$, 
\begin{equation}\label{X-modeRefractionNormal}
n_{X}^{2}= \varepsilon _{yy}-\frac{g^{2}}{ \varepsilon _{zz}}.
\end{equation}

It is obvious from  Eq.~(\ref{X-modeRefraction}) that the refractive index for an X-wave is strongly enhanced (singular in the absence of losses) when 
\begin{equation}\label{BulkPlasmonCondition}
\cos^{2} \theta  \varepsilon _{zz}+\sin^{2} \theta  \varepsilon _{yy}=0 
\end{equation}
which corresponds to the bulk plasmon excitation. Indeed, from Maxwell's equations in the Coulomb gauge  one can show that  $\vert \frac{1}{c}\frac{ \partial A}{ \partial t} \vert /\vert \nabla  \varphi  \vert \sim  \vert \frac{ \omega ^{2}}{ \omega ^{2}-c^{2}k^{2}} \vert  \vert \frac{j_{\perp}}{j_{\parallel}} \vert$, where  $\mathbold{j}=\mathbold{j}_{\perp}+\mathbold{j}_{\parallel}$,  $\nabla  \times \mathbold{j}_{\parallel}=0$,  $\nabla \cdot \mathbold{j}_{\perp}=0$. Therefore, if  $\vert \mathbold{j}_{\perp} \vert  \sim  \vert \mathbold{j}_{\parallel} \vert$, which corresponds to a general oblique propagation in an anisotropic medium, the wave is quasi-electrostatic at  $n^{2} \gg 1$. Eq.~(\ref{BulkPlasmonCondition}) corresponds to the condition  $\mathbold{n\cdot D}=0$  for  $\mathbold{E}=-\nabla  \varphi \parallel \mathbold{n}$ . If   $\varepsilon _{yy}= \varepsilon _{zz}= \varepsilon _{\perp}$ the dispersion equation for a plasmon propagating in the plane orthogonal to the $x$-axis has a simple form  $\varepsilon _{\perp}=0$.

%%%%%%%%%%%%%%

\begin{figure}[htb]
\begin{center}
\includegraphics[scale=0.4]{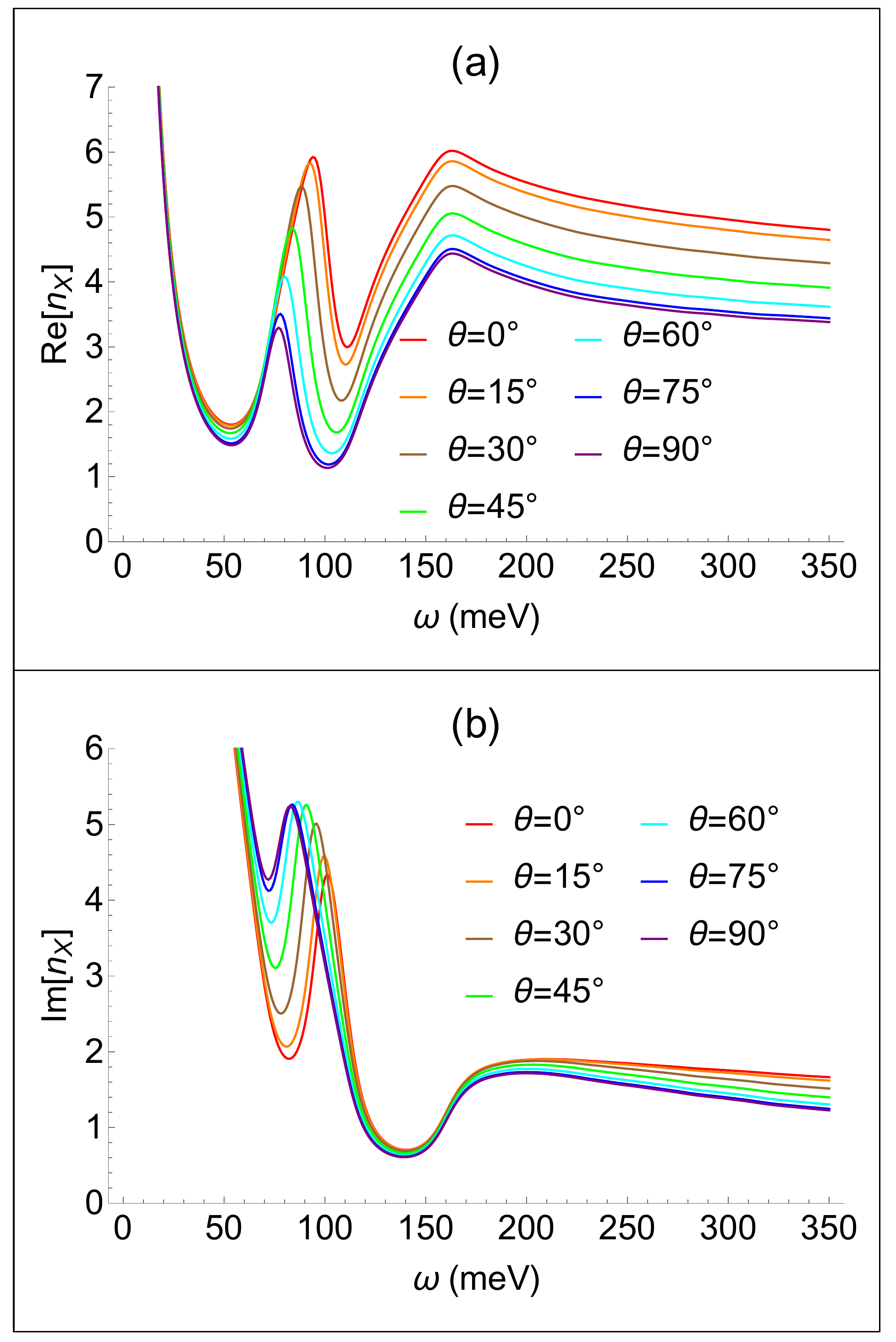}
\caption{Real and imaginary parts of the refractive index $n_X$ of an X-wave as a function of frequency for different values of the propagation angle $\theta$. Other parameters are $E_F = 80$ meV, $\hbar v_F b = 100$ meV, and dephasing rate $\gamma = 10$ meV.   }
\label{Fig:6}
\end{center}
\end{figure}

%%%%%%%%%%%%%%%

Figure 8 shows real and imaginary parts of the refractive index $n_X$ of an X-wave as a function of frequency for different values of the propagation angle $\theta$. Near the bulk plasmon resonance, i.e. around 100 meV for normal incidence, the value of $n_X^2$ becomes negative in the absence of losses according to Eq.~(\ref{X-modeRefractionNormal})This corresponds to a non-propagating photonic gap. Since we include significant loss rate $\gamma = 10$ meV in all simulations, the real part of $n_X$ does not go all the way to zero, but there is a strong absorption peak in the imaginary part of $n_X$.  We will later see that this spectral region leads to a telltale change of phase in reflection.  The second feature in all plots is an onset of interband transitions at $2 E_F = 160$ meV. 

%%%%%%%%%%%%%

\begin{figure}[htb]
\begin{center}
\includegraphics[scale=0.4]{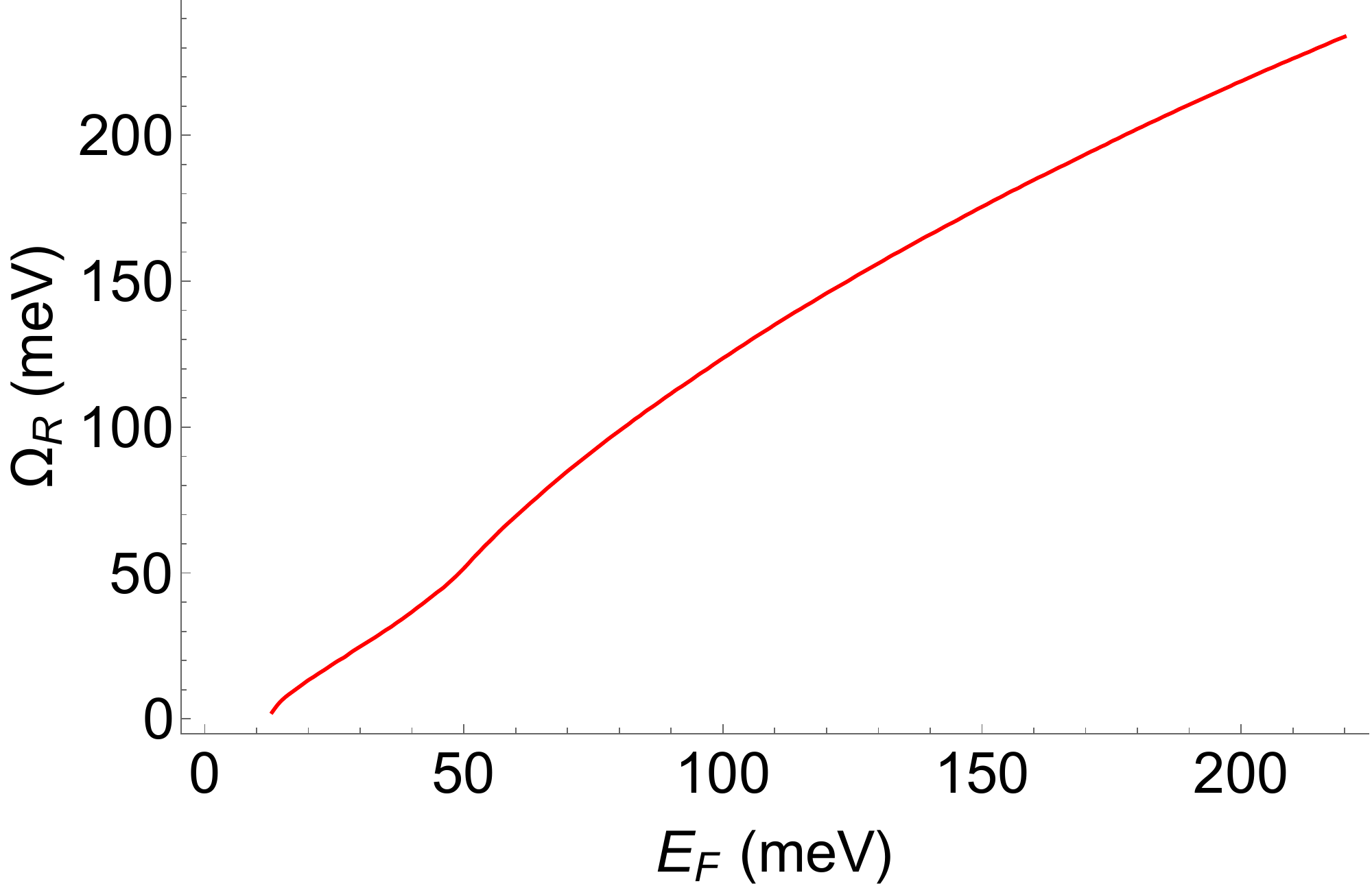}
\caption{Real part of the bulk plasmon resonance frequency at normal incidence $\theta = 0$ as a function of the Fermi energy.   }
\label{Fig:7}
\end{center}
\end{figure}

%%%%%%%%%%%%%%%

The real part of the bulk plasmon resonance frequency at normal incidence as a function of the Fermi energy is shown in Fig.~9. Note that according to Eq.~(\ref{X-modeRefractionNormal}) the magnitude of the refractive index at frequencies around plasmon resonance is determined by the value of the off-diagonal component of the dielectric tensor $g$. Therefore, measurements of the transmission and reflection provide a sensitive measure of the Weyl node separation. 

The same is true about the polarization effects. From the third row of Eqs.~(\ref{WaveEqMatrixForm}) one can get the expression for the polarization coefficient: 
\begin{equation}\label{PolarizationCoefficientKX}
K_{X}=\frac{E_{z}}{E_{y}}=\frac{ig-n_{X}^{2}\sin \theta \cos \theta }{ \varepsilon _{zz}-n_{X}^{2}\sin^{2} \theta }.
\end{equation}
Substituting Eq.~(\ref{X-modeRefraction}) into Eq.~(\ref{PolarizationCoefficientKX}) we get
\begin{equation}\label{ExpandedPolCoefKX}
K_{X}=\frac{ig \left( \cos^{2} \theta  \varepsilon _{zz}+\sin^{2} \theta  \varepsilon _{yy} \right) - \left(  \varepsilon _{yy} \varepsilon _{zz}-g^{2} \right) \sin \theta \cos \theta }{ \varepsilon _{zz} \left( \cos^{2} \theta  \varepsilon _{zz}+\sin^{2} \theta  \varepsilon _{yy} \right) - \left(  \varepsilon _{yy} \varepsilon _{zz}-g^{2} \right) \sin^{2} \theta }.
\end{equation}

At the resonant plasmon frequency defined by  $\cos^{2} \theta  \varepsilon _{zz}+\sin^{2} \theta  \varepsilon _{yy}=0$  we obtain  $K_{X}=\frac{1}{\tan \theta }$, which is expected. If we set  $\theta =0$, which corresponds to normal incidence, $K_{X}=\frac{ig}{ \varepsilon _{zz}}$, i.e. again proportional to $g$. In this case, the plasmon frequency is given by  $\varepsilon _{zz}=0$, and  $K_{X} \rightarrow \infty$ in the absence of losses.  If $\varepsilon _{yy}= \varepsilon _{zz}= \varepsilon _{\perp}$, Eq.~(\ref{ExpandedPolCoefKX}) gives
\begin{equation}
K_{X}=\frac{ig \varepsilon _{\perp}- \left(  \varepsilon _{\perp}^{2}-g^{2} \right) \sin \theta \cos \theta }{ \varepsilon _{\perp}^{2}\cos^{2} \theta +g^{2}\sin^{2} \theta }.
\end{equation}

For an isotropic medium, when  $g^{2}=0$, the last expression gives  $K_{X}=- \tan \theta$, as it should be for a transverse wave in an isotropic medium. 
%%%
%%%
%%%%%%%%%%%%%%%%%

\subsection{Propagation transverse to the $y$-axis}

In this case $n_{y}=0$, $n^{2}=n_{x}^{2}+n_{z}^{2}$, $n_{x}=n\cos\phi$, $n_{z}=n\sin \phi$;
\begin{equation}
\begin{pmatrix}
 \varepsilon _{xx}-n_{z}^{2}  &  0  &  n_{x}n_{z}\\
0  &   \varepsilon _{yy}-n^{2}  &  ig\\
n_{z}n_{x}  &  -ig  &   \varepsilon _{zz}-n_{x}^{2}\\
\end{pmatrix}
 \begin{pmatrix}
E_{x}\\
E_{y}\\
E_{z}\\
\end{pmatrix}
=0 \label{PropPerpToYaxis}
\end{equation}
\begin{align}
\left( \sin^{2} \phi  \varepsilon _{zz}+\cos^{2} \phi  \varepsilon _{xx} \right) n^{4}-n^{2} \left[  \varepsilon _{xx} \varepsilon _{zz}+ \varepsilon _{yy} \left( \sin^{2} \phi  \varepsilon _{zz}+\cos^{2} \phi  \varepsilon _{xx} \right) -\sin^{2} \phi g^{2} \right]&\nonumber\\
+ \varepsilon _{xx} \left(  \varepsilon _{yy} \varepsilon _{zz}-g^{2} \right) = 0.&\label{DetPropPerpToYaxis}
\end{align}
Note that the solution of Eq.~(\ref{DetPropPerpToYaxis}) at $\phi =\frac{ \pi }{2}$ corresponds to the normal incidence propagation along $z$ and therefore should coincide with Eqs.~(\ref{O-modeRefraction}), (\ref{X-modeRefraction}) at $\theta =0$. Indeed, from Eq.~(\ref{DetPropPerpToYaxis}) for $\phi =\frac{ \pi }{2}$ we obtain
\begin{align}
&\left( n^{2}- \varepsilon _{xx} \right)  \left[ n^{2}- \left(  \varepsilon _{yy}-\frac{g^{2}}{ \varepsilon _{zz}} \right)  \right] =0;
\label{FactoredDetPropPerpToYaxis}
\end{align}
from which $n_{O}^{2}= \varepsilon _{xx}$, $n_{X}^{2}= \varepsilon _{yy}-\frac{g^{2}}{ \varepsilon _{zz}}$, as expected.

The case $n^{2} \rightarrow \infty$ in the absence of losses, when
\begin{equation}\label{PlasmonConditionPropPerpToYaxis}
\sin^{2} \phi  \varepsilon _{zz}+\cos^{2} \phi  \varepsilon _{xx}=0 
\end{equation}
corresponds to the condition $\mathbold{n\cdot D}=0$  where $\mathbold{E}=-\nabla  \varphi \parallel \mathbold{n}$.  From Eq.~(\ref{DetPropPerpToYaxis}) we obtain
\begin{align}
&n_{O,X}^{2}=\frac{ \varepsilon _{xx} \varepsilon _{zz}+ \varepsilon _{yy} \left( \sin^{2} \phi  \varepsilon _{zz}+\cos^{2} \phi  \varepsilon _{xx} \right) -\sin^{2} \phi g^{2}}{2 \left( \sin^{2} \phi  \varepsilon _{zz}+\cos^{2} \phi  \varepsilon _{xx} \right) } \pm\nonumber\\
&\frac{\sqrt{ \left[  \varepsilon _{xx} \varepsilon _{zz}+ \varepsilon _{yy} \left( \sin^{2} \phi  \varepsilon _{zz}+\cos^{2} \phi  \varepsilon _{xx} \right) -\sin^{2} \phi g^{2} \right] ^{2}-4 \left( \sin^{2} \phi  \varepsilon _{zz}+\cos^{2} \phi  \varepsilon _{xx} \right)  \varepsilon _{xx} \left(  \varepsilon _{yy} \varepsilon _{zz}-g^{2} \right) }}{2 \left( \sin^{2} \phi  \varepsilon _{zz}+\cos^{2} \phi  \varepsilon _{xx} \right) } \label{nOandnX}
\end{align}
In Eq.~(\ref{nOandnX}) the signs $\pm$ are chosen for $n_{O,X}^{2}$ according to the limiting case $\phi =\frac{ \pi }{2}$.

%%%
%%%

%%%%%%%%%%%%%%%%
\begin{figure}[htb]
\begin{center}
\includegraphics[scale=0.4]{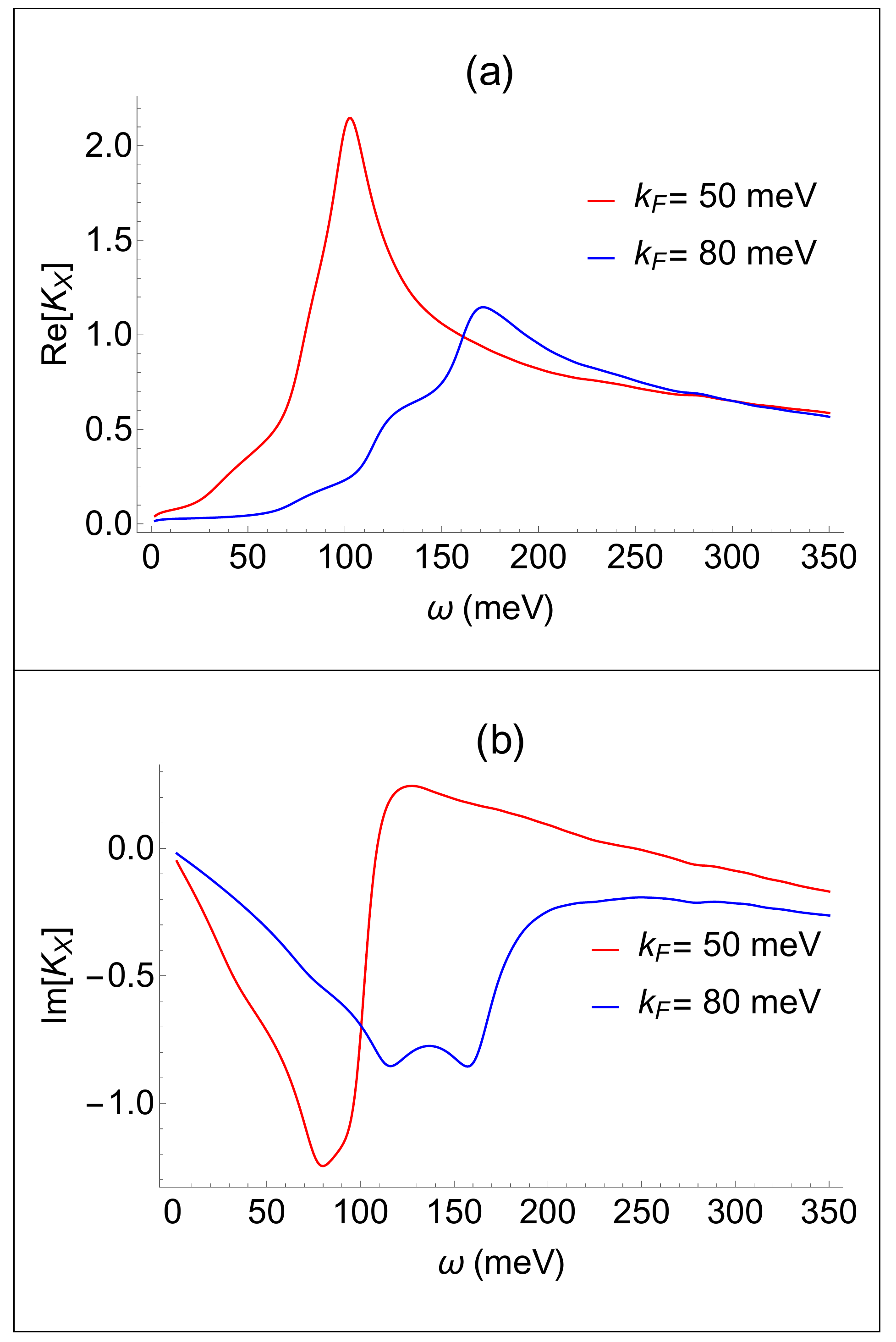}
\caption{Spectra of real and imaginary parts of the polarization coefficient $K_X = E_z/E_y$ for an incident wave linearly polarized in y-direction after traversing a 1-$\mu$m film in x-direction.  }
\label{Fig:8}
\end{center}
\end{figure}

%%%%%%%%%%%%%%%%% 

For the propagation along the $x$-axis of anisotropy, when $\phi =0$, Eq.~(\ref{DetPropPerpToYaxis}) gives
\begin{equation}\label{nOandnXPhiEqualZero}
n_{O,X}^{2}=\frac{ \varepsilon _{zz}+ \varepsilon _{yy}}{2} \pm \sqrt{ \left( \frac{ \varepsilon _{zz}- \varepsilon _{yy}}{2} \right) ^{2}+g^{2}}
\end{equation}
Note that the $x$-axis is also a gyrotropy axis related to the Weyl node separation along $x$. Therefore, the propagation along $x$ is similar to the Faraday  geometry in a magnetic field. In our case  the normal modes are elliptically polarized, and an incident linearly polarized wave experiences Faraday rotation and gains ellipticity after traversing a sample in $x$-direction. To quantify the effect, Fig.~10 shows the polarization coefficient $K_X = E_z/E_y$ after traversing a 1-$\mu$m thick film for a wave initially linearly polarized in $y$-direction. The real part of $K_X$ is a measure of the polarization rotation whereas its imaginary part is a measure of ellipticity. Clearly, a rotation by $\sim \pi/2$ by very thin (0.5-1 $\mu$m) Weyl semimetal films is possible at frequencies near the interband absorption edge. This is a giant Faraday rotation, comparable to the one observed at THz frequencies in narrow-gap semiconductors in the vicinity of a cyclotron resonance in Tesla-strength magnetic fields; see e.g. \cite{arikawa} for the review. Note that in our case no magnetic field is needed and the effect is controlled by the Weyl node separation and by the Fermi level. Previously Faraday rotation and nonreciprocity in light propagation associated with it was studied in  \cite{kargarian2015, kotov2018} using the model with an axion $\theta$-term in the electromagnetic field action. 

\subsection{Oblique propagation of bulk polaritons }

In the general case the direction of the wave vector is determined by two angles $\theta$  and $\phi$:
\begin{equation}\nonumber
n_{x}=n\cos \phi  \,,  n_{z}=n\sin \phi \cos \theta  \,,  n_{y}=n\sin \phi \sin \theta .
\end{equation}
The general expression for $n_{O,X}^{2}$ is quite cumbersome. At the same time, in the particular case of  $\varepsilon _{yy}= \varepsilon _{zz}= \varepsilon _{\perp}$,  the result should not depend on the angle  $\theta$  and should coincide with the one for a magnetized plasma: 
\begin{align}
&n_{O,X}^{2}=\frac{ \varepsilon _{\perp} \left[  \varepsilon _{xx} \left( 1+\cos^{2} \phi  \right) +\sin^{2} \phi  \varepsilon _{\perp} \right] -\sin^{2} \phi g^{2}}{2 \left( \sin^{2} \phi  \varepsilon _{\perp}+\cos^{2} \phi  \varepsilon _{xx} \right) } \pm\nonumber\\ 
&\frac{\sqrt{ \left(  \varepsilon _{\perp} \left[  \varepsilon _{xx} \left( 1+\cos^{2} \phi  \right) +\sin^{2} \phi  \varepsilon _{\perp} \right] -\sin^{2} \phi g^{2} \right) ^{2}-4 \varepsilon _{xx} \left( \sin^{2} \phi  \varepsilon _{\perp}+\cos^{2} \phi  \varepsilon _{xx} \right)  \left(  \varepsilon _{\perp}^{2}-g^{2} \right) }}{2 \left( \sin^{2} \phi  \varepsilon _{\perp}+\cos^{2} \phi  \varepsilon _{xx} \right) }\label{nOandnXOblique}
\end{align}

The condition $\mathbold{n\cdot D}=0$  at $\mathbold{E}=-\nabla  \varphi \parallel \mathbold{n}$  in the case of an oblique propagation gives 
\begin{equation}\label{ObliquePlasmonCondition}
\varepsilon _{xx}\cos^{2} \phi +\sin^{2} \phi  \left( \sin^{2} \theta  \varepsilon _{yy}+\cos^{2} \theta  \varepsilon _{zz} \right) =0.
\end{equation}
Therefore, Eq.~(\ref{ObliquePlasmonCondition}) determines the frequencies of bulk plasmons in the general case. Under the condition  $\varepsilon _{yy}= \varepsilon _{zz}= \varepsilon _{\perp}$  the plasmon dispersion equation takes a form similar to plasmons in a magnetized plasma: 
\begin{equation}\label{ObliquePlasmonConditionAlt}
\varepsilon _{xx}\cos^{2} \phi +\sin^{2} \phi  \varepsilon _{\perp}=0.
\end{equation}
%%%
%%%
%%%%%%%%%%%%%%%%%%

\section{Boundary conditions}

So far we considered propagation and transmission of electromagnetic waves in bulk samples. Now we turn to effects of reflection and surface wave propagation that are equally sensitive to the electronic structure of WSMs. Moreover, in many situations they are easier to observe than bulk propagation effects. 

We start with the derivation of the boundary conditions at $z = 0$ surface. Assume that there is an isotropic dielectric medium  with dielectric constant  $n_{up}^{2}= \varepsilon _{up}$  above a WSM.  The boundary conditions include:   

(i) Gauss' law for the normal components of the electric induction vector: 
\begin{equation}\label{B.C.fromGaussLaw}
\varepsilon _{up}E_{z} \left( z=+0 \right) -D_{z} \left( z=-0 \right) =4 \pi  \rho ^{S}=-i\frac{4 \pi }{ \omega } \left( \frac{ \partial }{ \partial x}j_{x}^{S}+\frac{ \partial }{ \partial y}j_{y}^{S} \right)  
\end{equation}
where  $\rho ^{S}$, $j_{x}^{S}$ and $j_{y}^{S}$ are the surface charge and components of the surface current that are connected by the continuity equation. For the wave field we have $\frac{ \partial }{ \partial x, \partial y} \rightarrow ik_{x,y}$.

(ii) Equations for the magnetic field components:
\begin{align}
&B_{z} \left( z=-0 \right) =B_{z} \left( z=+0 \right),\label{MagneticFieldZ-Component}\\
&B_{y} \left( z=+0 \right) -B_{y} \left( z=-0 \right) =-\frac{4 \pi }{c}j_{x}^{S},\label{MagneticFieldY-Component}\\   
&B_{x} \left( z=+0 \right) -B_{x} \left( z=-0 \right) =\frac{4 \pi }{c}j_{y}^{S}. \label{MagneticFieldX-Component}
\end{align}
Due to the presence of the components of the surface conductivity   $\sigma _{zz}^{S}$  and  $\sigma _{zy}^{S}=- \sigma _{yz}^{S}$ a surface dipole layer is formed at the boundary between the two media. Its dipole moment is  
\begin{align}
&\mathbold{d}=\Re \left[ \mathbold{z_{0}}d_{z}e^{-i \omega t+ik_{x}x+ik_{y}y} \right],\nonumber\\
&d_{z}=\frac{i}{ \omega } \left[  \sigma _{zy}^{S}E_{y} \left( z=-0 \right) + \sigma _{zz}^{S}E_{z} \left( z=-0 \right)  \right].\label{SurfaceDipoleMoment}
\end{align}
Note that when dealing with a surface response, we will always choose the fields  at $z=-0$  in Eq.~(\ref{SurfaceDipoleMoment}) and similar relationships. The presence of the dipole layer changes the boundary conditions for the tangential field components of $\mathbold{E}$. Consider Maxwell's equations 
\begin{equation}
 \frac{ \partial E_{z}}{ \partial y}-\frac{ \partial E_{y}}{ \partial z}=i\frac{ \omega }{c}B_{x},\, \frac{ \partial E_{x}}{ \partial z}-\frac{ \partial E_{z}}{ \partial x}=i\frac{ \omega }{c}B_{y} \nonumber.
\end{equation}
For convenience, let's assume that the dipole layer has a small but finite thickness $L$:
\[  \vert k_{x,y} \vert L \ll 1\quad \text{and} \quad  \frac{ \omega }{c}L \ll 1. \]
Using  $\frac{ \partial }{ \partial x, \partial y} \rightarrow ik_{x,y} $  and integrating   $  \int _{-\frac{L}{2}}^{\frac{L}{2}} \ldots  dz $ , we obtain
\begin{equation}\label{EzandExRelationship}
ik_{x,y} \int _{-\frac{L}{2}}^{\frac{L}{2}}E_{z}\, dz=E_{x,y} \left( z=\frac{L}{2} \right) -E_{x,y} \left( z=-\frac{L}{2} \right)
\end{equation}
We neglect the integral over the magnetic field components assuming that  $ \frac{ \omega }{c}L \rightarrow 0 $. Next we use Gauss' law under the condition  $  \vert k_{x,y} \vert L \rightarrow 0$, which will yield in the region of the dipole layer: 
\[ \frac{ \partial E_{z}}{ \partial z}=4 \pi  \rho  \left( z \right),\quad \rho  \left( z \right) =- \left( \frac{ \partial P_{z}}{ \partial z}+\frac{ \partial p_{z}}{ \partial z} \right).\]
Here  $ P_{z} $  is a component of the volume polarization whereas  $ p_{z} $  describes the distribution of the polarization in the dipole layer, so that 
\[ \int _{-\frac{L}{2}}^{\frac{L}{2}}\frac{ \partial p_{z}}{ \partial z}\, dz=0\quad  \text{and}\quad  \int _{-\frac{L}{2}}^{\frac{L}{2}}p_{z}\, dz=d_{z}.\]
Substituting  $ E_{z}=-4 \pi  \left( P_{z}+p_{z} \right)  $   into Eq.~(\ref{EzandExRelationship})  and integrating over $dz$ at  $  \vert k_{x,y} \vert L \rightarrow 0 $  and finite  $ P_{z} $ , we obtain

\begin{equation}\label{TangentialEandDipole}
E_{x,y} \left( z=\frac{L}{2} \right) -E_{x,y} \left( z=-\frac{L}{2} \right) =-i4 \pi k_{x,y}d_{z} 
\end{equation}
The boundary condition Eq.~(\ref{TangentialEandDipole}) looks  unusual but it can be easily deduced from the radiation field of an individual dipole. 
%%%
%%%

Figures 11-14 show spectra of the surface conductivity components  for different values of the Fermi momentum. Note that the surface conductivity in Gaussian units has a dimension of velocity and its value is normalized by $e^2/(2 \pi \hbar)\simeq 3.5 \times 10^7$ cm/s in all plots. In contrast with the bulk conductivity, the surface conductivity had a Drude-like behavior at low frequencies only for the $yy$-component because of the surface state dispersion $E = - \hbar v_F k_y$. The surface optical response decreases with increasing Fermi energy and vanishes when all surface states within $k_x^2 + k_y^2 < b^2$ are occupied. 
%%%%%%%%%%%%%%%%%%%%

\begin{figure}[htb]
\begin{center}
\includegraphics[scale=0.4]{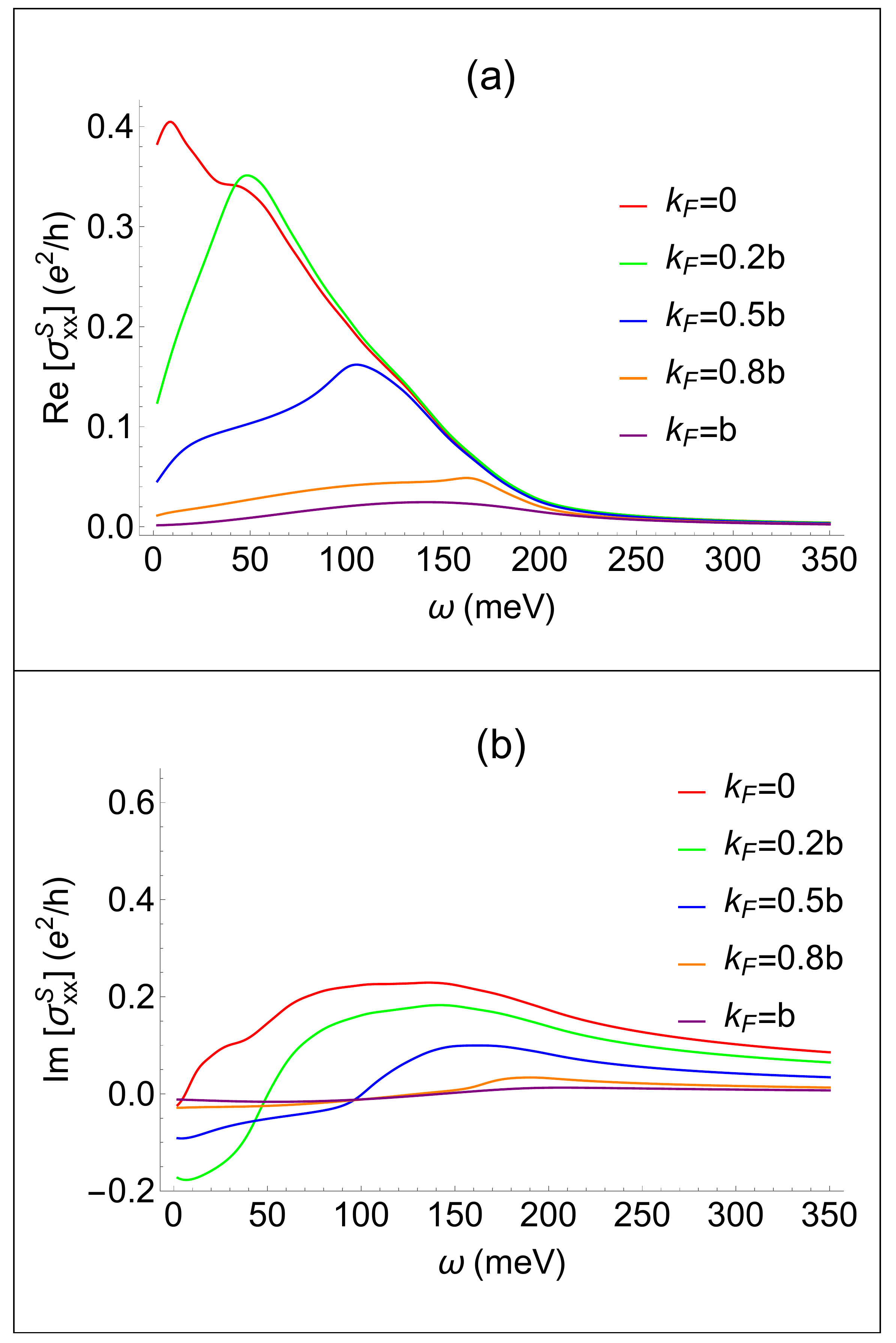}
\caption{Spectra of the real and imaginary parts of the $xx$ component of the surface conductivity at several values of the Fermi momentum for $\hbar v_F b = 100$ meV and dephasing rate $\gamma = 10$ meV.   }
\label{Fig:9}
\end{center}
\end{figure}

\begin{figure}[htb]
\begin{center}
\includegraphics[scale=0.4]{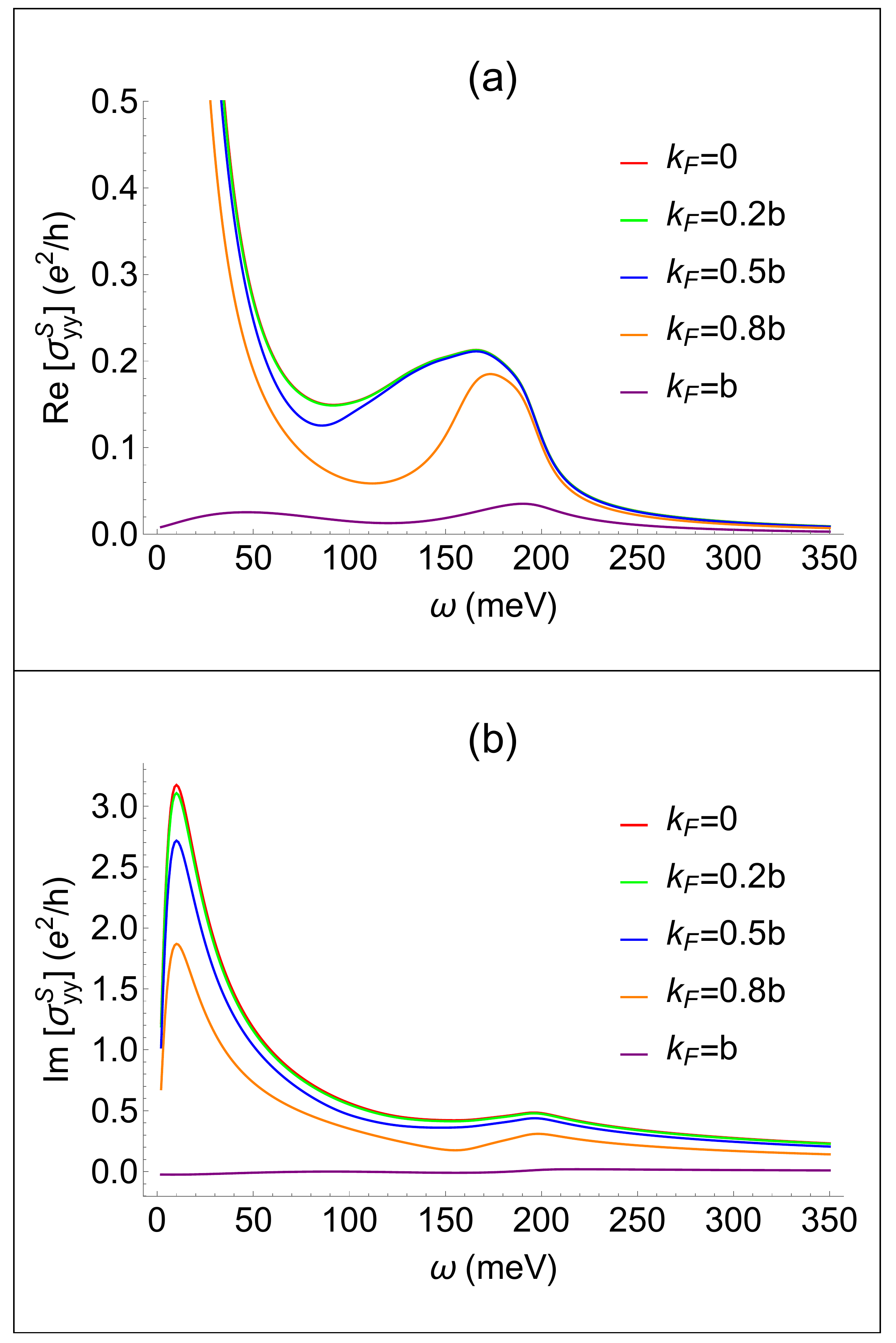}
\caption{Spectra of the real and imaginary parts of the $yy$ component of the surface conductivity at several values of the Fermi momentum for $\hbar v_F b = 100$ meV and dephasing rate $\gamma = 10$ meV.   }
\label{Fig:10}
\end{center}
\end{figure}

\begin{figure}[htb]
\begin{center}
\includegraphics[scale=0.4]{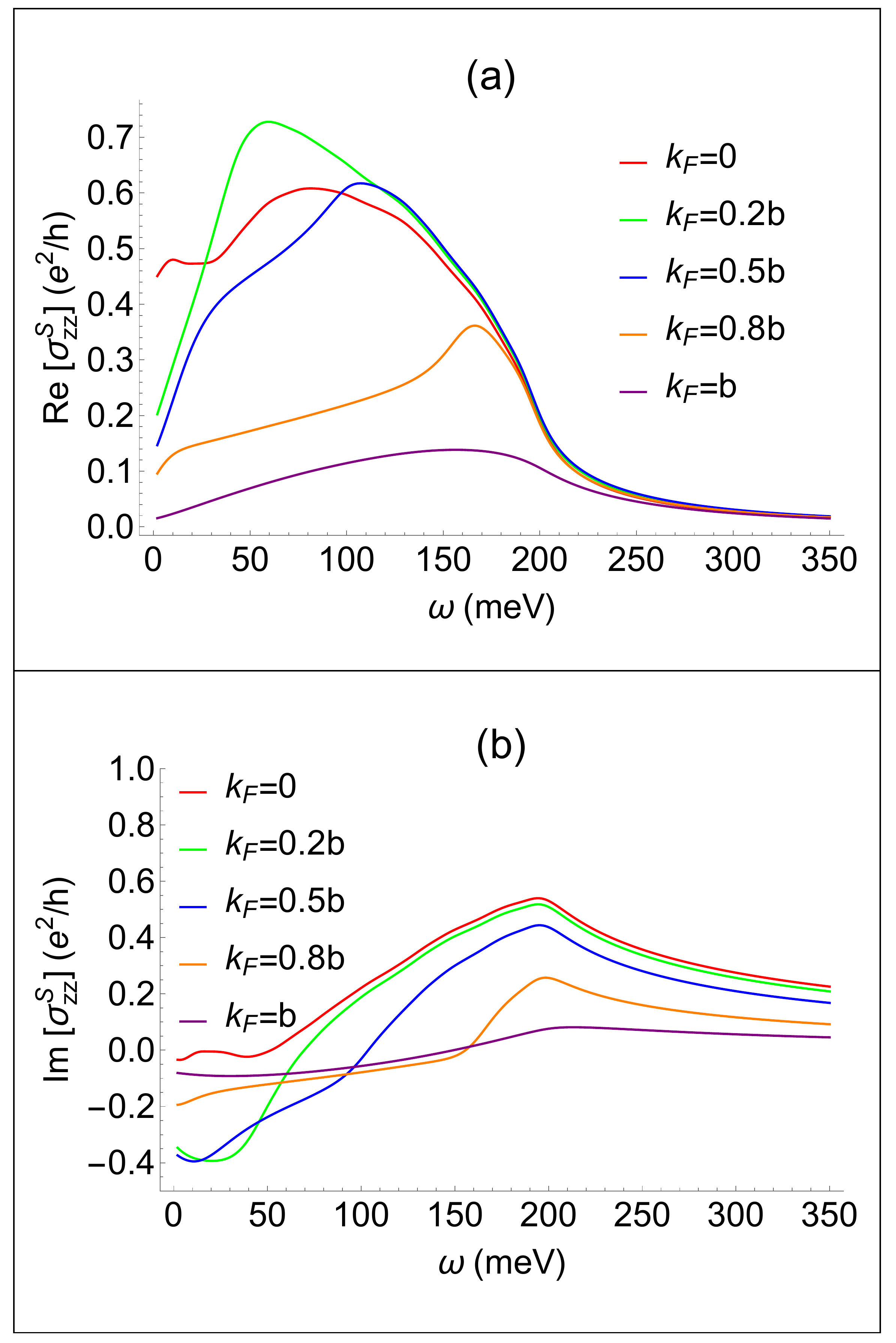}
\caption{Spectra of the real and imaginary parts of the $zz$ component of the surface conductivity at several values of the Fermi momentum for $\hbar v_F b = 100$ meV and dephasing rate $\gamma = 10$ meV.   }
\label{Fig:11}
\end{center}
\end{figure}

\begin{figure}[htb]
\begin{center}
\includegraphics[scale=0.4]{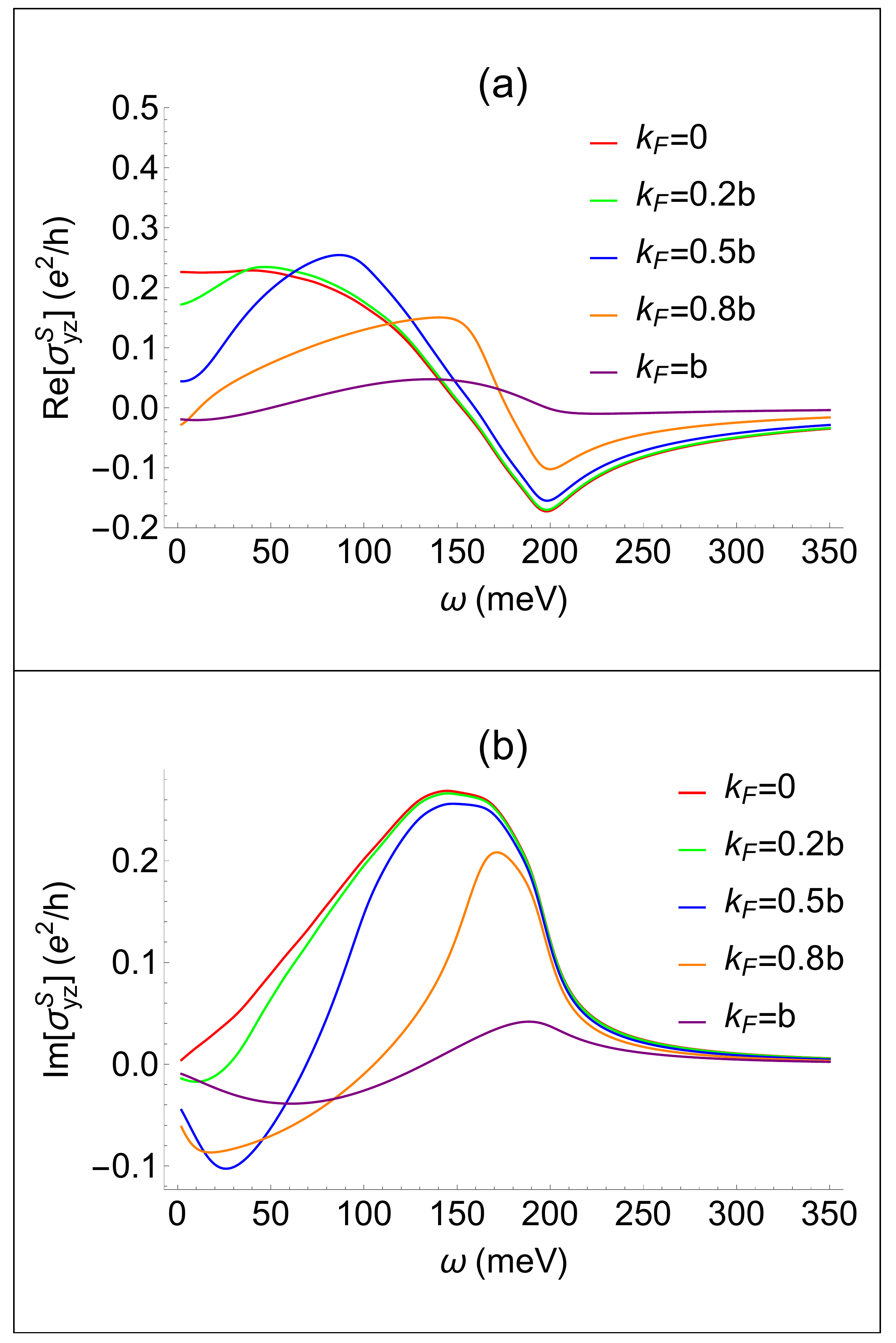}
\caption{Spectra of the real and imaginary parts of the $yz$ component of the surface conductivity at several values of the Fermi momentum for $\hbar v_F b = 100$ meV and dephasing rate $\gamma = 10$ meV.   }
\label{Fig:12}
\end{center}
\end{figure}

%%%%%%%%%%%%%%%%%%%%%%%%%%%%%%

\section{Reflection from the surface of a Weyl semimetal}

Consider radiation incident from a medium with refractive index $n_{up}$ on a WSM at an angle $\theta$ between the wavevector of the wave and the normal  to a WSM. For simplicity consider the propagation transverse to the $x$-axis. The reflection spectra provide information about both bulk and surface conductivity components. Here we will pay particular attention to the case when the contribution of the surface states becomes significant or dominant, thus allowing one to probe surface states by optical means. 
%%%
%%%

\subsection{Reflection with excitation of an $O$-mode}

In this geometry, the complex amplitudes of the electric field of the incident $E_1$, reflected $E_2$, and transmitted $E_O$ wave are parallel to the $x$-axis. The refractive index of the transmitted wave is  $ n_{O}^{2}= \varepsilon _{xx}= \varepsilon _{xx}^{ \left( 0 \right) }+i\frac{4 \pi }{ \omega } \sigma _{xx}^{B} $  (see Eq.~(\ref{O-modeRefraction})).

Applying Maxwell's equations with standard boundary conditions including the surface current, we arrive at 
\begin{align}
R=\frac{E_{2}}{E_{1}}&= -\frac{\cos \theta _{O}\sqrt{ \varepsilon _{xx}^{ \left( 0 \right) }+i\frac{4 \pi }{ \omega } \sigma _{xx}^{B}}+\frac{4 \pi }{c} \sigma _{xx}^{S}-\cos \theta n_{up}}{\cos \theta _{O}\sqrt{ \varepsilon _{xx}^{ \left( 0 \right) }+i\frac{4 \pi }{ \omega } \sigma _{xx}^{B}}+\frac{4 \pi }{c} \sigma _{xx}^{S}+\cos \theta n_{up}} \label{O-modeReflection}
\end{align}
where  $ n_{up}\sin \theta =n_{O}\sin \theta _{O} $.  Assuming  $\sigma _{xx}^{S}=0 $  we obtain  $ R=\frac{E_{2}}{E_{1}}=\frac{\cos \theta n_{up}-\cos \theta _{O}n_{O}}{\cos \theta _{O}n_{O}+\cos \theta n_{up}} $, which is a standard Fresnel formula.

For the same magnitude of  $  \sigma _{xx}^{S}$,  the relative contribution of surface states to the reflected field depends on the parameter  $ \frac{ \vert  \varepsilon _{xx}^{ \left( 0 \right) } \vert }{4 \pi  \vert  \sigma _{xx}^{B} \vert/ \omega} $. If  $ \frac{ \omega  \vert  \varepsilon _{xx}^{ \left( 0 \right) } \vert }{4 \pi  \vert  \sigma _{xx}^{B} \vert } \gg 1 $, the relative contribution of surface states is determined by the expression:  $ \frac{2 \omega  \vert  \sigma _{xx}^{S} \vert /c}{\vert  \sigma _{xx}^{B} \vert /\vert  \varepsilon _{xx}^{ \left( 0 \right) } \vert } $. If  $ \frac{ \omega  \vert  \varepsilon _{xx}^{ \left( 0 \right) } \vert }{4 \pi  \vert  \sigma _{xx}^{B} \vert } \ll 1 $, one needs to evaluate the ratio  $ \frac{2\sqrt{ \pi } \sigma _{xx}^{S}/c}{\sqrt{ \sigma _{xx}^{B}/\omega }} $.
%%%
%%%

\subsection{Reflection with excitation of an $X$-mode}

In this geometry, the complex Fourier harmonics for the incident and reflected waves are $$ 
 \left( \mathbold{y_{0} \mp z_{0}}\tan \theta  \right) E_{1,2}e^{ \mp i\frac{ \omega }{c}n_{up}\cos \theta z-i\frac{ \omega }{c}n_{up}\sin \theta y-i \omega t}. $$ 
The transmitted wave is 
$$
\left(   \mathbold{y_{0}+z_{0}}K_{X} \right) E_{X}e^{-i\frac{ \omega }{c}n_{X}\cos \theta _{X}z-i\frac{ \omega }{c}n_{X}\sin \theta _{X}y-i \omega t}, $$
where  $ n_{X}^{2} $  and  $ K_{X} $  are given by Eqs.~(\ref{X-modeRefraction}) and (\ref{PolarizationCoefficientKX}), in which one should substitute  $  \theta  \rightarrow  \theta _{X} $. 
The corresponding complex amplitudes of the magnetic field are
 $ B_{1x}=\frac{n_{up}}{\cos \theta }E_{1}$, 
$ B_{2x}=-\frac{n_{up}}{\cos \theta }E_{2}$, 
$ B_{ \left( X \right) x}=n_{X} \left( \cos \theta _{X}-\sin \theta _{X}K_{X} \right) E_{X}$. 

At the plasmon frequency, when  $ K_{X}=\frac{1}{\tan \theta _{X}} $, the last equation gives  $ B_{ \left( X \right) x}=0 $, as should be expected. For an isotropic medium, when  $ K_{X}=- \tan \theta _{X} $, we obtain  $ B_{ \left( X \right) x}=\frac{n_{X}}{\cos \theta _{X}}E_{X} $  which is also expected for a transverse wave (note that  $ E_{X} $  is an amplitude of the y-component of the extraordinary (X-)mode).

We will use the boundary conditions
\begin{equation}\label{X-modeB.C.1}
E_{1}+E_{2}-E_{X}=i \omega \frac{4 \pi }{c}n_{up}\sin \theta d_{z},\quad d_{z}=\frac{i}{ \omega } \left(  \sigma _{zy}^{S}+ \sigma _{zz}^{S}K_{X} \right) E_{X}
\end{equation}
\begin{equation}\label{X-modeB.C.2}
\frac{n_{up}}{\cos \theta } \left( E_{1}-E_{2} \right) -n_{X} \left( \cos \theta _{X}-\sin \theta _{X}K_{X} \right) E_{X}=\frac{4 \pi }{c}j_{y}^{S},\quad j_{y}^{S}= \left(  \sigma _{yy}^{S}+ \sigma _{yz}^{S}K_{X} \right) E_{X} 
\end{equation}
to obtain
\begin{align}
R&=\frac{E_{2}}{E_{1}}\nonumber\\
&=\frac{n_{up} \left[ 1-\frac{4 \pi }{c}n_{up}\sin \theta  \left(  \sigma _{zy}^{S}+ \sigma _{zz}^{S}K_{X} \right)  \right] -n_{X}\cos \theta  \left( \cos \theta _{X}-\sin \theta _{X}K_{X} \right) +\frac{4 \pi }{c}\cos^{2} \theta  \left(  \sigma _{yy}^{S}+ \sigma _{yz}^{S}K_{X} \right) }{n_{X}\cos \theta  \left( \cos \theta _{X}-\sin \theta _{X}K_{X} \right) +\frac{4 \pi }{c}\cos^{2} \left(  \sigma _{yy}^{S}+ \sigma _{yz}^{S}K_{X} \right) +n_{up} \left[ 1-\frac{4 \pi }{c}n_{up}\sin \theta  \left(  \sigma _{zy}^{S}+ \sigma _{zz}^{S}K_{X} \right)  \right] } \label{X-modeReflection}
\end{align}
where  $ n_{up}\sin \theta =n_{X}\sin \theta _{X} $. In the limit of an isotropic medium, where  $ K_{X}=- \tan \theta _{X} $,  $  \sigma _{ij}^{S}=0 $, we obtain  $ R=\frac{E_{2}}{E_{1}}=\frac{n_{up}\cos \theta _{X}-n_{X}\cos \theta }{n_{X}\cos \theta +n_{up}\cos \theta _{X}} $  which is a standard Fresnel equation.

For the normal incidence the expressions are simplified: 
\begin{equation}\nonumber
n_{X}^{2}= \varepsilon _{yy}-\frac{g^{2}}{ \varepsilon _{zz}}= \varepsilon _{yy}^{ \left( 0 \right) }+i\frac{4 \pi }{ \omega } \sigma _{yy}^{B} - \frac{ \left( \frac{4 \pi  \sigma _{yz}^{B}}{ \omega } \right) ^{2}}{ \varepsilon _{zz}^{ \left( 0 \right) }+i\frac{4 \pi }{ \omega } \sigma _{zz}^{B}},\quad K_{X}=\frac{ig}{ \varepsilon _{zz}}=i\frac{\frac{4 \pi  \sigma _{yz}^{B}}{ \omega }}{ \varepsilon _{zz}^{ \left( 0 \right) }+i\frac{4 \pi }{ \omega } \sigma _{zz}^{B}},
\end{equation}
which gives
\begin{equation}\label{X-modeReflectionNormalIncidence}
R=\frac{n_{up}-n_{X}+\frac{4 \pi }{c} \left(  \sigma _{yy}^{S}+i \sigma _{yz}^{S}\frac{g}{ \varepsilon _{zz}} \right) }{n_{up}+n_{X}+\frac{4 \pi }{c} \left(  \sigma _{yy}^{S}+i \sigma _{yz}^{S}\frac{g}{ \varepsilon _{zz}} \right) } 
\end{equation}

The contribution of surface states is less trivial for X-mode excitation as compared to the excitation of an $O$-mode. For normal incidence (see Eq.~(\ref{X-modeReflectionNormalIncidence})) one can see that at the plasmon resonance frequency, when  $  \varepsilon _{zz} \rightarrow 0 $ in the absence of losses, the contribution of the surface conductivity can become dominant. Indeed, in Eq.~(\ref{X-modeReflectionNormalIncidence}) the term  $  \sigma _{yz}^{S}\frac{g}{ \varepsilon _{zz}} $  diverges as  $ \frac{1}{ \varepsilon _{zz}} $, whereas the refractive index  $ n_{X} $  diverges weaker, as  $ \frac{1}{\sqrt{ \varepsilon _{zz}}}$. When  $  \sigma _{ij}^{S}=0 $  while  $ n_{X} \gg n_{up} $  we have  $ R=-1 $  (we take into account that the magnitude of  $ n_{X} $  is large at the plasmon frequency). In the opposite case, when the contribution of the surface conductivity dominates, i.e.  $ \frac{4 \pi }{c} \vert  \sigma _{yz}^{S}\frac{g}{ \varepsilon _{zz}} \vert  \gg  \vert n_{X} \vert  \approx \frac{g}{\sqrt{ \vert  \varepsilon _{zz} \vert }} $ , we obtain  $ R=+1 $ , i.e. the phase of the reflected field is rotated by 180 degrees.

The enhanced contribution of the surface conductivity at normal incidence in the vicinity of the bulk plasmon resonance is expected. Indeed, at plasmon resonance the $z$-component  $ E_{z} $  of the field in the medium becomes very large, which leads to a dominant contribution of the surface current  $ j_{y}^{S}= \sigma _{yz}^{S}E_{z} $.

For oblique incidence  $   \theta  \neq 0 $  and small losses the calculations of the reflection in the vicinity of plasmon resonance have a technical subtlety, related to the presence of the term  $ n_{X}\cos \theta  \left( \cos \theta _{X}-\sin \theta _{X}K_{X} \right)  $  in Eq.~(\ref{X-modeReflection}). Indeed, at the plasmon frequency  $ n_{X} \rightarrow \infty $ as losses $\gamma \rightarrow 0$; however, for a plasmon we also have   $ K_{X} \rightarrow \frac{1}{\tan \theta _{X}} $, i.e. $  \left( \cos \theta _{X}-\sin \theta _{X}K_{X} \right)  \rightarrow 0 $. One needs to treat the resulting uncertainty of the product with caution.  The details are presented in Appendix F. 

The main result is that 
the contribution of surface states to the reflected wave is determined by the ratio 
\[\frac{\vert  \sigma _{yz}^{S} \vert}{c\sqrt{ \vert  \varepsilon _{zz} \vert }/4 \pi  }\]
and therefore becomes significant or dominant  at the plasmon resonance frequency, when $  \varepsilon _{zz}= \varepsilon _{zz}^{ \left( 0 \right) }+i\frac{4 \pi }{ \omega } \sigma _{zz}^{B} \rightarrow 0 $. When the bulk contribution dominates the reflection coefficient  $ R$ is close to $-1$. When the surface contribution dominates, $ R $ is close to $+1$ i.e.~the phase of the reflected field flips. 

%%%
%%%

\section{Surface plasmon-polaritons}

Surface plasmon-polaritons can be supported by both bulk and surface electron states. Here we derive dispersion relations for surface waves including both bulk and surface conductivity for several specific cases. Emphasis is placed on the situations where the dispersion is significantly affected or dominated by surface states and can therefore be used for diagnostics of surface states and Fermi arcs.  Previously, surface plasmons in WSMs have been considered in the low-frequency limit within a semiclassical description of particle motion with added ad hoc anomalous Hall term \cite{song2017} and with a quantum-mechanical description \cite{andolina2018} based on the Hamiltonian in \cite{okugawa2014}. Both studies indicated strong anisotropy and dispersion of surface plasmons. 

%%%
%%%
\subsection{Quasielectrostatic approximation}

Within the quasielectrostatic approximation the electric field can be defined through the scalar potential:  
$$\mathcal{\vec{E}}=\Re \left[ \vec{E} \left( z \right) e^{ik_{x}x+ik_{y}y-i \omega t} \right] =-\nabla \mathcal{F},\quad   \mathcal{F}=\Re \left[  \Phi  \left( z \right) e^{ik_{x}x+ik_{y}y-i \omega t} \right]. $$
We introduce the vector of electric induction, $ \mathcal{\vec{D}}=\Re \left[ \vec{D} \left( z \right) e^{ik_{x}x+ik_{y}y-i \omega t} \right] = \hat{\varepsilon} \mathcal{\vec{E}} $ 
and use Gauss' law for each halfspace: 
\begin{equation}\label{GaussLawElectricInductionQ-E.A.}
\nabla \cdot \mathcal{\vec{D}}=0.
\end{equation}

In general, there can be an electric dipole layer at the boundary between the two media. The dipole layer has a jump in the scalar potential  $  \Phi  \left( z \right)  $,  
\begin{equation}\label{SPPScalarPotentialB.C.}
\Phi  \left( z=+0 \right) - \Phi  \left( z=-0 \right) =4 \pi d_{z},  
\end{equation}
where  $ d_{z} $  is determined by Eqs.~(\ref{SurfaceDipoleMoment}).

Next, we define the potential  $  \Phi  \left( z \right)  $ for the surface mode as
\[
\Phi  \left( z>0 \right) =  \Phi _{up}e^{- \kappa _{up}z},\quad \Phi  \left( z<0 \right) =  \Phi _{W}e^{+ \kappa _{W}z}.
\]
Using Eq.~(\ref{GaussLawElectricInductionQ-E.A.}) in each halfspace, we obtain
\begin{align}
&k_{x}^{2}+k_{y}^{2}- \kappa _{up}^{2}=0, \label{FromGaussLawElectricInductionQ-E.A.Upper}\\
&k_{x}^{2} \varepsilon _{xx}+k_{y}^{2} \varepsilon _{yy}- \kappa _{W}^{2} \varepsilon _{zz}=0.\label{FromGaussLawElectricInductionQ-E.A.Lower}
\end{align}
Using the boundary condition Eq.~(\ref{B.C.fromGaussLaw}) we get
\[
n_{up}^{2} \kappa _{up} \Phi _{up}- \left[  \varepsilon _{zz} \left( - \kappa _{W} \Phi _{W} \right) + \varepsilon _{zy} \left( -ik_{y} \Phi _{W} \right)  \right] =-i\frac{4 \pi }{ \omega } \left( \frac{ \partial }{ \partial x}j_{x}^{S}+\frac{ \partial }{ \partial y}j_{y}^{S} \right) 
\]
which gives 
\begin{equation}\label{B.C.FromGaussLawForSPPCase}
n_{up}^{2} \kappa _{up} \Phi _{up}+ \left[  \kappa _{W} \left(  \varepsilon _{zz}+\frac{4 \pi }{ \omega }k_{y} \sigma _{yz}^{S} \right) +gk_{y}+i\frac{4 \pi }{ \omega } \left( k_{x}^{2} \sigma _{xx}^{S}+k_{y}^{2} \sigma _{yy}^{S} \right)  \right]  \Phi _{W}=0
\end{equation}
where  $  \varepsilon _{yz}=- \varepsilon _{zy}=ig= i\frac{4 \pi  \sigma _{yz}^{B}}{ \omega } $. Using also the boundary condition Eq.~(\ref{SPPScalarPotentialB.C.}) together with Eqs.~(\ref{SurfaceDipoleMoment}), we obtain
\begin{equation}\label{DiscontinuityInSPPScalarPotential}
\Phi _{up}+ \left( i\frac{4 \pi }{ \omega } \kappa _{W} \sigma _{zz}^{S}-\frac{4 \pi }{ \omega }k_{y} \sigma _{zy}^{S}-1 \right)  \Phi _{W}=0
\end{equation}
From these relationships one can get the dispersion equation for surface waves.  Note that the confinement constants $\kappa _{W}$ and $\kappa_{up}$ are generally complex-valued. Their imaginary parts give rise to a Poynting flux away from the surface which contributes to surface wave attenuation. 

%%%
%%%

\subsubsection{Neglecting surface states}

First, we neglect the surface conductivity to consider surface plasmons supported by bulk carriers only. 
In this case from Eqs.~(\ref{FromGaussLawElectricInductionQ-E.A.Upper}), (\ref{DiscontinuityInSPPScalarPotential}) we get $  \kappa _{up}=\sqrt{k_{x}^{2}+k_{y}^{2}} $,  $  \Phi _{up}= \Phi _{W} $. Denoting  $ k_{x}^{2}+k_{y}^{2}=k^{2} $,  $ k_{x}=k\cos \phi  $,  $ k_{y}=k\sin \phi  $, we obtain from Eq.~(\ref{FromGaussLawElectricInductionQ-E.A.Lower}) 
\begin{equation}\label{NoDiscontinuityScalarPotential}
\kappa _{W}=k\sqrt{\frac{\cos^{2} \phi  \varepsilon _{xx}+\sin^{2} \phi  \varepsilon _{yy}}{ \varepsilon _{zz}}}.
\end{equation}
Furthermore, from Eq.~(\ref{B.C.FromGaussLawForSPPCase}) for $  \kappa _{up}=k $ and $  \Phi _{up}= \Phi _{W} $ we have
\begin{equation}\label{B.C.FromGaussLawForNoDiscontinuityInScalarPotential}
n_{up}^{2}k+ \kappa _{W} \varepsilon _{zz}+gk\sin \phi =0,
\end{equation}
where  $  \varepsilon _{yz}=ig= i\frac{4 \pi  \sigma _{yz}^{B}}{ \omega } $. Substituting Eq.~(\ref{NoDiscontinuityScalarPotential}) into Eq.~(\ref{B.C.FromGaussLawForNoDiscontinuityInScalarPotential}), we obtain the dispersion relation
\begin{equation}\label{DispersionForOmegaVsAnglePhi}
D \left(  \omega , \phi  \right) =n_{up}^{2}+ \varepsilon _{zz}\sqrt{\frac{\cos^{2} \phi  \varepsilon _{xx}+\sin^{2} \phi  \varepsilon _{yy}}{ \varepsilon _{zz}}}+g\sin \phi =0.
\end{equation}

The dispersion equation Eq.~(\ref{DispersionForOmegaVsAnglePhi}) gives the dependence $\omega  \left(  \phi  \right)  $, but does not have any  dependence on the magnitude of $k$. This situation is similar to the dispersion relation for bulk plasmons in the quasielectrostatic approximation, Eq.~(\ref{ObliquePlasmonCondition}). It is also similar to waves in classical magnetized plasmas. Of course the range of values of $k$ is constrained by the validity of the quasielectrostatic approximation.

%%%
%%%

\subsubsection{Including surface states}

If we now include the surface conductivity, Eqs.~(\ref{FromGaussLawElectricInductionQ-E.A.Upper})-(\ref{DiscontinuityInSPPScalarPotential}) give 
\begin{multline}\label{DispersionWithSurfaceEffects} 
D \left(  \omega , \phi  \right) - \frac{4 \pi }{ \omega }k \bigg[ \sqrt{\frac{\cos^{2} \phi  \varepsilon _{xx}+\sin^{2} \phi  \varepsilon _{yy}}{ \varepsilon _{zz}}} \left( in_{up}^{2} \sigma _{zz}^{S}-\sin \phi  \sigma _{yz}^{S} \right) \\
-n_{up}^{2}\sin \phi  \sigma _{yz}^{S}-i \left( \cos^{2} \phi  \sigma _{xx}^{S}+\sin^{2} \phi  \sigma _{yy}^{S} \right)  \bigg] =0 
\end{multline}
where the function  $ D \left(  \omega , \phi  \right)  $ is determined by Eq.~(\ref{DispersionForOmegaVsAnglePhi}). As we see, taking the surface conductivity into account  brings the dependence on the magnitude  of the wave vector $k$ into the dispersion relation.  Therefore, measuring the frequency dispersion of the surface plasmon resonance provides a direct characterization of surface states.

%%%%%%%%%%%%%%

\begin{figure}[htb]
\begin{center}
\includegraphics[scale=0.4]{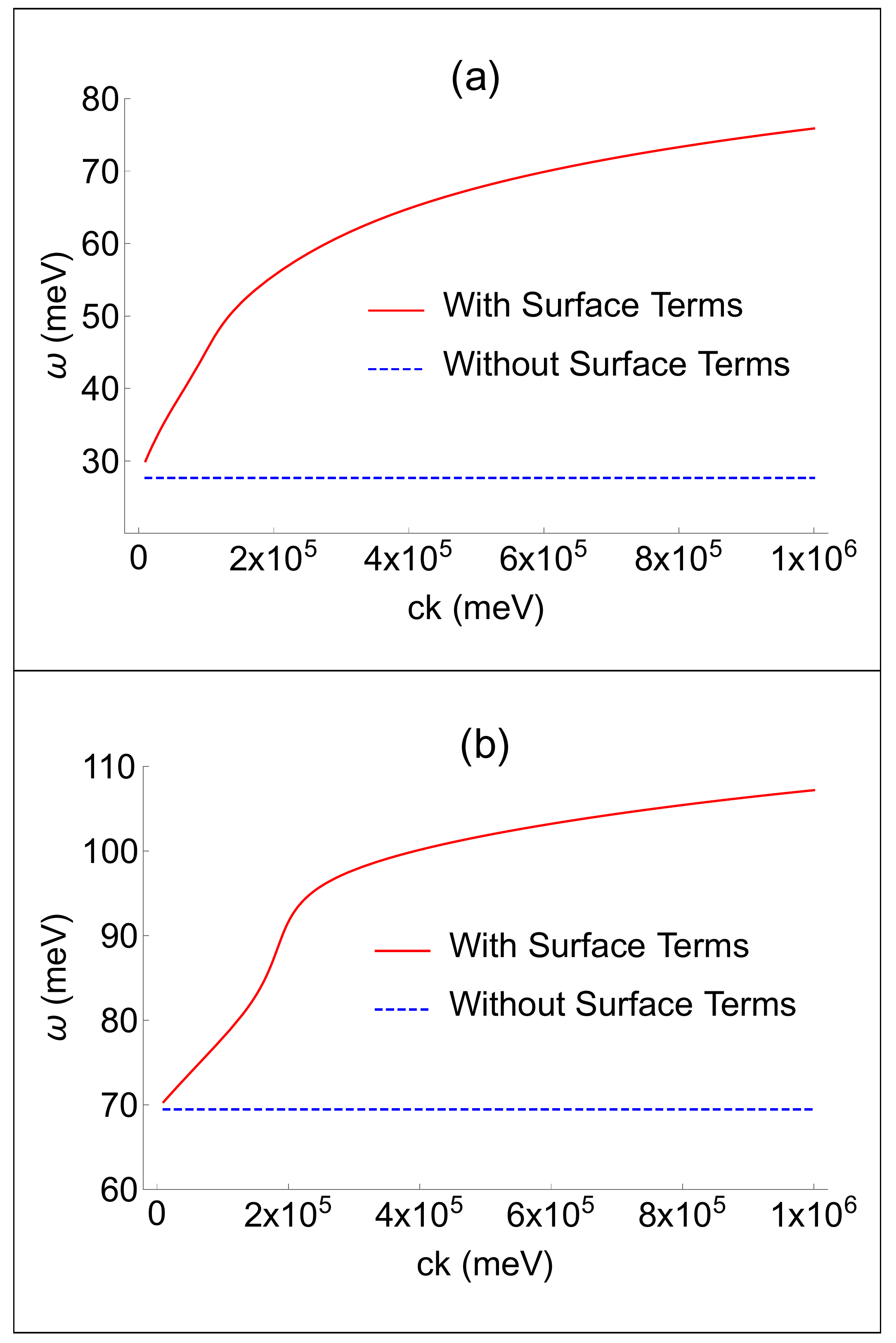}
\caption{Real part of the surface plasmon frequency as a function of real plasmon wavenumber obtained as a solution to the dispersion equation Eq.~(\ref{DispersionWithSurfaceEffects})  for $\phi = \pi/2$, $\hbar v_F b = 100$ meV and two values of the electron Fermi momentum $k_F = 0.5b$ and $0.8b$. The surface plasmon frequency neglecting surface conductivity contribution is shown as a dashed line.  }
\label{Fig:13}
\end{center}
\end{figure}

%%%%%%%%%%%%%%%%%%%

Figure 15 shows the surface plasmon dispersion for propagation along $y$, i.e.~transverse to the gyrotropy $x$-axis, for two values of the Fermi momentum.  The real part of the surface plasmon frequency ignoring the contribution of the surface conductivity is shown as a dashed horizontal line for each value of $k_F$. Clearly, the contribution of surface electron states is important everywhere, except maybe in a narrow region of small wavenumbers $k$ where the quasistatic approximation breaks down. The plot has a horizontal axis $ck$ in units of meV in order to directly compare with frequencies. The inequality $ck \gg \omega$ is satisfied almost everywhere. 

The fact that the contribution of the surface current is so important can be understood from the structure of Eq.~(\ref{DispersionWithSurfaceEffects}). Clearly, the relative contribution of the bulk and surface terms can be estimated by comparing the magnitudes of $|\sigma^B|$ and $|k\sigma^S|$ where $\sigma^B$ and $\sigma^S$ are appropriate  components of bulk and surface conductivity tensors and $k$ is a wavenumber of a given electromagnetic mode. This is true not only for surface modes but also for other electromagnetic wave processes at the boundary such as reflection. In the mid/far-infrared spectral region of interest to us, $|k\sigma^S| \ll |\sigma^B|$ for vacuum wavelengths $ck \sim \omega$. However, for large surface plasmon wavenumbers shown in Fig.~15 the opposite condition  $|k\sigma^S| \geq |\sigma^B|$ is satisfied.   

Note the dispersion in Fig.~15  is stronger (the slope is steeper) at frequencies corresponding to Re$[\epsilon_{zz}] \approx 0$, i.e. near the resonance for bulk plasmons propagating along $z$. This follows from Eq.~(\ref{DispersionWithSurfaceEffects}) where the surface terms contain a factor $1/\sqrt{\epsilon_{zz}}$. Physically, this is expected: indeed, as we already commented, at the plasmon resonance the $z$-component  $ E_{z} $  of the field in the medium becomes very large, which leads to an enhanced contribution of the surface current  $ j_{y}^{S}= \sigma _{yz}^{S}E_{z} $.

%%%
%%%

\subsection{Surface waves beyond the quasielectrostatic approximation}

For small wavenumbers  the quasielectrostatic approximation is no longer valid. On the other hand, in this case one can neglect the surface conductivity as we pointed out in the previous paragraph. This is not an interesting limit as far as the spectroscopy of surface states is concerned, but we still derive the resulting dispersion relation for completeness. 
For the electric field of a surface mode in the upper halfspace with the refractive index $ n_{up}$, 
$$
\vec{\cal{E}}_{up}=\Re \left[ \vec{E}_{up}e^{ik_{x}x+ik_{y}y- \kappa _{up}z-i \omega t} \right],  
$$
the Maxwell's equation for $\nabla \times \vec{\cal{E}}$ gives
\begin{equation}\label{FaradaysLawUpperHalfSpace}
k_{y}E_{z}-i \kappa _{up}E_{y}=\frac{ \omega }{c}B_{x},\quad k_{x}E_{z}-i \kappa _{up}E_{x}=-\frac{ \omega }{c}B_{y},\quad  k_{x}E_{y}-k_{y}E_{x}=\frac{ \omega }{c}B_{z}.
\end{equation}
For the field in the Weyl semimetal,
$$
\vec{\cal{E}}_{W}=\Re \left[ \vec{E}_{W}e^{ik_{x}x+ik_{y}y+ \kappa _{W}z-i \omega t} \right]
$$
the same equation gives, after replacing $  \kappa _{up} \rightarrow - \kappa _{W} $ in Eq.~(\ref{FaradaysLawUpperHalfSpace}),
\begin{equation}\label{FaradaysLawLowerHalfSpace}
k_{y}E_{z}+i \kappa _{W}E_{y}=\frac{ \omega }{c}B_{x},\quad k_{x}E_{z}+i \kappa _{W}E_{x}=-\frac{ \omega }{c}B_{y},\quad k_{x}E_{y}-k_{y}E_{x}=\frac{ \omega }{c}B_{z}.
\end{equation}
The inverse decay length for the field in the upper halfspace is given by $
\kappa _{up}^{2}=k^{2}-n_{up}^{2}\frac{ \omega ^{2}}{c^{2}}$.

In a WSM we can use a  version of Eq.~(\ref{WaveEqMatrixForm}) after replacing $ k_{z} \rightarrow -i \kappa _{W} $:
\begin{equation}\label{ScreenedWaveEqMatrixForm}
\begin{pmatrix}
\frac{ \omega ^{2}}{c^{2}} \varepsilon _{xx}-k_{y}^{2}+ \kappa _{W}^{2}  &  k_{x}k_{y}  &  -ik_{x} \kappa _{W}\\
k_{y}k_{x}  &  \frac{ \omega ^{2}}{c^{2}} \varepsilon _{yy}-k_{x}^{2}+ \kappa _{W}^{2}  &  i\frac{ \omega ^{2}}{c^{2}}g-ik_{y} \kappa _{W}\\
-ik_{x} \kappa _{W}  &  -i\frac{ \omega ^{2}}{c^{2}}g-ik_{y} \kappa _{W}  &  \frac{ \omega ^{2}}{c^{2}} \varepsilon _{zz}-k^{2}\\
\end{pmatrix}
 \begin{pmatrix}
E_{x}\\
E_{y}\\
E_{z}\\
\end{pmatrix}
=0, 
\end{equation}
where $ k^{2}=k_{x}^{2}+k_{y}^{2} $.

Consider again a surface wave propagating transverse to the anisotropy axis ($k_{x}=0$). 
In this case, there are two solutions to the dispersion equation Eq.~(\ref{ScreenedWaveEqMatrixForm}), an O-wave and an X-wave. However, one can show that an O-wave with $E_x \neq 0$ does not exist as a surface wave. Moreover, this statement remains true even with the surface current taken into account. Only the X-wave with $ E_{y,z} \neq 0 $ can exist as a surface wave. Its inverse confinement length in the Weyl semimetal is given by
\begin{equation}\label{SPPX-modeDispersion}
\kappa _{W}^{2}=\frac{ \varepsilon _{yy}}{ \varepsilon _{zz}} \left( k^{2}-n_{X}^{2}\frac{ \omega ^{2}}{c^{2}} \right)  
\end{equation}
where
$$ 
n_{X}^{2}= \varepsilon _{zz}-\frac{g^{2}}{ \varepsilon _{yy}} 
$$
is the refractive index of an extraordinary wave propagating in the volume in the $y$-direction (see Eq.~(\ref{X-modeRefraction}) for $  \theta =\frac{ \pi }{2} $). The polarization of an extraordinary wave is determined by 
\begin{equation}\label{DeterminsPolarizatinOfX-modeWave}
i \left( \frac{ \omega ^{2}}{c^{2}}g+k \kappa _{W} \right) E_{y}= \left( \frac{ \omega ^{2}}{c^{2}} \varepsilon _{zz}-k^{2} \right) E_{zW} 
\end{equation}
which follows from Eq.~(\ref{ScreenedWaveEqMatrixForm}). After some straightforward algebra, we obtain
 the dispersion relation for a surface wave:
\begin{equation}\label{DispersionOfSurfaceWave}
\left( k^{2}-\frac{ \omega ^{2}}{c^{2}}n_{up}^{2} \right)  \left( gk+ \varepsilon _{zz}\; \sqrt[]{\frac{ \varepsilon _{yy}}{ \varepsilon _{zz}}}\sqrt[]{k^{2}-\frac{ \omega ^{2}}{c^{2}}n_{X}^{2}} \right) +\sqrt{k^{2}-\frac{ \omega ^{2}}{c^{2}}n_{up}^{2}} \left( k^{2}-\frac{ \omega ^{2}}{c^{2}} \varepsilon _{zz} \right) n_{up}^{2}=0. 
\end{equation}
In the limit of large wavenumbers $k$ this equation becomes the quasielectrostatic dispersion relation Eq.~(\ref{DispersionForOmegaVsAnglePhi}) at $  \phi =\frac{ \pi }{2} $. 

For the propagation in $x$-direction, one can repeat the above analysis for the case $k_y = 0$ and obtain that there are no surface wave solutions when the surface conductivity is neglected. 

One interesting solution of the dispersion equation Eq.~(\ref{DispersionOfSurfaceWave}) is a strongly
nonelectrostatic case when the surface mode is weakly localized
in a medium above the WSM surface, e.g.~in the air. The energy of
this wave is mostly contained in an ambient medium above the WSM surface
where there is no absorption. Therefore, such surface waves can have
a long propagation length; see e.g.~\cite{barnes,homola,
agran}.

To find this solution we assume $n_{up}^{2}=1$ and introduce the
notation $\frac{\omega}{c}=k_{0}$. A weak localization outside a
WSM means that $|\kappa_{up}|\ll k_{0}$. Then, assuming $k\simeq k_{0}+\delta k$, where $k_0 \gg |\delta k|$, we obtain $\kappa_{up}\simeq\sqrt{2k_{0}\delta k}$. From
Eqs.~(\ref{DispersionOfSurfaceWave}) and (\ref{SPPX-modeDispersion}) in the first order with respect to $\sqrt{\frac{\delta k}{k_{0}}}$
we get
\begin{equation} 
\label{80}
\delta k \simeq \frac{k_{0}}{2}\frac{\left(\varepsilon_{zz}-1\right)^{2}}{\left[g+\sqrt{\varepsilon_{zz}\varepsilon_{yy}\left(1-\varepsilon_{zz}+\frac{g^{2}}{\varepsilon_{yy}}\right)}\right]^{2}},
\end{equation}
\begin{equation}
\label{81}
{\rm Re}\kappa_{W}^{2}\simeq {\rm Re} \left[ k_{0}^{2}\frac{\varepsilon_{yy}}{\varepsilon_{zz}}\left(1-\varepsilon_{zz}+\frac{g^{2}}{\varepsilon_{yy}}\right) \right].
\end{equation}

%%%%%%%%%%%%%%

\begin{figure}[htb]
\begin{center}
\includegraphics[scale=0.4]{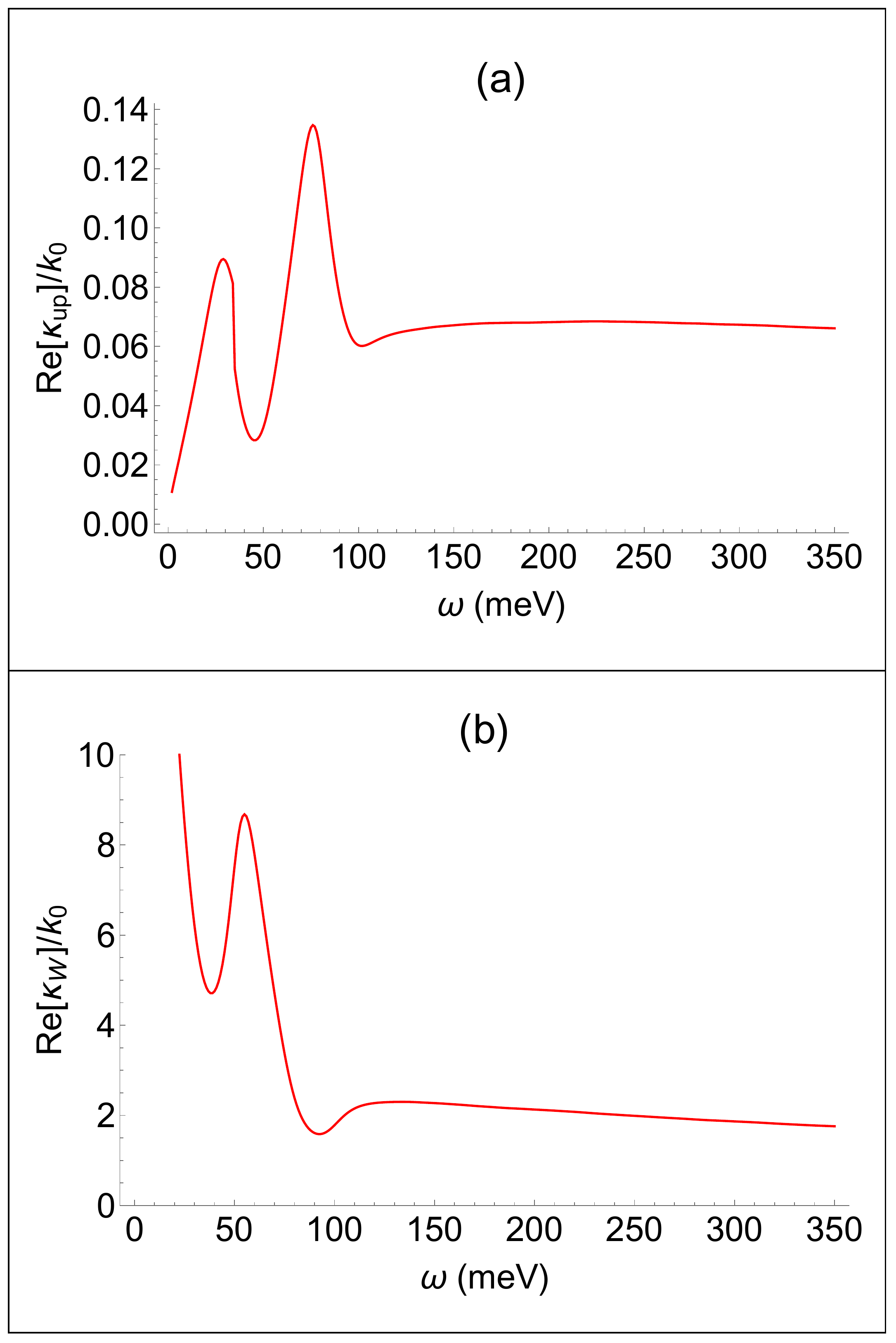}
\caption{Normalized confinement constants (a) Re$[\kappa_{up}]/k_0 \simeq {\rm Re}[\sqrt{2 \delta k/k_0}]$ and (b) Re$[\kappa_W]/k_0$ as functions of frequency, for the Fermi momentum $k_F = 0.5b$. Other parameters are $\hbar v_F b = 100$ meV and $\gamma = 10$ meV. }
\label{Fig:16}
\end{center}
\end{figure}

%%%%%%%%%%%%%%%%%%%

This solution describes surface waves if Re$[\kappa_W] > 0$ and Re$[\kappa_{up}] >0$. In addition, $|\delta k| \ll k_{0}$ has to be satisfied. We checked that all three inequalities are satisfied for the numerical parameters chosen to calculate the conductivity tensor. As an example, Fig.~16 shows normalized confinement constants Re$[\kappa_W]/k_0$ and Re$[\kappa_{up}]/k_0 \simeq {\rm Re}[\sqrt{2 \delta k/k_0}]$ as functions of frequency, for the Fermi momentum $k_F = 0.5b$. Clearly, the solution describes a surface wave which is weakly confined in the air and strongly confined in the WSM. The spectra remain qualitatively the same with increasing Fermi momentum, but the oscillating feature moves to higher energies, roughly following the spectral region where the real parts of $\varepsilon_{zz}$ and  $\varepsilon_{yy}$ cross zero. We note again that the confinement constants $\kappa _{W}$ and $\kappa_{up}$ are complex-valued. Their imaginary parts give rise to a Poynting flux away from the surface which contributes to surface wave attenuation. 

%%%%%%%%%%%%%%%%%%%
%%%%%%%%%%%%%%%%%%%%%%%
%%%%%%%%%%%%%%%%%%%%%%%%

\section{Summary and conclusions}

We presented systematic studies of the optical properties and  electromagnetic modes of Weyl semimetals.   Both bulk and surface conductivity tensors are derived from a single microscopic Hamiltonian.  The presence of separated Weyl nodes and associated surface states gives rise to distinct signatures in the transmission, reflection, and polarization of bulk and surface electromagnetic waves.  These signatures can be used for quantitative characterization of electronic structure of Weyl semimetals. Particularly sensitive spectroscopic probes of bulk electronic properties include strong anisotropy in propagation of both bulk and surface modes, birefringent dispersion and absorption spectra of ordinary and extraordinary normal modes, the frequency of bulk plasmon resonance as a function of incidence angle and doping level, and the polarization rotation and ellipticity for incident linearly polarized light. The sensitive characterization of surface electronic states can be achieved by measuring the phase change of the reflection coefficient of incident plane waves, the frequency dispersion of surface plasmon-polariton modes, and strong anisotropy of surface plasmon-polaritons with respect to their propagation direction and polarization. 

Potential optoelectronic applications of Weyl semimetal films in the mid-infrared and THz spectral regions will benefit from the strong anisotropy, gyrotropy, and birefringence of these materials, giant polarization rotation for light transmitted along the gyrotropy axis of submicron films,  and strongly localized surface plasmon-polariton modes. All effects are tunable  by doping. 

\begin{acknowledgments} 
The authors are grateful to I. Tokman and M. Erukhimova for useful discussions. This work has been supported by the Air Force Office for Scientific Research
through Grants No.~FA9550-17-1-0341 and FA9550-14-1-0376.
I.O. and M.T. acknowledge the support by the Ministry of Education Science of the Russian Federation contract No.~14.W03.31.0032. 
\end{acknowledgments}

%%%%%%%%%%%%%%%%%%%%%%%%%%%%%

\appendix

\section{Evaluation of the matrix elements of the current density operator}

 We  denote everywhere the bulk states by 
Latin letters, and the surface states by Greek letters, i.e. $\left|n\right\rangle =\left|B\right\rangle ,\left|\mu\right\rangle =\left|S\right\rangle $.
In this section we evaluate the matrix elements 
of the current density operator that enter Eq.~(\ref{kubo}) for the components of bulk and surface conductivity tensors.

\begin{align}
 & (j_{x})_{nm}=\left\langle n\right|\hat{j}_{x}\left|m\right\rangle \nonumber \\
 & =\frac{ev_{F}}{\hbar b}\text{\ensuremath{\int}}d\text{\textthreesuperior}r\left(\text{\textgreek{Y}}_{{\displaystyle {\textstyle {\scriptstyle \boldsymbol{k_{\textrm{n}}}}}},s_{n}}^{B}(\boldsymbol{r})\right)^{\dagger}\left(-i\text{\ensuremath{\hbar\partial}}_{x}\right)\hat{\sigma}_{x}\text{\textgreek{Y}}_{{\displaystyle {\textstyle {\scriptstyle \boldsymbol{k_{\textrm{m}}}}}},s_{m}}^{B}(\boldsymbol{r})\nonumber \\
 & =\frac{ev_{F}}{2b}k_{nx}\delta_{\boldsymbol{k}_{n},\boldsymbol{k}_{m}}\nonumber \\
 & \times\left[s_{m}\sqrt{\left(1+s_{m}\cos\theta_{\boldsymbol{k_{\textrm{n}}}}\right)\left(1-s_{n}\cos\theta_{\boldsymbol{k}_{\boldsymbol{\textrm{n}}}}\right)}e^{i\phi_{\boldsymbol{k}_{n}}}+s_{n}\sqrt{\left(1+s_{n}\cos\theta_{\boldsymbol{k_{\textrm{n}}}}\right)\left(1-s_{m}\cos\theta_{\boldsymbol{k}_{\boldsymbol{\textrm{n}}}}\right)}e^{-i\phi_{\boldsymbol{k}_{n}}}\right]
\end{align}

\begin{align}
& (j_{x})_{\mu\nu} =\left\langle \mu\right|\hat{j}_{x}\left|\nu\right\rangle 
=\frac{ev_{F}}{\hbar b}\text{\ensuremath{\int}}d\text{\textthreesuperior}r\left(\text{\textgreek{Y}}_{{\displaystyle {\textstyle {\scriptstyle \boldsymbol{k_{\mu}}}}}}^{S}(\boldsymbol{r})\right)^{\dagger}\left(-i\text{\ensuremath{\hbar\partial}}_{x}\right)\hat{\sigma}_{x}\text{\textgreek{Y}}_{{\displaystyle {\textstyle {\scriptstyle \boldsymbol{k_{\nu}}}}}}^{S}(\boldsymbol{r})  =0,
\end{align}

\begin{align}
(j_{x})_{\mu m} & =\left\langle \mu\right|\hat{j}_{x}\left|m\right\rangle 
 =\frac{ev_{F}}{\hbar b}\text{\ensuremath{\int}}d\text{\textthreesuperior}r\left(\text{\textgreek{Y}}_{{\displaystyle {\textstyle {\scriptstyle \boldsymbol{k_{\mu}}}}}}^{S}(\boldsymbol{r})\right)^{\dagger}\left(-i\text{\ensuremath{\hbar\partial}}_{x}\right)\hat{\sigma}_{x}\text{\textgreek{Y}}_{{\displaystyle {\textstyle {\scriptstyle \boldsymbol{k_{m}}}}},s_{m}}^{B}(\boldsymbol{r})\nonumber \\
 & =\frac{2ev_{F}s_{m}k_{mx}k_{mz}}{ib(\kappa_{m}^{2}+k_{mz}^{2})}\sqrt{\frac{\kappa_{m}\left(1+s_{m}\cos\theta_{\boldsymbol{k}_{m}}\right)}{L_{z}}}\delta_{k_{mx},k_{\mu x}}\delta_{k_{my},k_{\mu y}}
\end{align}

\begin{align}
(j_{y})_{nm} & =\frac{ev_{F}}{\hbar b}\text{\ensuremath{\int}}d\text{\textthreesuperior}r\left(\text{\textgreek{Y}}_{{\displaystyle {\textstyle {\scriptstyle \boldsymbol{k_{\textrm{n}}}}}},s_{n}}^{B}(\boldsymbol{r})\right)^{\dagger}\left(-i\text{\ensuremath{\hbar\partial}}_{y}\right)\hat{\sigma}_{x}\text{\textgreek{Y}}_{{\displaystyle {\textstyle {\scriptstyle \boldsymbol{k_{\textrm{m}}}}}},s_{m}}^{B}(\boldsymbol{r})-ev_{F}\text{\ensuremath{\int}}d\text{\textthreesuperior}r\left(\text{\textgreek{Y}}_{{\displaystyle {\textstyle {\scriptstyle \boldsymbol{k_{n}}}}},s_{n}}^{B}(\boldsymbol{r})\right)^{\dagger}\hat{\sigma}_{z}\text{\textgreek{Y}}_{{\displaystyle {\textstyle {\scriptstyle \boldsymbol{k_{m}}}}},s_{m}}^{B}(\boldsymbol{r}) \nonumber \\
 & =\frac{ev_{F}}{2b}k_{ny}\delta_{\boldsymbol{k}_{n},\boldsymbol{k}_{m}}\nonumber \\
 & \times\left[s_{m}\sqrt{\left(1+s_{m}\cos\theta_{\boldsymbol{k_{\textrm{n}}}}\right)\left(1-s_{n}\cos\theta_{\boldsymbol{k}_{\boldsymbol{\textrm{n}}}}\right)}e^{i\phi_{\boldsymbol{k}_{n}}}+s_{n}\sqrt{\left(1+s_{n}\cos\theta_{\boldsymbol{k_{\textrm{n}}}}\right)\left(1-s_{m}\cos\theta_{\boldsymbol{k}_{\boldsymbol{\textrm{n}}}}\right)}e^{-i\phi_{\boldsymbol{k}_{n}}}\right]\nonumber \\
 & +\frac{ev_{F}}{2}\delta_{\boldsymbol{k}_{n},\boldsymbol{k}_{m}}\left[s_{n}s_{m}\sqrt{\left(1+s_{n}\cos\theta_{\boldsymbol{k_{\textrm{n}}}}\right)\left(1+s_{m}\cos\theta_{\boldsymbol{k}_{n}}\right)}-\sqrt{\left(1-s_{n}\cos\theta_{\boldsymbol{k}_{n}}\right)\left(1-s_{m}\cos\theta_{\boldsymbol{k}_{\boldsymbol{\textrm{n}}}}\right)}\right]
\end{align}
\begin{align}
(j_{y})_{\mu\nu} 
 & =\frac{ev_{F}}{\hbar b}\text{\ensuremath{\int}}d\text{\textthreesuperior}r\left(\text{\textgreek{Y}}_{{\displaystyle {\textstyle {\scriptstyle \boldsymbol{k_{\mu}}}}}}^{S}(\boldsymbol{r})\right)^{\dagger}\left(-i\text{\ensuremath{\hbar\partial}}_{y}\right)\hat{\sigma}_{x}\text{\textgreek{Y}}_{{\displaystyle {\textstyle {\scriptstyle \boldsymbol{k_{\nu}}}}}}^{S}(\boldsymbol{r})-ev_{F}\text{\ensuremath{\int}}d\text{\textthreesuperior}r\left(\text{\textgreek{Y}}_{{\displaystyle {\textstyle {\scriptstyle \boldsymbol{k_{\mu}}}}}}^{S}(\boldsymbol{r})\right)^{\dagger}\hat{\sigma}_{z}\text{\textgreek{Y}}_{{\displaystyle {\textstyle {\scriptstyle \boldsymbol{k_{\nu}}}}}}^{S}(\boldsymbol{r})\nonumber \\
  & =-ev_{F}\delta_{k_{\mu x},k_{\nu x}}\delta_{k_{\mu y},k_{\nu y}}
\end{align}

\begin{align}
  (j_{y})_{\mu m} & =\frac{ev_{F}}{\hbar b}\text{\ensuremath{\int}}d\text{\textthreesuperior}r\left(\text{\textgreek{Y}}_{{\displaystyle {\textstyle {\scriptstyle \boldsymbol{k_{\mu}}}}}}^{S}(\boldsymbol{r})\right)^{\dagger}\left(-i\text{\ensuremath{\hbar}}\partial_{y}\right)\hat{\sigma}_{x}\text{\textgreek{Y}}_{{\displaystyle {\textstyle {\scriptstyle \boldsymbol{k_{m}}}}},s_{m}}^{B}(\boldsymbol{r})-ev_{F}\text{\ensuremath{\int}}d\text{\textthreesuperior}r\left(\text{\textgreek{Y}}_{{\displaystyle {\textstyle {\scriptstyle \boldsymbol{k_{\mu}}}}}}^{S}(\boldsymbol{r})\right)^{\dagger}\hat{\sigma}_{z}\text{\textgreek{Y}}_{{\displaystyle {\textstyle {\scriptstyle \boldsymbol{k_{m}}}}},s_{m}}^{B}(\boldsymbol{r})\nonumber \\
 & =\frac{2ev_{F}s_{m}k_{my}k_{mz}}{ib(\kappa_{m}^{2}+k_{mz}^{2})}\sqrt{\frac{\kappa_{m}\left(1+s_{m}\cos\theta_{\boldsymbol{k}_{m}}\right)}{L_{z}}}\delta_{k_{mx},k_{\mu x}}\delta_{k_{my},k_{\mu y}};
\end{align}

\begin{align}
(j_{z})_{nm} & =ev_{F}\text{\ensuremath{\int}}d\text{\textthreesuperior}r\left(\text{\textgreek{Y}}_{{\displaystyle {\textstyle {\scriptstyle \boldsymbol{k_{n}}}}},s_{n}}^{B}(\boldsymbol{r})\right)^{\dagger}\hat{\sigma}_{y}\text{\textgreek{Y}}_{{\displaystyle {\textstyle {\scriptstyle \boldsymbol{k_{m}}}}},s_{m}}^{B}(\boldsymbol{r})
  =i\frac{ev_{F}}{2}\delta_{\boldsymbol{k}_{n},\boldsymbol{k}_{m}}\nonumber \\
 & \times\left[s_{n}\sqrt{\left(1+s_{n}\cos\theta_{\boldsymbol{k_{\textrm{n}}}}\right)\left(1-s_{m}\cos\theta_{\boldsymbol{k}_{\boldsymbol{\textrm{n}}}}\right)}e^{-i\phi_{\boldsymbol{k}_{n}}}-s_{m}\sqrt{\left(1+s_{m}\cos\theta_{\boldsymbol{k_{\textrm{n}}}}\right)\left(1-s_{n}\cos\theta_{\boldsymbol{k}_{\boldsymbol{\textrm{n}}}}\right)}e^{i\phi_{\boldsymbol{k}_{n}}}\right]
\end{align}

\begin{align}
(j_{z})_{\mu\nu}  =ev_{F}\text{\ensuremath{\int}}d\text{\textthreesuperior}r\left(\text{\textgreek{Y}}_{{\displaystyle {\textstyle {\scriptstyle \boldsymbol{k_{\mu}}}}}}^{S}(\boldsymbol{r})\right)^{\dagger}\hat{\sigma}_{y}\text{\textgreek{Y}}_{{\displaystyle {\textstyle {\scriptstyle \boldsymbol{k_{\nu}}}}}}^{S}(\boldsymbol{r}) =0,
\end{align}

\begin{align}
(j_{z})_{\mu m}  & =ev_{F}\text{\ensuremath{\int}}d\text{\textthreesuperior}r\left(\text{\textgreek{Y}}_{{\displaystyle {\textstyle {\scriptstyle \boldsymbol{k_{\mu}}}}}}^{S}(\boldsymbol{r})\right)^{\dagger}\hat{\sigma}_{y}\text{\textgreek{Y}}_{{\displaystyle {\textstyle {\scriptstyle \boldsymbol{k_{m}}}}},s_{m}}^{B}(\boldsymbol{r})\nonumber \\
 & =-\frac{2ev_{F}s_{m}k_{mz}}{\kappa_{m}^{2}+k_{mz}^{2}}\sqrt{\frac{\kappa_{m}\left(1+s_{m}\cos\theta_{\boldsymbol{k}_{m}}\right)}{L_{z}}}\delta_{k_{mx},k_{\mu x}}\delta_{k_{my},k_{\mu y}},
\end{align}
where we have used $\kappa=\frac{b^{2}-\left(k_{x}^{2}+k_{y}^{2}\right)}{2b}$.

%%%%%%%%%%%%%%%%%%%%%%%%%%%%%%%

\section{Calculation of the bulk optical conductivity tensor}

The 3D integrals over electron momenta cannot be evaluated analytically in most cases, even in the zero temperature limit. Whenever the integrals remain in the final expression, they were evaluated numerically for the plots in the main text. 

\subsection{Contribution of intraband transitions ($s=+1\rightarrow s=+1$)}

 In this case the matrix elements $\boldsymbol{j}_{nm}^{(\boldsymbol{q})}$
of the current density operator reduce to
\begin{equation}
(j_{x})_{nn}=ev_{F}s_{n}\frac{k_{nx}}{b}\left|\sin\theta_{\boldsymbol{k}_{n}}\right|\cos\phi_{\boldsymbol{k}_{n}},
\end{equation}

\begin{equation}
(j_{y})_{nn}=ev_{F}s_{n}\left(\frac{k_{ny}}{b}\left|\sin\theta_{\boldsymbol{k}_{n}}\right|\cos\phi_{\boldsymbol{k}_{n}}+\cos\theta_{\boldsymbol{k}_{n}}\right),
\end{equation}

\begin{equation}
(j_{z})_{nn}=ev_{F}s_{n}\left|\sin\theta_{\boldsymbol{k}_{n}}\right|\sin\phi_{\boldsymbol{k}_{n}}.
\end{equation}
Therefore, we obtain
\begin{align}
 & \sigma_{xx}^{intra}\left(\omega\right)=g\frac{i\hbar}{V}\sum_{mn}\left(\frac{f_{n}-f_{m}}{E_{m}-E_{n}}\right)\frac{\left|\left\langle n\right|\hat{j}_{x}\left|m\right\rangle \right|^{2}}{\hbar(\omega+i\gamma)+(E_{n}-E_{m})}\nonumber \\
 & =\frac{ige^{2}v_{F}^{2}}{b^{2}(\omega+i\gamma)}\frac{1}{V}\sum_{n}\left(-\frac{\text{\ensuremath{\partial}}f_{n}}{\text{\ensuremath{\partial}}E_{n}}\right)k_{nx}^{2}\sin^{2}\theta_{\boldsymbol{k}_{n}}\cos^{2}\phi_{\boldsymbol{k}_{n}}\nonumber \\
 &  = \frac{ige^{2}v_{F}^{2}}{b^{2}(\omega+i\gamma)}\int_{\infty}\frac{d^{3}k}{\left(2\pi\right)^{3}}\delta(E_{B}-E_{F})k_{x}^{2}\sin^{2}\theta_{\boldsymbol{k}}\cos^{2}\phi_{\boldsymbol{k}}\nonumber \\
\nonumber \\
  & =\frac{ige^{2}v_{F}}{4\pi^{3}b^{2}k_{F}\hbar(\omega+i\gamma)}\int_{-\infty}^{\infty}dk_{x}\int_{-\infty}^{\infty}dk_{y}\frac{k_{x}^{2}K_{x}^{2}\Theta\left(k_{F}-\sqrt{K_{x}^{2}+k_{y}^{2}}\right)}{\sqrt{k_{F}^{2}-\left(K_{x}^{2}+k_{y}^{2}\right)}}
\end{align}

Similarly, 
\begin{align}
 & \sigma_{yy}^{intra}\left(\omega\right) = \frac{ige^{2}v_{F}}{4\pi^{3}b^{2}k_{F}\hbar(\omega+i\gamma)}\int_{-\infty}^{\infty}dk_{x}\int_{-\infty}^{\infty}dk_{y}\frac{k_{y}^{2}\left(K_{x}+b\right)^{2}\Theta\left(k_{F}-\sqrt{K_{x}^{2}+k_{y}^{2}}\right)}{\sqrt{k_{F}^{2}-\left(K_{x}^{2}+k_{y}^{2}\right)}}
\end{align}

\begin{align}
 & \sigma_{zz}^{intra}\left(\omega\right) = \frac{ige^{2}v_{F}}{4\pi^{3}k_{F}\hbar(\omega+i\gamma)}\int_{-\infty}^{\infty}dk_{x}\int_{-\infty}^{\infty}dk_{y}\Theta\left(k_{F}-\sqrt{K_{x}^{2}+k_{y}^{2}}\right)\sqrt{k_{F}^{2}-\left(K_{x}^{2}+k_{y}^{2}\right)}
\end{align}

Here $\Theta(k)$ is the step function and we have used $\cos\theta_{\boldsymbol{k}}=\frac{k_{y}}{\sqrt{K_{x}^{2}+k_{y}^{2}+k_{z}^{2}}}$,
$e^{i\phi_{\boldsymbol{k}}}=\frac{K_{x}+ik_{z}}{\sqrt{K_{x}^{2}+k_{z}^{2}}}$
, $K_{x}\equiv\frac{\left(k_{x}^{2}+k_{y}^{2}\right)-b^{2}}{2b}$, 
and $k_{F}\equiv\frac{E_{F}}{\hbar v_{F}}$.

\begin{align}
  \sigma_{xy}^{intra}\left(\omega\right) = \sigma_{xz}^{intra}\left(\omega\right) =  \sigma_{yz}^{intra}\left(\omega\right) = 0.  
   \end{align}

%%%%%%%%%%%%%%%%%%%%%%%%%

\subsection{Contribution of interband transitions ($s\rightarrow-s,\,\left|B\right\rangle \leftrightarrow\left|S\right\rangle $)}

 In this case, i.e. $s_{m}=-s_{n}=\pm1,\,n\neq m,$ the matrix
elements $\boldsymbol{j}_{nm}^{(\boldsymbol{q})}$ of the current density
operator reduce to

\begin{equation}
(j_{x})_{nm}=ev_{F}s_{n}\delta_{\boldsymbol{k}_{n},\boldsymbol{k}_{m}}\frac{k_{nx}}{b}\left(s_{n}\cos\theta_{\boldsymbol{k}_{n}}\cos\phi_{\boldsymbol{k}_{n}}-i\sin\phi_{\boldsymbol{k}_{n}}\right),
\end{equation}

\begin{equation}
(j_{y})_{nm}=ev_{F}s_{n}\delta_{\boldsymbol{k}_{n},\boldsymbol{k}_{m}}\left[\frac{k_{ny}}{b}\left(s_{n}\cos\theta_{\boldsymbol{k}_{n}}\cos\phi_{\boldsymbol{k}_{n}}-i\sin\phi_{\boldsymbol{k}_{n}}\right)-s_{n}\left|\sin\theta_{\boldsymbol{k}_{n}}\right|\right],
\end{equation}

\begin{equation}
(j_{z})_{nm}=ev_{F}s_{n}\delta_{\boldsymbol{k}_{n},\boldsymbol{k}_{m}}\left(i\cos\phi_{\boldsymbol{k}_{n}}+s_{n}\cos\theta_{\boldsymbol{k}_{n}}\sin\phi_{\boldsymbol{k}_{n}}\right),
\end{equation}
where $n\neq m.$ Therefore, we obtain

\begin{align}
 & \sigma_{xx}^{inter}\left(\omega\right)=g\frac{i\hbar}{V}\sum_{s=\pm1}\sum_{mn}\left(\frac{f_{n\left(-s\right)}-f_{m\left(s\right)}}{E_{m\left(s\right)}-E_{n\left(-s\right)}}\right)\frac{\left|\left\langle -sn\right|\hat{j}_{x}\left|ms\right\rangle \right|^{2}}{\hbar(\omega+i\gamma)+(E_{n\left(-s\right)}-E_{m\left(s\right)})}\nonumber \\
 & = i\hbar g\sum_{s=\pm1}\int_{\infty}\frac{d^{3}k}{\left(2\pi\right)^{3}}\left(\frac{f_{\boldsymbol{k}\left(-s\right)}-f_{\boldsymbol{k}\left(s\right)}}{E_{\boldsymbol{k}\left(s\right)}-E_{\boldsymbol{k}\left(-s\right)}}\right)\frac{e^{2}v_{F}^{2}k_{x}^{2}\left(\cos^{2}\theta_{\boldsymbol{k}}\cos^{2}\phi_{\boldsymbol{k}}+\sin^{2}\phi_{\boldsymbol{k}}\right)}{b^{2}\left[\hbar(\omega+i\gamma)+(E_{\boldsymbol{k}\left(-s\right)}-E_{\boldsymbol{k}\left(s\right)})\right]}\nonumber \\
  & =\frac{ige^{2}\left(\omega+i\gamma\right)}{8\pi^{3}b^{2}\hbar v_{F}}\int_{-\infty}^{\infty}dk_{x}\int_{-\infty}^{\infty}dk_{y}
  \times \left[ \text{\textgreek{J}}\left(k_{F}-\sqrt{K_{x}^{2}+k_{y}^{2}}\, \right) \right. \nonumber \\
 &\left. \times2k_{x}^{2}\left(\frac{K_{x}^{2}\sqrt{k_{F}^{2}-K_{x}^{2}-k_{y}^{2}}}{k_{F}\left(\frac{\omega+i\gamma}{v_{F}}\right)^{2}\left(K_{x}^{2}+k_{y}^{2}\right)}+\frac{\left[\left(\frac{\omega+i\gamma}{v_{F}}\right)^{2}-4K_{x}^{2}\right]\arctan\left[\frac{\left(\frac{\omega+i\gamma}{v_{F}}\right)\sqrt{k_{F}^{2}-K_{x}^{2}-k_{y}^{2}}}{k_{F}\sqrt{4\left(K_{x}^{2}+k_{y}^{2}\right)-\left(\frac{\omega+i\gamma}{v_{F}}\right)^{2}}}\right]}{\left(\frac{\omega+i\gamma}{v_{F}}\right)^{3}\sqrt{4\left(K_{x}^{2}+k_{y}^{2}\right)-\left(\frac{\omega+i\gamma}{v_{F}}\right)^{2}}}\right) \right. \nonumber \\
 & \left. -\text{\textgreek{J}}\left(K-\sqrt{K_{x}^{2}+k_{y}^{2}}\, \right) \right. \nonumber \\
 &\left. \times2k_{x}^{2}\left(\frac{K_{x}^{2}\sqrt{K^{2}-K_{x}^{2}-k_{y}^{2}}}{K\left(\frac{\omega+i\gamma}{v_{F}}\right)^{2}\left(K_{x}^{2}+k_{y}^{2}\right)}+\frac{\left[\left(\frac{\omega+i\gamma}{v_{F}}\right)^{2}-4K_{x}^{2}\right]\arctan\left[\frac{\left(\frac{\omega+i\gamma}{v_{F}}\right)\sqrt{K^{2}-K_{x}^{2}-k_{y}^{2}}}{K\sqrt{4\left(K_{x}^{2}+k_{y}^{2}\right)-\left(\frac{\omega+i\gamma}{v_{F}}\right)^{2}}}\right]}{\left(\frac{\omega+i\gamma}{v_{F}}\right)^{3}\sqrt{4\left(K_{x}^{2}+k_{y}^{2}\right)-\left(\frac{\omega+i\gamma}{v_{F}}\right)^{2}}}\right) \right]
\end{align}
where we have used $K_{x}\equiv\frac{\left(k_{x}^{2}+k_{y}^{2}\right)-b^{2}}{2b}=-\kappa,$
$\cos\theta_{\boldsymbol{k}}\left(-k_{x}\right)=\cos\theta_{\boldsymbol{k}}\left(k_{x}\right),$$\sin\theta_{\boldsymbol{k}}\left(-k_{x}\right)=\sin\theta_{\boldsymbol{k}}\left(k_{x}\right)$
$\cos\phi_{\boldsymbol{k}}\left(-k_{x}\right)=\cos\phi_{\boldsymbol{k}}\left(k_{x}\right),$
and $\sin\phi_{\boldsymbol{k}}\left(-k_{x}\right)=\sin\phi_{\boldsymbol{k}}\left(k_{x}\right).$

Similarly, 
\begin{align}
 & \sigma_{yy}^{inter}\left(\omega\right) = \frac{ige^{2}(\omega+i\gamma)}{4\pi^{3}b^{2}\hbar v_{F}}\int_{-\infty}^{\infty}dk_{x}\int_{-\infty}^{\infty}dk_{y} \times \left[ \text{\textgreek{J}}\left(k_{F}-\sqrt{K_{x}^{2}+k_{y}^{2}}\, \right) \times  \right.\nonumber \\
 & \left. \left(\frac{\left(b+K_{x}\right)^{2}k_{y}^{2}\sqrt{k_{F}^{2}-K_{x}^{2}-k_{y}^{2}}}{k_{F}\left(\frac{\omega+i\gamma}{v_{F}}\right)^{2}\left(K_{x}^{2}+k_{y}^{2}\right)}+\frac{\left[\left(\frac{\omega+i\gamma}{v_{F}}\right)^{2}\left(b^{2}+k_{y}^{2}\right)-4\left(b+K_{x}\right)^{2}k_{y}^{2}\right]\arctan\left[\frac{\left(\frac{\omega+i\gamma}{v_{F}}\right)\sqrt{k_{F}^{2}-K_{x}^{2}-k_{y}^{2}}}{k_{F}\sqrt{4\left(K_{x}^{2}+k_{y}^{2}\right)-\left(\frac{\omega+i\gamma}{v_{F}}\right)^{2}}}\right]}{\left(\frac{\omega+i\gamma}{v_{F}}\right)^{3}\sqrt{4\left(K_{x}^{2}+k_{y}^{2}\right)-\left(\frac{\omega+i\gamma}{v_{F}}\right)^{2}}}\right) \right. \nonumber \\
 &\left. -\text{\textgreek{J}}\left(K-\sqrt{K_{x}^{2}+k_{y}^{2}}\, \right) \times \right. \nonumber \\
 & \left. \left(\frac{\left(b+K_{x}\right)^{2}k_{y}^{2}\sqrt{K^{2}-K_{x}^{2}-k_{y}^{2}}}{K\left(\frac{\omega+i\gamma}{v_{F}}\right)^{2}\left(K_{x}^{2}+k_{y}^{2}\right)}+\frac{\left[\left(\frac{\omega+i\gamma}{v_{F}}\right)^{2}\left(b^{2}+k_{y}^{2}\right)-4\left(b+K_{x}\right)^{2}k_{y}^{2}\right]\arctan\left[\frac{\left(\frac{\omega+i\gamma}{v_{F}}\right)\sqrt{K^{2}-K_{x}^{2}-k_{y}^{2}}}{K\sqrt{4\left(K_{x}^{2}+k_{y}^{2}\right)-\left(\frac{\omega+i\gamma}{v_{F}}\right)^{2}}}\right]}{\left(\frac{\omega+i\gamma}{v_{F}}\right)^{3}\sqrt{4\left(K_{x}^{2}+k_{y}^{2}\right)-\left(\frac{\omega+i\gamma}{v_{F}}\right)^{2}}}\right)\right]
\end{align}

\begin{align}
 & \sigma_{zz}^{inter}\left(\omega\right) = \frac{ige^{2}\left(\omega+i\gamma\right)}{8\pi^{3}\hbar v_{F}}\int_{-\infty}^{\infty}dk_{x}\int_{-\infty}^{\infty}dk_{y}\left(K_{x}^{2}+k_{y}^{2}\right) \left[ \text{\textgreek{J}}\left(K-\sqrt{K_{x}^{2}+k_{y}^{2}}\, \right) \right. \nonumber \\
 & \left. \times\left(\frac{2\sqrt{K^{2}-K_{x}^{2}-k_{y}^{2}}}{K\left(\frac{\omega+i\gamma}{v_{F}}\right)^{2}\left(K_{x}^{2}+k_{y}^{2}\right)}-\frac{8\left[\left(\frac{\omega+i\gamma}{v_{F}}\right)^{2}-4K_{x}^{2}\right]\arctan\left[\frac{\left(\frac{\omega+i\gamma}{v_{F}}\right)\sqrt{K^{2}-K_{x}^{2}-k_{y}^{2}}}{K\sqrt{4\left(K_{x}^{2}+k_{y}^{2}\right)-\left(\frac{\omega+i\gamma}{v_{F}}\right)^{2}}}\right]}{\left(\frac{\omega+i\gamma}{v_{F}}\right)^{3}\sqrt{4\left(K_{x}^{2}+k_{y}^{2}\right)-\left(\frac{\omega+i\gamma}{v_{F}}\right)^{2}}}\right) \right. \nonumber \\
 & \left. -\text{\textgreek{J}}\left(k_{F}-\sqrt{K_{x}^{2}+k_{y}^{2}}\, \right) \right. \nonumber \\
 & \left. \times\left(\frac{2\sqrt{k_{F}^{2}-K_{x}^{2}-k_{y}^{2}}}{k_{F}\left(\frac{\omega+i\gamma}{v_{F}}\right)^{2}\left(K_{x}^{2}+k_{y}^{2}\right)}-\frac{8\left[\left(\frac{\omega+i\gamma}{v_{F}}\right)^{2}-4K_{x}^{2}\right]\arctan\left[\frac{\left(\frac{\omega+i\gamma}{v_{F}}\right)\sqrt{k_{F}^{2}-K_{x}^{2}-k_{y}^{2}}}{k_{F}\sqrt{4\left(K_{x}^{2}+k_{y}^{2}\right)-\left(\frac{\omega+i\gamma}{v_{F}}\right)^{2}}}\right]}{\left(\frac{\omega+i\gamma}{v_{F}}\right)^{3}\sqrt{4\left(K_{x}^{2}+k_{y}^{2}\right)-\left(\frac{\omega+i\gamma}{v_{F}}\right)^{2}}}\right) \right].
\end{align}

The only nonzero off-diagonal element is $\sigma_{zy}^{inter}(\omega)=-\sigma_{yz}^{inter}(\omega)$, as expected: 

\begin{align}
 & \sigma_{yz}^{inter}\left(\omega\right)  =  \frac{-ge^{2}}{4\pi^{3}b\hbar}\int_{-\infty}^{\infty}dk_{x}\int_{-\infty}^{\infty}dk_{y}\left(k_{y}^{2}-bK_{x}\right)\nonumber \\
 & \times\left( \text{\textgreek{J}}\left(k_{F}-\sqrt{K_{x}^{2}+k_{y}^{2}}\, \right)\frac{2\arctan\left[\frac{\left(\frac{\omega+i\gamma}{v_{F}}\right)\sqrt{k_{F}^{2}-K_{x}^{2}-k_{y}^{2}}}{k_{F}\sqrt{4\left(K_{x}^{2}+k_{y}^{2}\right)-\left(\frac{\omega+i\gamma}{v_{F}}\right)^{2}}}\right]}{\left(\frac{\omega+i\gamma}{v_{F}}\right)\sqrt{4\left(K_{x}^{2}+k_{y}^{2}\right)-\left(\frac{\omega+i\gamma}{v_{F}}\right)^{2}}} \right.  \nonumber \\
 & \left. -\text{\textgreek{J}}\left(K-\sqrt{K_{x}^{2}+k_{y}^{2}}\, \right)\frac{2\arctan\left[\frac{\left(\frac{\omega+i\gamma}{v_{F}}\right)\sqrt{K^{2}-K_{x}^{2}-k_{y}^{2}}}{K\sqrt{4\left(K_{x}^{2}+k_{y}^{2}\right)-\left(\frac{\omega+i\gamma}{v_{F}}\right)^{2}}}\right]}{\left(\frac{\omega+i\gamma}{v_{F}}\right)\sqrt{4\left(K_{x}^{2}+k_{y}^{2}\right)-\left(\frac{\omega+i\gamma}{v_{F}}\right)^{2}}}\right)
\end{align}

Here we have introduced a cutoff at $k = K$ in the integration over electron momenta  in order to regularize the divergent integral
$\int\frac{d^{3}k}{\left(2\pi\right)^{3}}$ which comes from $\frac{1}{V}\sum_{n}\rightarrow\int\frac{d^{3}k}{\left(2\pi\right)^{3}}$. The divergence is an artifact of the effective Hamiltonian Eq.~(\ref{Eq:Hamiltonian}) which has a ``bottomless''
valence band with electrons occupying all states to $k\rightarrow\infty$. The regularization
makes the 
valence band  bounded from below. We chose the cutoff at the momentum corresponding to the energy of 2 eV, i.e. much higher than the range of interest to us near the Weyl nodes. In the numerical examples in the paper the value of half-separation between Weyl nodes $\hbar v_F b$ is chosen to be 100 meV. We have verified that an exact value of the cutoff has a negligible effect on the low-energy optical response below 350 meV, as long as $K$ is large enough.  

%%%%%%%%%%%%%%%%%%%%%%%%%%%%%%

\section{Calculation of the surface electrical conductivity }

\subsection{Surface-to-surface states intraband transitions }

\begin{align}
 & \sigma_{yy}^{intra}\left(\omega\right)=g\frac{i\hbar}{S}\sum_{\mu\nu}\left(\frac{f_{\mu}-f_{\nu}}{E_{\nu}-E_{\mu}}\right)\frac{\left|\left\langle \mu\right|\hat{j}_{y}\left|\nu\right\rangle \right|^{2}}{\hbar(\omega+i\gamma)+(E_{\mu}-E_{\nu})}\nonumber \\
 & =\frac{ig\hbar e^{2}v_{F}^{2}}{S}\sum_{\mu}\left(-\frac{\text{\ensuremath{\partial}}f_{\mu}}{\text{\ensuremath{\partial}}E_{\mu}}\right)\frac{1}{\hbar(\omega+i\gamma)}
  =\Theta\left(b-k_{F}\right)\frac{ige^{2}v_{F}\sqrt{b^{2}-k_{F}^{2}}}{2\pi^{2}\hbar\left(\omega+i\gamma\right)}.
\end{align}
All other tensor components are equal to zero. 

%%%%%%%%%%%%%%%%%%%%%%%%%%%%

\subsection{Surface-to-bulk states  transitions}
\begin{align}
 & \sigma_{xx}^{inter}\left(\omega\right)=g\frac{i\hbar}{S}\sum_{s=\pm1}\sum_{m\mu}\left(\frac{f_{\mu}-f_{m\left(s\right)}}{E_{m\left(s\right)}-E_{\mu}}\right)\frac{\left|\left\langle \mu\right|\hat{j}_{x}\left|ms\right\rangle \right|^{2}}{\hbar(\omega+i\gamma)+(E_{\mu}-E_{m\left(s\right)})}\nonumber \\
 & = \frac{i4ge^{2}v_{F}^{2}\hbar}{b^{2}}\sum_{s=\pm1}\int_{\infty}\frac{d^{3}k}{\left(2\pi\right)^{3}}\Theta\left[b^{2}-\left(k_{x}^{2}+k_{y}^{2}\right)\right]\Theta\left(k_{z}\right)\nonumber \\
 & \times\left(\frac{f_{\boldsymbol{k}}^{S}-f_{\boldsymbol{k}\left(s\right)}}{E_{\boldsymbol{k}\left(s\right)}-E_{\boldsymbol{k}}^{S}}\right)\frac{k_{x}^{2}k_{z}^{2}\kappa\left(1+s\cos\theta_{\boldsymbol{k}}\right)}{(\kappa^{2}+k_{z}^{2})^{2}\left[\hbar(\omega+i\gamma)+(E_{\boldsymbol{k}}^{S}-E_{\boldsymbol{k}\left(s\right)})\right]}\nonumber \\
 & = \frac{ige^{2}}{h} \int_{0}^{\infty}dk_{z}\int_{-\infty}^{\infty}dk_{x}\int_{-\infty}^{\infty}dk_{y}\Theta\left[b^{2}-\left(k_{x}^{2}+k_{y}^{2}\right)\right]\frac{k_{z}^{2}k_{x}^{2}K_{x}}{\pi^{2}(K_{x}^{2}+k_{z}^{2})^{2}b^{2}}\nonumber \\
 & \times \left[ \frac{\Theta\left(k_{F}-\sqrt{K_{x}^{2}+k_{y}^{2}+k_{z}^{2}}\right)-\Theta\left(k_{F}+k_{y}\right)}{\sqrt{K_{x}^{2}+k_{y}^{2}+k_{z}^{2}}\left[(\frac{\omega+i\gamma}{v_{F}}-k_{y})-\sqrt{K_{x}^{2}+k_{y}^{2}+k_{z}^{2}}\right]} \right. \nonumber \\
 & \left.  -\frac{\Theta\left(-k_{F}-k_{y}\right)}{\sqrt{K_{x}^{2}+k_{y}^{2}+k_{z}^{2}}\left[(\frac{\omega+i\gamma}{v_{F}}-k_{y})+\sqrt{K_{x}^{2}+k_{y}^{2}+k_{z}^{2}}\right]} \right] 
 \label{c2} 
 \end{align}

Similarly,  
\begin{align}
 & \sigma_{yy}^{inter}\left(\omega\right) = \frac{ige^{2}}{h} \int_{0}^{\infty}dk_{z}\int_{-\infty}^{\infty}dk_{x}\int_{-\infty}^{\infty}dk_{y}\Theta\left[b^{2}-\left(k_{x}^{2}+k_{y}^{2}\right)\right]\frac{k_{z}^{2}k_{y}^{2}K_{x}}{\pi^{2}(K_{x}^{2}+k_{z}^{2})^{2}b^{2}}\nonumber \\
 & \times \left[ \frac{\Theta\left(k_{F}-\sqrt{K_{x}^{2}+k_{y}^{2}+k_{z}^{2}}\right)-\Theta\left(k_{F}+k_{y}\right)}{\sqrt{K_{x}^{2}+k_{y}^{2}+k_{z}^{2}}\left[(\frac{\omega+i\gamma}{v_{F}}-k_{y})-\sqrt{K_{x}^{2}+k_{y}^{2}+k_{z}^{2}}\right]} \right. \nonumber \\
 & \left. -\frac{\Theta\left(-k_{F}-k_{y}\right)}{\sqrt{K_{x}^{2}+k_{y}^{2}+k_{z}^{2}}\left[(\frac{\omega+i\gamma}{v_{F}}-k_{y})+\sqrt{K_{x}^{2}+k_{y}^{2}+k_{z}^{2}}\right]}\right] \label{c3} 
 \end{align}
 
 \begin{align}
 & \sigma_{zz}^{inter}\left(\omega\right) = \frac{ige^{2}}{h} \int_{0}^{\infty}dk_{z}\int_{-\infty}^{\infty}dk_{x}\int_{-\infty}^{\infty}dk_{y}\Theta\left[b^{2}-\left(k_{x}^{2}+k_{y}^{2}\right)\right]\frac{k_{z}^{2}K_{x}}{\pi^{2}(K_{x}^{2}+k_{z}^{2})^{2}}\nonumber \\
 & \times\left[ \frac{\Theta\left(k_{F}-\sqrt{K_{x}^{2}+k_{y}^{2}+k_{z}^{2}}\right)-\Theta\left(k_{F}+k_{y}\right)}{\sqrt{K_{x}^{2}+k_{y}^{2}+k_{z}^{2}}\left[(\frac{\omega+i\gamma}{v_{F}}-k_{y})-\sqrt{K_{x}^{2}+k_{y}^{2}+k_{z}^{2}}\right]}\right. \nonumber \\
 & \left. -\frac{\Theta\left(-k_{F}-k_{y}\right)}{\sqrt{K_{x}^{2}+k_{y}^{2}+k_{z}^{2}}\left[(\frac{\omega+i\gamma}{v_{F}}-k_{y})+\sqrt{K_{x}^{2}+k_{y}^{2}+k_{z}^{2}}\right]}\right].
 \label{c4} 
\end{align}

The only nonzero off-diagonal element is 
\begin{align}
 & \sigma_{yz}^{inter}\left(\omega\right) =\frac{-ge^{2}}{h} \int_{0}^{\infty}dk_{z}\int_{-\infty}^{\infty}dk_{x}\int_{-\infty}^{\infty}dk_{y}\Theta\left[b^{2}-\left(k_{x}^{2}+k_{y}^{2}\right)\right]\frac{k_{z}^{2}k_{y}K_{x}}{\pi^{2}(K_{x}^{2}+k_{z}^{2})^{2}b}\nonumber \\
 & \times \left[ \frac{\Theta\left(k_{F}-\sqrt{K_{x}^{2}+k_{y}^{2}+k_{z}^{2}}\right)-\Theta\left(k_{F}+k_{y}\right)}{\sqrt{K_{x}^{2}+k_{y}^{2}+k_{z}^{2}}\left[(\frac{\omega+i\gamma}{v_{F}}-k_{y})-\sqrt{K_{x}^{2}+k_{y}^{2}+k_{z}^{2}}\right]} \right. \nonumber \\
 & \left. -\frac{\Theta\left(-k_{F}-k_{y}\right)}{\sqrt{K_{x}^{2}+k_{y}^{2}+k_{z}^{2}}\left[(\frac{\omega+i\gamma}{v_{F}}-k_{y})+\sqrt{K_{x}^{2}+k_{y}^{2}+k_{z}^{2}}\right]} \right].
 \label{c5} 
\end{align}

In Eqs.~(\ref{c2})-(\ref{c5}) the integral over $k_z$ can be carried out analytically in terms of elementary functions, leading however to very lengthy expressions which we do not present here. The remaining integration was carried out numerically. All integrals are finite, i.e. no cutoff is necessary. 

%%%%%%%%%%%%%%%%%%%%%%%%%%%

\section{Drude-like low-frequency limit }

 In the limit when the frequency and the Fermi energy are much smaller than $\hbar v_F b$, only the electron momenta close to the corresponding Weyl point $k_x = \pm b$ matter. Therefore, 
we introduce $\delta k_{x}=k_{x}-b$ for electron states near one Weyl point and replace the degeneracy factor
by $2\times g$ to account for the contribution from the second Weyl point. In this case,
$K_{x}\sim\frac{\left(k_{x}-b\right)\left(k_{x}+b\right)}{2b}\approx\delta k_{x},\,k_{x}$$=b+\delta k_{x},$
and all diagonal intraband components have the same Drude form: 
\begin{equation}
\sigma_{xx}^{intra}\left(\omega\right)=\sigma_{yy}^{intra}\left(\omega\right)=\sigma_{zz}^{intra}\left(\omega\right)=\frac{ge^{2}v_{F}k_{F}^{2}}{3\pi^{2}\hbar(-i\omega+\gamma)}.
\end{equation}
All off-diagonal conductivity elements are zero in this limit. 

%%%%%%%%%%%%%%%%%%%%%%%%%%%%

\section{Small $b$ Expansion}

In the limit $b\ll1$, we can expand the conductivity in powers of
$b$ to the leading order: $b\ll1$, $\frac{1}{b}\gg1,$ $K_{x}=\frac{\left(k_{x}^{2}+k_{y}^{2}\right)-b^{2}}{2b}\sim\frac{\left(k_{x}^{2}+k_{y}^{2}\right)}{2b}\sim\frac{\left(k_{x}^{2}+k_{y}^{2}+k_{z}^{2}\right)}{2b}\gg k_{x,y,z},\frac{\omega}{v_{F}}$
for $k_{x,y,z}\neq0$. Then we obtain 
\begin{align}
 & \sigma_{yz}^{B}\left(\omega\right) \approx \frac{-ge^{2}}{3\sqrt{2}\pi^{2}\hbar} \frac{b^{3/2}}{k_F^{1/2}} 
 \label{syzB} 
\end{align}

\begin{align}
 & \sigma_{xx}^{B}\left(\omega\right)\approx\frac{ge^{2}k_{F}^{2}v_{F}}{3\pi^{2}\hbar(-i\omega+\gamma)}+\frac{2\sqrt{2}ge^{2}(-i\omega+\gamma)}{45\pi^{2}\hbar v_{F}} \frac{b^{3/2}}{k_F^{3/2}}
\end{align}

\begin{align}
 & \sigma_{yy}^{B}\left(\omega\right)\approx\frac{ge^{2}k_{F}^{2}v_{F}}{3\pi^{2}\hbar(-i\omega+\gamma)}+\frac{7\sqrt{2}ge^{2}(-i\omega+\gamma)}{360\pi^{2}\hbar v_{F}} \frac{b^{3/2}}{k_F^{3/2}}
\end{align}

\begin{align}
\sigma_{zz}^{B}\left(\omega\right)\approx\frac{ge^{2}k_{F}^{2}v_{F}}{3\pi^{2}\hbar(-i\omega+\gamma)}+\frac{ge^{2}(-i\omega+\gamma)}{6\sqrt{2}\pi^{2}\hbar v_{F}} \frac{b^{3/2}}{k_F^{3/2}}
\end{align}

\begin{align}
 & \sigma_{xx}^{S}\left(\omega\right)=\sigma_{yy}^{S}\left(\omega\right)=\sigma_{zz}^{S}\left(\omega\right)\approx\frac{ge^{2}v_{F}}{2\sqrt{2k_{F}}\pi^{3}\hbar(-i\omega+\gamma)}b^{\frac{3}{2}}.
\end{align}

All off-diagonal surface terms are zero.

\section{Reflection in the vicinity of plasmon resonance}

For oblique incidence  $   \theta  \neq 0 $ and small losses the calculations of the reflection in the vicinity of plasmon resonance have a technical subtlety, related to the presence of the term  $ n_{X}\cos \theta  \left( \cos \theta _{X}-\sin \theta _{X}K_{X} \right)  $  in Eq.~(\ref{X-modeReflection}). Indeed, at the plasmon frequency  $ n_{X} \rightarrow \infty $ as losses $\gamma \rightarrow 0$; however, for a plasmon we also have   $ K_{X} \rightarrow \frac{1}{\tan \theta _{X}} $, i.e. $  \left( \cos \theta _{X}-\sin \theta _{X}K_{X} \right)  \rightarrow 0 $. One needs to treat the resulting uncertainty of the product with caution.

We substitute the relationship  $\sin \theta _{X}=\frac{n_{up}\sin \theta }{n_{X}} $  into the expression for the refractive index of an extraordinary wave:
\[
n_{X}^{2}=\frac{ \varepsilon _{yy} \varepsilon _{zz}-g^{2}}{\cos^{2} \theta _{X} \varepsilon _{zz}+\sin^{2} \theta _{X} \varepsilon _{yy}}=\frac{ \varepsilon _{yy} \varepsilon _{zz}-g^{2}}{ \varepsilon _{zz}-\sin^{2} \theta  \left( \frac{n_{up}}{n_{X}} \right) ^{2} \left(  \varepsilon _{zz }-  \varepsilon _{yy} \right) },\]
which gives
\begin{equation}\label{ObliqueXmodeRefraction}
n_{X}^{2}= \varepsilon _{yy}-\frac{g^{2}}{ \varepsilon _{zz}}+\sin^{2} \theta n_{up}^{2} \left( 1-\frac{ \varepsilon _{yy}}{ \varepsilon _{zz}} \right)
\end{equation}
In the case  $  \varepsilon _{yy}= \varepsilon _{zz}= \varepsilon _{\perp} $, Eq.~(\ref{ObliqueXmodeRefraction}) for an arbitrary angle  $  \theta  $  leads to the familiar expression  $ n_{X}^{2}= \varepsilon _{\perp}-\frac{g^{2}}{ \varepsilon _{\perp}} $.
Next we use Eq.~(\ref{PolarizationCoefficientKX}):
\[
K_{X}=\frac{ig-n_{X}^{2}\sin \theta _{X}\cos \theta _{X}}{ \varepsilon _{zz}-n_{X}^{2}\sin^{2} \theta _{X}}=\frac{ig-n_{up}\sin \theta n_{X}\sqrt{1- \left( \frac{\sin \theta n_{up}}{n_{X}} \right) ^{2}}}{ \varepsilon _{zz}-\sin^{2} \theta n_{up}^{2}}. 
\]
Consider the expression $n_{X}\cos \theta  \left( \cos \theta _{X}-\sin \theta _{X}K_{X} \right)  $:
\begin{align} 
n_{X}\cos \theta  &\left( \cos \theta _{X}-\sin \theta _{X}K_{X} \right)\nonumber\\
&=n_{X}\cos \theta  \left( \cos \theta _{X}-\frac{ig\sin \theta _{X}-\sin \theta _{X}n_{up}\sin \theta n_{X}\sqrt{1- \left( \frac{\sin \theta n_{up}}{n_{X}} \right) ^{2}}}{ \varepsilon _{zz}-\sin^{2} \theta n_{up}^{2}} \right)\nonumber\\
&=n_{X}\cos \theta  \left( \sqrt{1- \left( \frac{\sin \theta n_{up}}{n_{X}} \right) ^{2}}-\frac{ig\frac{\sin \theta n_{up}}{n_{X}}-\sin^{2} \theta n_{up}^{2}\sqrt{1- \left( \frac{\sin \theta n_{up}}{n_{X}} \right) ^{2}}}{ \varepsilon _{zz}-\sin^{2} \theta n_{up}^{2}} \right).  \nonumber
\end{align}
The condition  $ \frac{n_{X}}{n_{up}} \gg 1 $, which is satisfied at the plasmon frequency, allows one to simplify the above expressions for any angle of incidence  $  \theta  $
\begin{equation}\label{PolarizationCoefficientKXAtPlasmonFrequency}
K_{X}=\frac{ig-n_{X}^{2}\sin \theta _{X}\cos \theta _{X}}{ \varepsilon _{zz}-n_{X}^{2}\sin^{2} \theta _{X}} \approx \frac{ig-n_{X}n_{up}\sin \theta }{ \varepsilon _{zz}-\sin^{2} \theta n_{up}^{2}}
\end{equation}

\begin{equation}\label{X-modeRefractionAtPlasmonFrequency}
n_{X}cos \theta  \left( \cos \theta _{X}-\sin \theta _{X}K_{X} \right)  \approx n_{X}\cos \theta  \left( 1-\frac{ig\frac{\sin \theta n_{up}}{n_{X}}-\sin^{2} \theta n_{up}^{2}}{ \varepsilon _{zz}-\sin^{2} \theta n_{up}^{2}} \right)  
\end{equation}

Since for  $ \frac{n_{X}}{n_{up}} \gg 1 $  we always have  $ \sin \theta _{X} \ll 1 $, the plasmon frequency always corresponds to  $  \vert  \varepsilon _{zz} \vert  \ll 1 $  (at normal incidence,  $  \varepsilon _{zz}=0  $ exactly). Taking into account Eq.~(\ref{ObliqueXmodeRefraction}), we obtain  $ 1 \gg  \vert  \varepsilon _{zz} \vert  \sim n_{X}^{-2} $.

Now let us consider the range of incidence angles close to normal incidence, when  $ \sin^{2} \theta  \ll 1 $. Two cases need to be treated separately:  $  \vert  \varepsilon _{zz} \vert  \ll \sin^{2} \theta n_{up}^{2} \ll 1 $  and  $ \sin^{2} \theta n_{up}^{2} \ll  \vert  \varepsilon _{zz} \vert  \ll 1 $.  \newline{}
\textbf{(i)}  $  \vert  \varepsilon _{zz} \vert  \ll sin^{2} \theta n_{up}^{2} \ll 1 $\newline{}
In this case
\begin{equation}\label{nXandKXforThetaGreatherThanEpsilonzz}
n_{X}^{2} \approx  \varepsilon _{yy}-\frac{g^{2}}{ \varepsilon _{zz}},\quad  K_{X} \approx \frac{n_{X}}{n_{up}sin \theta }
\end{equation}

\begin{equation}\label{nXandKXForThetaGreatherThanEpsilonzzAlt}
n_{X}\cos \theta  \left( 1-\frac{ig\frac{\sin \theta n_{up}}{n_{X}}-\sin^{2} \theta n_{up}^{2}}{ \varepsilon _{zz}-\sin^{2} \theta n_{up}^{2}} \right)  \approx \frac{ig}{sin \theta n_{up}}
\end{equation}
where  $ g=\frac{4 \pi  \sigma _{yz}^{B}}{ \omega } $,
\begin{equation}\label{X-modeReflectionForThetaGreatherThanEpsilonzz}
R \approx \frac{n_{up}^{2}\sin \theta  - i\frac{4 \pi  \sigma _{yz}^{B}}{ \omega }+\frac{4 \pi }{c} \sigma _{yz}^{S}n_{X}}{n_{up}^{2}\sin \theta  + i\frac{4 \pi  \sigma _{yz}^{B}}{ \omega }+\frac{4 \pi }{c} \sigma _{yz}^{S}n_{X}}. 
\end{equation}
For real  $  \sigma _{yz}^{ \left( B,S \right) } $  we always have  $  \vert R \vert =1 $; however, the phase of the reflected field depends on the contribution of surface states. Since in the vicinity of plasmon resonance  $ n_{X} \sim \frac{1}{\sqrt{ \vert  \varepsilon _{zz} \vert }} \gg 1 $, at these frequencies the contribution of surface states may become important. This is especially clear in the limit of small enough angles, when  $ n_{up}^{2}sin \theta   \ll  \vert \frac{4 \pi  \sigma _{yz}^{B}}{ \omega } \vert  $. In this case
\begin{equation}\label{X-modeReflectionSmallAngleSigmaYZDominant}
R \approx \frac{- i\frac{4 \pi  \sigma _{yz}^{B}}{ \omega }+\frac{4 \pi }{c} \sigma _{yz}^{S}n_{X}}{+ i\frac{4 \pi  \sigma _{yz}^{B}}{ \omega }+\frac{4 \pi }{c} \sigma _{yz}^{S}n_{X}}.
\end{equation}\par

When the bulk contribution dominates we have  $ R=-1 $, whereas if the surface contribution dominates we obtain  $ R=+1 $, i.e. the phase of the reflected field flips.

The relative contribution of surface states is determined by the ratio  $ \frac{ \vert  \sigma _{yz}^{S}n_{X} \vert }{\frac{c}{ \omega } \vert  \sigma _{yz}^{B} \vert } $. Taking into account that  $  \vert n_{X} \vert  \approx \frac{ \vert g \vert }{\sqrt{ \vert  \varepsilon _{zz} \vert }} $  and  $  \vert g \vert =\frac{4 \pi  \vert  \sigma _{yz}^{B} \vert }{ \omega } $, the above ratio can be reduced to  $ \frac{\frac{4 \pi  \vert  \sigma _{yz}^{S} \vert }{c}}{\sqrt{ \vert  \varepsilon _{zz} \vert }} $.\newline{}

\textbf{(ii)}  $ \sin^{2} \theta n_{up}^{2} \ll  \vert  \varepsilon _{zz} \vert  \ll 1 $\newline{}
This case is similar to the one at  $  \theta =0 $. Indeed, for this range of parameters we obtain
\begin{equation}\label{nXandKXForEpsilonzzGreatherThanTheta}
n_{X}^{2} \approx  \varepsilon _{yy}-\frac{g^{2}}{ \varepsilon _{zz}},\quad  K_{X} \approx \frac{ig}{ \varepsilon _{zz}}
\end{equation}

\begin{equation}\label{nXandKXForEpsilonzzGreatherThanThetaAlt}
n_{X}cos \theta  \left( 1-\frac{ig\frac{sin \theta n_{up}}{n_{X}}-sin^{2} \theta n_{up}^{2}}{ \varepsilon _{zz}} \right)  \approx n_{X}.
\end{equation}

\begin{equation}\label{X-modeReflectionForEpsilonzzGreatherThanTheta}
R \approx \frac{-n_{X}+\frac{4 \pi }{c} \sigma _{yz}^{S}\frac{ig}{ \varepsilon _{zz}}}{n_{X}+\frac{4 \pi }{c} \sigma _{yz}^{S}\frac{ig}{ \varepsilon _{zz}}} 
\end{equation}
Eqs.~(\ref{nXandKXForEpsilonzzGreatherThanTheta}), (\ref{nXandKXForEpsilonzzGreatherThanThetaAlt}) are the same as for the normal incidence. Eq.~(\ref{X-modeReflectionForEpsilonzzGreatherThanTheta}) can be obtained from the normal incidence formula Eq. (\ref{X-modeReflectionNormalIncidence}) if  $  \vert  \sigma _{yy}^{S} \vert  \ll  \vert  \sigma _{yz}^{S}\frac{g}{ \varepsilon _{zz}} \vert  $  and  $ n_{X} \gg n_{up} $; the latter inequalities are valid near the plasmon resonance, where  $ n_{X} \sim \frac{1}{\sqrt[]{ \vert  \varepsilon _{zz} \vert }} \rightarrow \infty $.

For real values of  $\sigma _{yz}^{ \left( S \right) }$  we always have  $  \vert R \vert =1 $, but the phase of the reflected field depends on the contribution of surface states. Again, when the bulk contribution dominates we have  $ R=-1 $, whereas if the surface contribution dominates we obtain  $ R=+1 $.

The relative contribution of surface states is determined by the ratio $ \frac{\frac{4 \pi }{c} \vert  \sigma _{yz}^{S}\frac{g}{ \varepsilon _{zz}} \vert }{ \vert n_{X} \vert  } $. Again taking into account $  \vert n_{X} \vert  \approx \frac{ \vert g \vert }{\sqrt{ \vert  \varepsilon _{zz} \vert }} $ and $  \vert g \vert =\frac{4 \pi  \vert  \sigma _{yz}^{B} \vert }{ \omega } $ we obtain that the above ratio is reduced to exactly the same expression as before: $ \frac{4 \pi  \vert  \sigma _{yz}^{S} \vert /c}{\sqrt{ \vert  \varepsilon _{zz} \vert }} $.

To summarize, the effect of surface states on the reflected wave is determined by the ratio 
\[\frac{\vert  \sigma _{yz}^{S} \vert}{c\sqrt{ \vert  \varepsilon _{zz} \vert }/4 \pi  }\]
and therefore becomes significant or dominant  at the plasmon resonance frequency, when $  \varepsilon _{zz}= \varepsilon _{zz}^{ \left( 0 \right) }+i\frac{4 \pi }{ \omega } \sigma _{zz}^{B} \rightarrow 0 $.
%%%
%%%

\end{document}